\documentclass[10pt,twocolumn,letterpaper]{article}

\usepackage[pagenumbers]{cvpr} 

\usepackage{todonotes}
\usepackage{booktabs}
\usepackage{array}
\newcolumntype{C}[1]{>{\centering\arraybackslash}p{#1}}
\definecolor{cvprblue}{rgb}{0.21,0.49,0.74}
\usepackage[pagebackref,breaklinks,colorlinks,allcolors=cvprblue]{hyperref}

\def\paperID{*****} 
\def\confName{CVPR}
\def\confYear{2026}

\title{Neural Dynamic GI: Random-Access Neural Compression for Temporal Lightmaps in Dynamic Lighting Environments}

\author{
Jianhui Wu$^{1}$\quad
Jian Zhou$^{1}$\quad
Zhi Zhou$^{1}$\quad
Zhangjin Huang$^{1}$\thanks{Corresponding authors.}\quad
Chao Li$^{2}$\footnotemark[1]\\[4pt]
$^{1}$University of Science and Technology of China\quad
$^{2}$Zhejiang University\\[4pt]
\url{https://magicdawnlab.github.io/}
}

\begin{document}
\twocolumn[{
  \renewcommand\twocolumn[1][]{#1}
  \maketitle
  \vspace{-20pt}
  \begin{center}
    \captionsetup{type=figure}
    \includegraphics[width=1.0\linewidth]{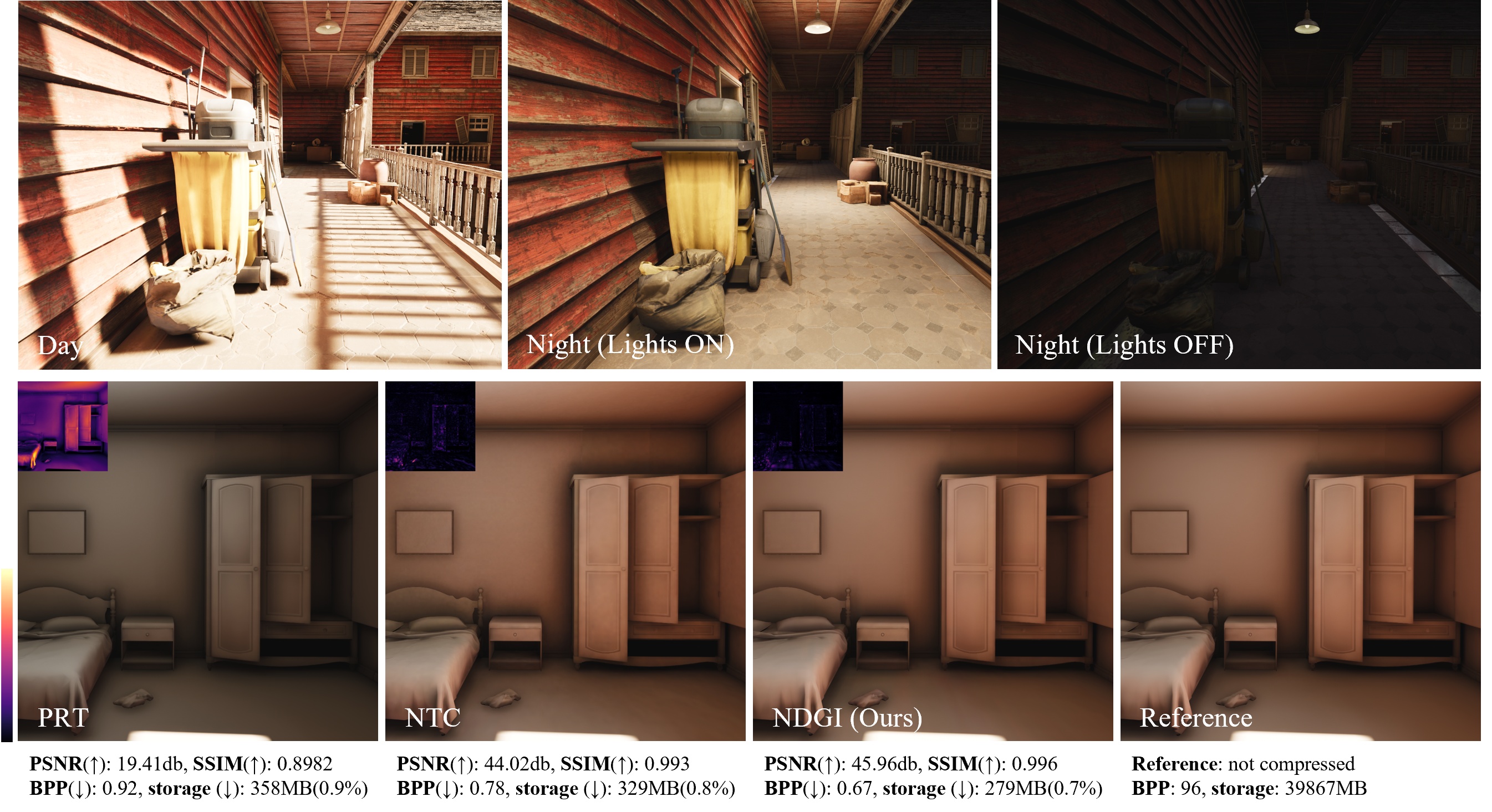}
    \captionof{figure}{We evaluate performance in the ``FarmLand" scene, which is derived from a real video game. The top row shows GI effects at different times. The bottom row presents lighting reconstructed by PRT~\cite{sloan2023precomputed}, NTC~\cite{Vaidyanathan_2023}, and our method. To enable a clearer comparison, textures are removed to emphasize illumination effects. The top-left part of the image shows the error relative to the reference. We report bits per pixel~(BPP) and compression ratio as evaluation metrics. Compared with PRT and NTC, our method achieves a higher compression ratio and superior lighting quality, while incurring lower real-time computational overhead and less noise than NTC.}
    \label{fig:head}
  \end{center}
}]

\begin{abstract}
High-quality global illumination~(GI) in real-time rendering is commonly achieved using precomputed lighting techniques, with lightmap as the standard choice. To support GI for static objects in dynamic lighting environments, multiple lightmaps at different lighting conditions need to be precomputed, which incurs substantial storage and memory overhead.

To overcome this limitation, we propose Neural Dynamic GI~(NDGI), a novel compression technique specifically designed for temporal lightmap sets. Our method utilizes multi-dimensional feature maps and lightweight neural networks to integrate the temporal information instead of storing multiple sets explicitly, which significantly reduces the storage size of lightmaps. Additionally, we introduce a block compression~(BC) simulation strategy during the training process, which enables BC compression on the final generated feature maps and further improves the compression ratio. To enable efficient real-time decompression, we also integrate a virtual texturing~(VT) system with our neural representation.  

Compared with prior methods, our approach achieves high-quality dynamic GI while maintaining remarkably low storage and memory requirements, with only modest real-time decompression overhead. To facilitate further research in this direction, we will release our temporal lightmap dataset precomputed in multiple scenes featuring diverse temporal variations.
\end{abstract}   
\section{Introduction}
\label{sec:intro}

Real-time rendering quality has advanced rapidly in recent years, driven by progress in global illumination (GI) algorithms including voxel-based approaches~\cite{villegas2016deferred}, real-time ray tracing~\cite{10.1145/3532720.3546895,ouyang2021restir}, and denoising methods~\cite{xing2020path}. Among various real-time GI solutions, precomputed lighting~\cite{10.1111/j.1467-8659.2012.02093.x} remains widely adopted due to its stable performance and scalability across hardware. In this paradigm, indirect illumination, static shadows, and selected direct lighting components are precomputed for static geometry and stored in structured forms such as lightmaps or spherical harmonic~(SH) probes. At runtime, these precomputed results enable efficient reproduction of complex lighting with minimal computational cost. In particular, lightmap-based techniques~\cite{10.1145/3283254.3283281, epic_lm} offer higher spatial fidelity than probe-based methods~\cite{epic_vlm}, capturing detailed surface illumination at fine resolutions.

\begin{figure}[!t]
\centering
\setlength{\tabcolsep}{2pt}
\begin{tabular}{cccc}
\includegraphics[width=0.24\linewidth]{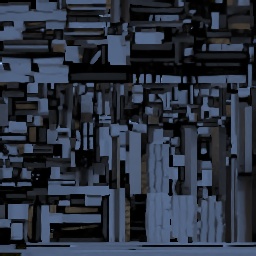} &
\includegraphics[width=0.24\linewidth]{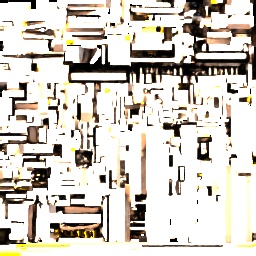} &
\includegraphics[width=0.24\linewidth]{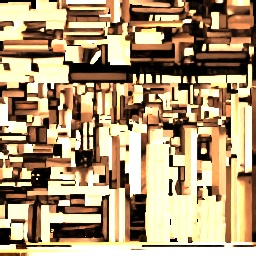} &
\includegraphics[width=0.24\linewidth]{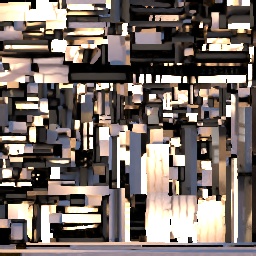} \\
{\scriptsize morning} & {\scriptsize noon} & {\scriptsize dusk} & {\scriptsize night} \\
\end{tabular}
\caption{Examples of lightmaps under different lighting conditions. During night scenes, certain local light sources are enabled, resulting in brighter illumination recorded in the lightmap.}
\label{fig:lmexample}
\end{figure}

However, extending lightmap-based techniques to dynamic lighting environments remains challenging. Realistic time-of-day~(TOD) effects or dynamic light changes require precomputing multiple sets of lightmaps corresponding to different lighting conditions~(Figure~\ref{fig:lmexample}), followed by real-time interpolation. This design leads to prohibitively large storage and memory consumption, making it unsuitable for large-scale scenes. As scene complexity and texture resolution continue to grow, efficient compression becomes crucial to enabling dynamic GI at scale.

Conventional GPU texture compression methods~\cite{10.5555/2383795.2383812, DXTC} are hardware efficient and support random access. However, maintaining quality often requires relatively high bitrates and leads to block artifacts, and processing textures independently fails to exploit temporal redundancy across multiple lightmap sets. Recent advances in neural compression~\cite{Vaidyanathan_2023} improve compression ratios and reconstruction quality, but most approaches rely on large decoders that hinder real-time performance. To our knowledge, no prior work addresses compression specifically for multiple lightmap sets, nor applies such compression to dynamic GI.

To address these challenges, we propose Neural Dynamic GI~(NDGI)—a neural compression framework specifically designed for temporal lightmap sets. Instead of storing multiple lightmap sets explicitly, NDGI encodes temporal illumination into hybrid feature maps and reconstructs lighting values on-demand through a lightweight neural decoder. During training, we introduce a block compression (BC) simulation strategy to ensure that the learned feature maps are compatible with standard GPU BC formats, enabling further compression without visible quality degradation. To make the system practical for real-time rendering, we adopt a tile-based training paradigm that integrates NDGI with a virtual texturing~(VT)~\cite{vt} system. The VT framework manages texture streaming and caching at the tile level, enabling on-demand loading and high-speed decompression during rendering. This design reduces both storage and memory simultaneously while maintaining high-quality dynamic GI with negligible runtime overhead.

Moreover, to support future research in this direction, we construct and release a temporal lightmap dataset precomputed across multiple scenes. These scenes include various temporal variations, such as gradual changes in sunlight and skylight, as well as the switching of local light sources. The dataset provides a practical benchmark for studying dynamic illumination compression and serves as a valuable resource for developing future neural rendering systems.

In summary, our main contributions are:
\begin{itemize}
\item We propose Neural Dynamic GI~(NDGI), the first neural compression framework tailored for temporal lightmap sets, achieving high-quality dynamic GI with extremely low storage and memory costs.
\item We introduce a BC simulation strategy that ensures compatibility with standard BC formats, enabling further compression without sacrificing quality.
\item We design a virtual texturing-integrated neural decoding system that supports on-demand decompression for real-time rendering.
\item We provide a temporal lightmap dataset comprising multiple scenes with diverse temporal variations to support future research in dynamic GI compression.
\end{itemize}

\section{Previous Work}
\subsection{Precomputed Lighting}

Precomputed lighting computes illumination in an offline stage and stores it in compact representations for efficient reuse. During rendering, lighting reconstruction requires only sampling and lightweight computations, which makes this approach widely adopted in real-time applications.

Lightmaps are one of the most common carriers for precomputed lighting. Research on lightmaps primarily focuses on packing, and compression~\cite{10.1145/3617683} in order to improve storage efficiency and reduce runtime memory usage. The Directional Lightmap technique~\cite{10.1145/3283254.3283281} extends the conventional formulation by storing the dominant incident lighting direction alongside irradiance values, which introduces view-dependent effects and increases storage cost. Precomputed Radiance Transfer~(PRT) methods~\cite{sloan2023precomputed} enable dynamic lighting by storing transfer data that describe how outgoing light responds to incident light. Such transfer data can be encoded into lightmaps or represented within probes depending on the desired trade-off between accuracy and storage efficiency.

\subsection{Traditional Texture Compression}

Traditional texture compression primarily relies on block-based encoding. The idea dates back to Block Truncation Coding~(BTC)~\cite{1094560} for grayscale images, followed by extensions that support color images and efficient hardware decoding. For example, Campbell~\etal~\cite{10.1145/15886.15910} proposed Color Cell Compression~(CCC), and Knittel~\etal~\cite{knittel1996hardware} enhanced it with GPU‑accelerated decompression. DirectX Texture Compression~(DXTC)~\cite{DXTC} spans BC1 through BC7 and stores quantized color endpoints with per block interpolation weights, providing a practical balance between quality, bitrate, and decoding efficiency, with variants such as BC5 for normal maps and BC6H for high dynamic range~(HDR) content. Adaptive Scalable Texture Compression~(ASTC)~\cite{10.5555/2383795.2383812} offers flexible bitrate configurations and broad texture support including 3D and HDR textures and is widely adopted on modern mobile platforms. The Ericsson Texture Compression family provides another mobile‑oriented branch, where ETC1~\cite{10.1145/1071866.1071877} and ETC2~\cite{:10.2312/EGGH/EGGH07/049-054} achieve efficient compression with backward compatibility and low decoding cost.

Despite their widespread use, these block‑based approaches yield modest compression ratios and lack mechanisms for multi‑lightmap or temporally varying data required by dynamic global illumination. Our method addresses these limitations by enabling high‑quality, temporally coherent lightmap compression with significantly reduced storage requirements.

\subsection{Neural Texture Compression}

Neural texture compression methods have emerged as promising alternatives to traditional compression formats. These approaches provide a flexible trade-off among reconstruction quality, storage cost, and computational overhead. Instant NGP~\cite{10.1145/3528223.3530127} introduced hash encoding that uses multi-resolution grids to index hash tables for feature retrieval and has been adapted for large-scale image compression to achieve very low bitrates. Another line of work~\cite{Vaidyanathan_2023} employs feature pyramid to encode multi-channel textures across multiple mipmap levels. This approach supports random access at arbitrary positions and resolutions and also enables real-time decompression. Furthermore, Belcour~\cite{laurent2025hardware} and Weinreich~\cite{weinreich2024real} integrate traditional block compression into feature maps. By directly optimizing endpoints and weights, they achieve additional bitrate reduction with minimal quality loss. However, neural network based approaches often incur substantial decoding costs that hinder runtime performance, while our method attains high‑fidelity compression under minimal runtime overhead.

\begin{figure*}[t]
  \centering
    \includegraphics[width=\linewidth]{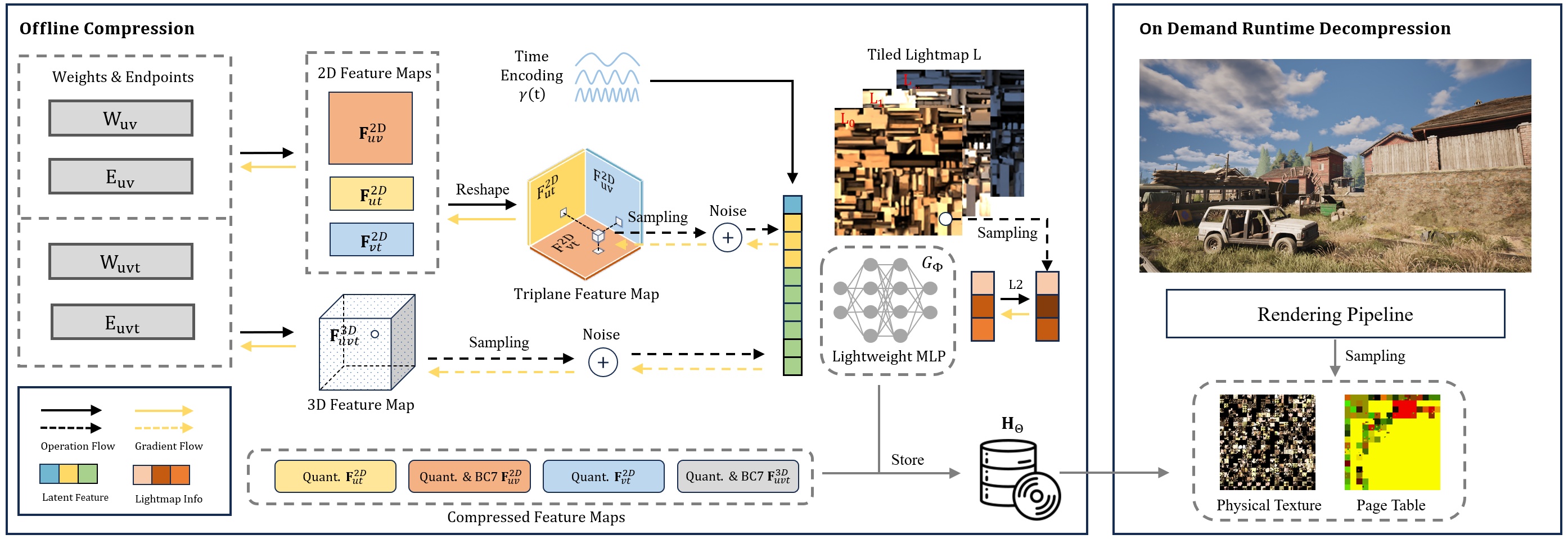}
    \label{fig:short-a}
  \hfill
  \caption{Method overview. Our method samples feature maps of different structures and feeds the features together with encoded time into an MLP to reconstruct the corresponding lighting~(Section~\ref{sec:hybrid feature map}). During training, we introduce random noise to simulate the quantization and use weights and endpoints to emulate BC compression~(Section~\ref{sec:quantization}). This approach allows us to further compress the feature map using quantization and BC7~\cite{DXTC}. The compressed feature maps and the MLP serve as the final parameters of our model. At runtime, they are streamed and decompressed on demand via a virtual texturing system to deliver high-quality dynamic global illumination~(Section~\ref{sec:decompression}).}
  \label{fig:method}
\end{figure*}

\section{Method}
\label{sec:method}

\subsection{Problem Statement}
Our compression target consists of multiple sets of lightmaps, which can be represented as: 
\begin{equation}
L=\{L_i|i=0,1,2..., n-1\},
\label{eq:eq1}
\end{equation}
where $n$ refers to the number of lightmap sets, and each $L_i$ is a multi‑channel lightmap of size $(H, W, C)$, representing lighting information at time $t_i$. For simplicity, we will assume there is only one lightmap in each set. Specifically, depending on the requirements of the scene, the distribution of time points $t_i$ may be non-uniform. For example, when a scene requires a rapid change in lighting at a certain moment (a pre-set switching event of a light source), multiple sets of lightmap may be computed around that time. 

In traditional methods, dynamic GI is achieved via runtime interpolation. The lighting $I$ at a specific texture coordinates $(u, v)$ and time $t\in(t_{i-1}, t_i)$ can be obtained as:

\begin{equation}
I(u, v, t) = \mathcal{I}_{[t_{i-1},t_i]}[\,L_{i-1}(u, v), L_{i}(u, v)\,](t).
\label{eq:eq2}
\end{equation}
Here, $\mathcal{I}$ denotes a general interpolation operator that estimates $I(u, v, t)$ based on $L_{i-1}(u, v)$ and $L_{i}(u, v)$. Depending on the specific configuration, $\mathcal{I}$ can represent different interpolation schemes (\eg linear, cubic, or nearest sampling). To implement the above interpolation algorithm, existing methods require substantial disk space and runtime memory to store the lightmap sets $L$, which is prohibitively expensive for performance-critical video games. 

To address this issue, our approach avoids directly storing $L$ by representing it as compact model $\mathbf{H}$ with parameter $\Theta$, such that the lighting can be directly represented as:
\begin{equation}
I(u, v, t) = \mathbf{H}_{\Theta}(u, v, t),
\label{eq:eq3}
\end{equation}
Furthermore, we employ several optimization techniques to achieve a balance among storage space, runtime memory, and real-time decompression performance, thereby enabling practical dynamic GI in real-time rendering. The details of our method will be introduced in subsequent sections.

\subsection{Method Overview}
Figure~\ref{fig:method} illustrates the overall framework of our approach. Given a set of precomputed lightmaps under varying lighting conditions, our method compresses them into a compact set of feature maps and a lightweight MLP for efficient decoding. At runtime, given specific texture coordinates and time $(u, v, t)$, the feature maps are sampled to obtain a latent vector, which is concatenated with a position-encoded time embedding and fed into the MLP to reconstruct lighting. The reconstructed results are compared with the original lightmaps during training to jointly optimize both the feature maps and the network.

To further enhance storage efficiency, we incorporate a block compression~(BC) simulation during training, enabling compatibility with hardware-supported BC formats. Noise is added to the latent vector to mitigate the impact of subsequent feature map quantization. In addition, a virtual texturing (VT)~\cite{vt} system is integrated at runtime to progressively stream lightmap regions on demand, significantly reducing decoding overhead and memory usage.

\begin{table}[t]
  \centering
  \caption{ Compressed representation of feature maps for a specific lightmap set comprising 26 lightmaps at $128 \times 128$ resolution.
 }
  \label{tab:compression_1}
  {\renewcommand{\arraystretch}{1.15}%
  { \small
    \begin{tabular}{c|c|c}
    \hline
    Feature Map & Resolution & Channels \\
    \midrule
    $\mathbf{F}^{2D}_{uv}$ & $128 \times 128$ & 4 \\
    $\mathbf{F}^{2D}_{ut}$ & $64 \times 24$ & 2 \\
    $\mathbf{F}^{2D}_{vt}$ & $64 \times 24$ & 2 \\
    $\mathbf{F}^{3D}_{uvt}$ & $32 \times 32 \times 12$ & 4 \\
    \hline
  \end{tabular}
  }
  }
\end{table}

\subsection{Hybrid Feature Map}
\label{sec:hybrid feature map}
Latent feature maps are commonly used in neural compression to store implicit information. To preserve the common spatial signal across the target lightmap set, we employ a 2D feature map $\mathbf{F}_{uv}^{2D}$ that has the same size as the target lightmap. Moreover, although lightmaps at different time points exhibit correlation, the variation is high frequency and more complex than those in material textures. As a result, a single 2D feature map is insufficient to represent temporal variation and leaves preserving temporal information completely to the lightweight MLP decoder. 

To address this challenge, we propose a hybrid feature map structure. We introduce an additional 3D feature map $\mathbf{F}^{3D}_{uvt}$ at a lower resolution to capture the luminance variations of the lightmap over time. Moreover, we extend the 2D feature map into triplane feature maps, denoted as $\mathbf{F}^{2D}_{uv}$, $\mathbf{F}^{2D}_{ut}$ and $\mathbf{F}^{2D}_{vt}$, to better capture fine details. During inference, we first sample $\mathbf{F}^{3D}_{uvt}$ using $(u, v, t)$ coordinates to get vector $V_{uvt}$. Then we project the coordinates onto three distinct planes, sample accordingly to get three feature vectors $V_{uv}, V_{ut}, V_{vt}$. Finally, we concatenate the four vectors with positional encoded time $\gamma(t)$ and input them into the decoder network $G_\Phi$, with parameter $\Phi$, to obtain the decompressed result $I(u, v, t)$:

{\small
\begin{equation}
\begin{aligned}
    \gamma(t) &= [\,\sin(2^0\pi t),\ \cos(2^0\pi t),\ \sin(2^1\pi t),\ \cos(2^1\pi t)\,], \\
    I(u,  & v,t)= G_\Phi(V_{uvt}, V_{uv}, V_{ut}, V_{vt}, \gamma(t)).
\end{aligned}
\label{eq:eq4}
\end{equation}
}
Decoder network $G_\Phi$ is a lightweight MLP, whose size can be adjusted to meet practical requirements and accommodate different hardware capabilities. In summary, the original lightmap set $L$ is compressed into the following parameter set $\Theta$:

\begin{equation}
    L \rightarrow \Theta = \{\mathbf{F}^{3D}_{uvt}, \mathbf{F}^{2D}_{uv}, \mathbf{F}^{2D}_{ut}, \mathbf{F}^{2D}_{vt}, \Phi\}.
\label{eq:eq5}
\end{equation}

This design enables the model to capture high-frequency information simultaneously in both the spatial and temporal domains. Table~\ref{tab:compression_1} illustrates the feature maps resolutions for a specific lightmap set comprising 26 lightmaps at $128 \times 128$ resolution, baked under varying lighting conditions. The resolution of these feature maps is adjustable and only needs to be compatible with the BC representation that will be introduced later~(Section~\ref{sec:quantization}). Furthermore, our method supports random access, enabling on demand decoding of individual regions without processing the rest.

\begin{figure}[!b]
\centering
\includegraphics[width=0.8\linewidth]{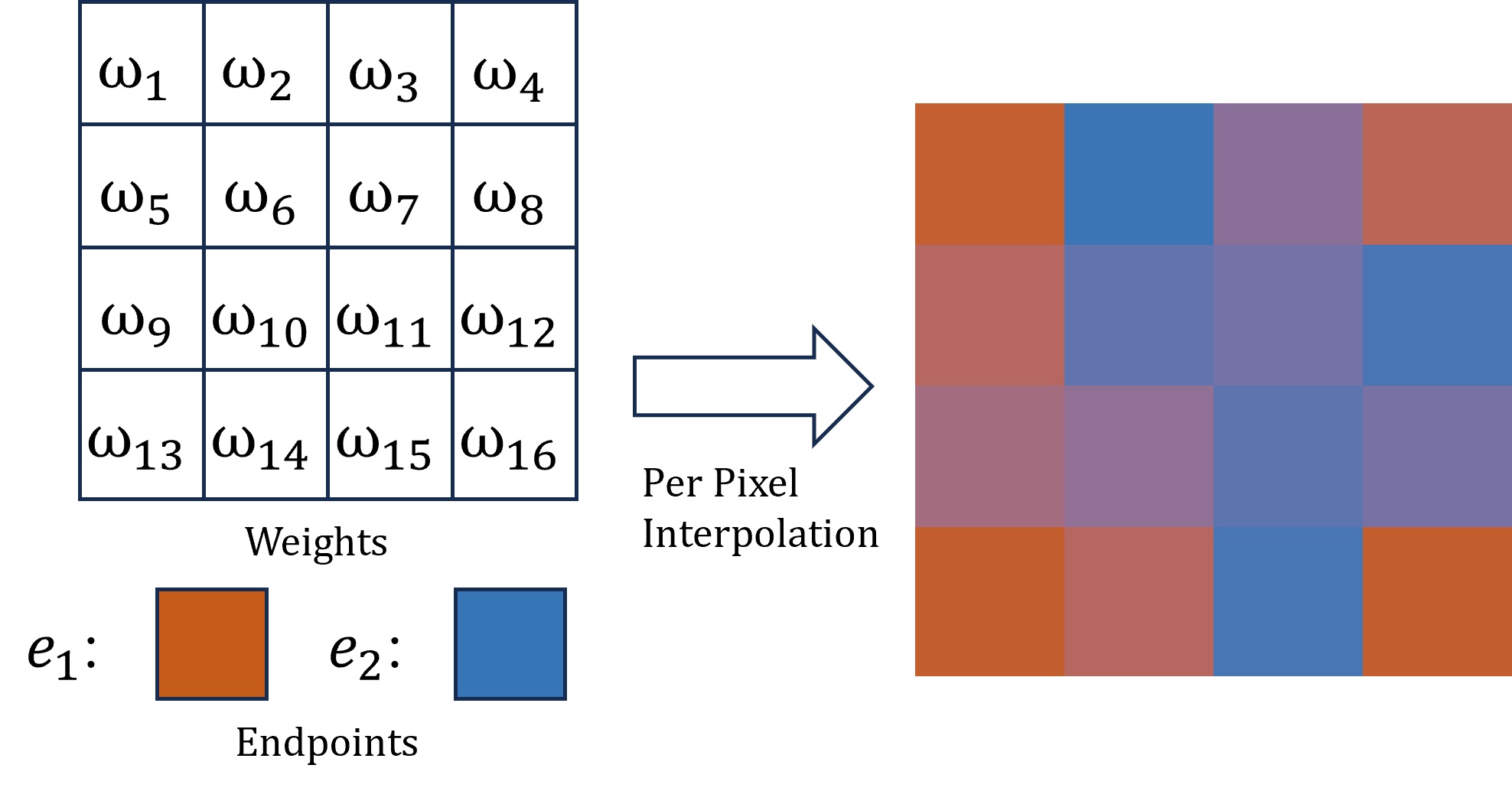}
\caption{Each $4 \times 4$ block is directly generated by per pixel interpolation between a pair of endpoints using a set of weights.}
\label{fig:bccompres}
\end{figure}

\subsection{Feature Map Compression}
\label{sec:quantization}

As shown in Figure~\ref{fig:method}, we adopt two strategies to further compress the four feature maps. For $\mathbf{F}^{2D}_{ut}$ and $\mathbf{F}^{2D}_{vt}$, we apply 8-bit post-training quantization and emulate quantization during training by injecting uniform noise~\cite{balle2016end, Vaidyanathan_2023}. Concretely, sampling at $(u,v,t)$ yields latent vectors $V_{ut}$ and $V_{vt}$, to which we add noise as follows:

\begin{equation}
\begin{aligned}
    V'_{ut} &= V_{ut} + \mathcal{U}(-0.5, 0.5) \cdot \alpha_{ut},  \\
    V'_{vt} &= V_{vt} + \mathcal{U}(-0.5, 0.5) \cdot \alpha_{vt}.
\end{aligned}
\end{equation}
Here, $\alpha_{ut}$ and $\alpha_{vt}$ are set to be $\frac{1}{256}$ in all our experiments, which corresponds to the 8-bit representation. For $\mathbf{F}^{3D}_{uvt}$ and $\mathbf{F}^{2D}_{uv}$, unlike $\mathbf{F}^{2D}_{ut}$ and $\mathbf{F}^{2D}_{vt}$, these two feature maps are much larger in size. Therefore, after training, we not only quantize them to 8-bit, but also apply the BC7 compression algorithm, which encodes each $4\times4$ texel block using multiple modes to the quantized results. For the 3D texture $\mathbf{F}^{3D}_{uvt}$, we slice it along the temporal dimension and compress each 2D slice independently. 

\begin{table}[t]
  \centering
  \caption{ The impact of quantization and BC compression on the BPP and compression ratio of our method.
 }
  \label{tab:compression}
  {\renewcommand{\arraystretch}{1.15}%
  {\small
    \begin{tabular}{l|c|c}
    \hline
    Compression Strategy & BPP & Compression Ratio \\
    \midrule
    Original & 4.62 & 4.8\% \\
    8-bit Quant. & 2.32 & 2.4\% \\
    8-bit Quant. \& BC & 0.68 & 0.7\% \\
    \hline
  \end{tabular}
  }

  }
\end{table}

However, when applying the BC algorithm to trained feature maps, significant information loss may occur. Inspired by previous work~\cite{1094560, laurent2025hardware}, we address this issue by adopting a BC-simulation strategy during the training of $\mathbf{F}^{3D}_{uvt}$ and $\mathbf{F}^{2D}_{uv}$, as illustrated in Figure~\ref{fig:bccompres}. Instead of directly updating their parameters, we divide each feature map into $4\times4$ blocks, where the feature values within each block are represented by a set of endpoints $\mathbf{E}$ and weights $\mathbf{W}$, as formulated below:

\begin{equation}
\left\{
\begin{array}{l}
\mathbf{E} = \{ e_1, e_2 \in [0,1]^k\}, \\ 
\mathbf{W} = \{w_1, \dots, w_{16}\},\quad w_p \in [0,1].
\end{array}
\right.
\label{eq:eq6}
\end{equation}

Each pixel $p$ in the $4\times4$ block is then reconstructed by linear interpolation between the two endpoints:

\begin{equation}
f_p \;=\; (1 - w_p)\,e_1 \;+\; w_p\,e_2,
\qquad p=1,\dots,16,
\label{eq:eq7}
\end{equation}
where $f_p \in [0,1]^k$ is the $k$ channels feature of pixel $p$. During training, we update the endpoints and weights of each block, use them to reconstruct $\mathbf{F}^{3D}_{uvt}$ and $\mathbf{F}^{2D}_{uv}$, and perform sampling on the reconstructed feature maps. Noise is then added to the sampled vector $V_{uvt}$, $V_{uv}$ to obtain the feature vectors $V'_{uvt}$, $V'_{uv}$, which are concatenated with $V'_{ut}$, $V'_{vt}$, and $\gamma(t)$ before being fed into the network. After training converges, we use the learned endpoints and weights to reconstruct $\mathbf{F}^{3D}_{uvt}$ and $\mathbf{F}^{2D}_{uv}$ once again, followed by quantization and BC7 compression to produce the final compressed results. Table~\ref{tab:compression} illustrates the impact of our quantization compression on the bitrates.

\subsection{Runtime Decompression}
\label{sec:decompression}
Since lighting varies with scene time, decompressing all lightmaps at load time is impractical. Decoding must be performed online. However, per frame per pixel decoding imposes a heavy computational burden, and adjacent frames typically change slowly, leading to redundant inference. 

To address this problem, we propose a method that leverages the caching mechanism of virtual texturing~(VT)~\cite{vt}. In the VT system, each lightmap is partitioned into fixed-size tiles, and at render time only the required tiles are streamed and written into a physical texture. To align training with runtime, we tile the lightmaps and train a separate model per tile. As illustrated in Figure~\ref{fig:method}, on each frame we fetch the parameters of the required tiles on demand, decompress them with a compute shader, and write the results to the physical texture. During shading, we first sample the page table to locate each tile within the physical texture, then sample the physical texture to obtain the final lighting. For tiles already resident in the physical texture, because lighting changes are negligible over short timescales, we can reuse the cached content for a period without redundant inference. This design substantially reduces online decoding cost, avoids redundant computation, and improves real-time performance.

\section{Implementation}
Our neural compression model is trained separately for each lightmap set using the PyTorch framework~\cite{paszke2019pytorch}. Before training, gamma correction is applied to enhance details in dark regions, and per channel mean normalization is performed at each time step. These mean values are later used during rendering to restore the original lightmap data. The trained model outputs are floating-point RGB values, which are quantized into an 8-bit 4-channel format to reduce GPU memory usage.

In large-scale scenes, thousands of independent NDGI models may need to be trained. To improve efficiency, we employ PyTorch’s batched matrix operations~\cite{BADDBMM} for parallel training. The tile-based structure of our approach also enables straightforward scaling to distributed training. The MLP has We use the Adam optimizer~\cite{kingma2014adam} with an initial learning rate of $10^{-3}$, a batch size of $2^{12}$, and use L2 loss for the loss function. The MLP network consists of two hidden layers with variable widths. GELU activation~\cite{hendrycks2016gaussian} is applied to hidden layers, with no activation at the output. In the final training stage, we freeze the feature maps and fine-tune the MLP under simulated quantization and BC compression. For real-time rendering, we utilize the virtual texturing system and compute shader in Unreal Engine~\cite{UE} to enable efficient lightmap reconstruction with minimal runtime overhead.

\begin{table}[b]
\centering
\caption{NDGI profiles with different BPP and decoder size. }
\label{tab:config}
\setlength{\tabcolsep}{6pt}
\renewcommand{\arraystretch}{1.15}
\resizebox{\columnwidth}{!}{%
\begin{tabular}{l|c|c|c}
\hline
Profile & $\mathbf{F}^{3D}_{uvt}$ Resolution & Hidden Size & BPP \\
\midrule
NDGI L. & $16 \times 16 \times 12 \times 4 $ & 16 & 0.50 \\
NDGI M. & $32 \times 32 \times 12 \times 4 $ & 16 & 0.68 \\
NDGI H. & $64 \times 64 \times 12 \times 4 $ & 16 & 1.39 \\
NDGI M.64 & $32 \times 32 \times 12 \times 4 $ & 64 & 0.86 \\
\hline
\end{tabular}%
}
\end{table}

\section{Experiment}
\begin{figure}[t]
  \centering
  \begin{subfigure}[t]{0.48\linewidth}
    \centering
    \includegraphics[width=1\linewidth]{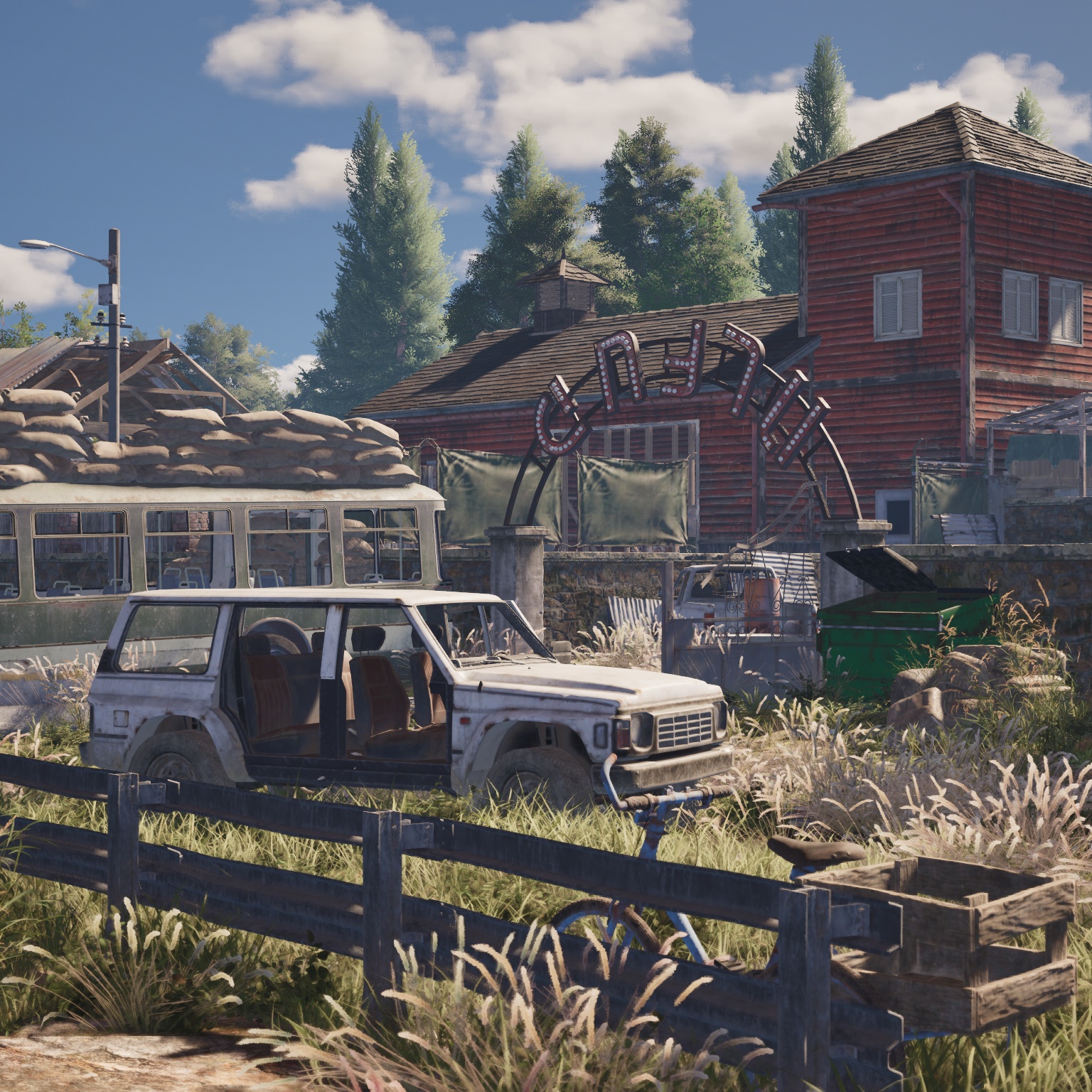}
    \caption{FarmLand}\label{subfig:a}
  \end{subfigure}\hfill
  \begin{subfigure}[t]{0.48\linewidth}
    \centering
    \includegraphics[width=1\linewidth]{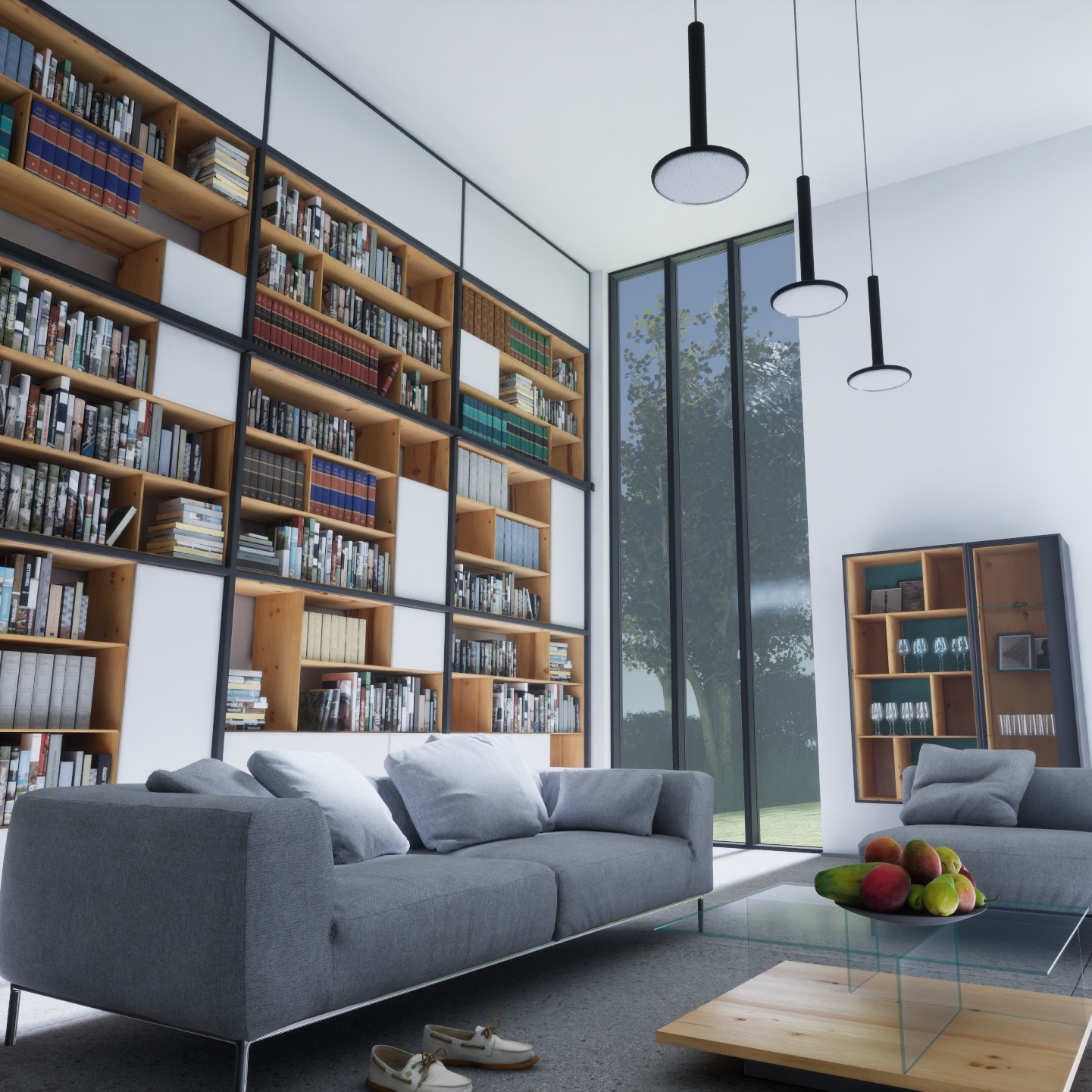}
    \caption{Room}\label{subfig:b}
  \end{subfigure}
  \par\smallskip
  \begin{subfigure}[t]{0.48\linewidth}
    \centering
    \includegraphics[width=1\linewidth]{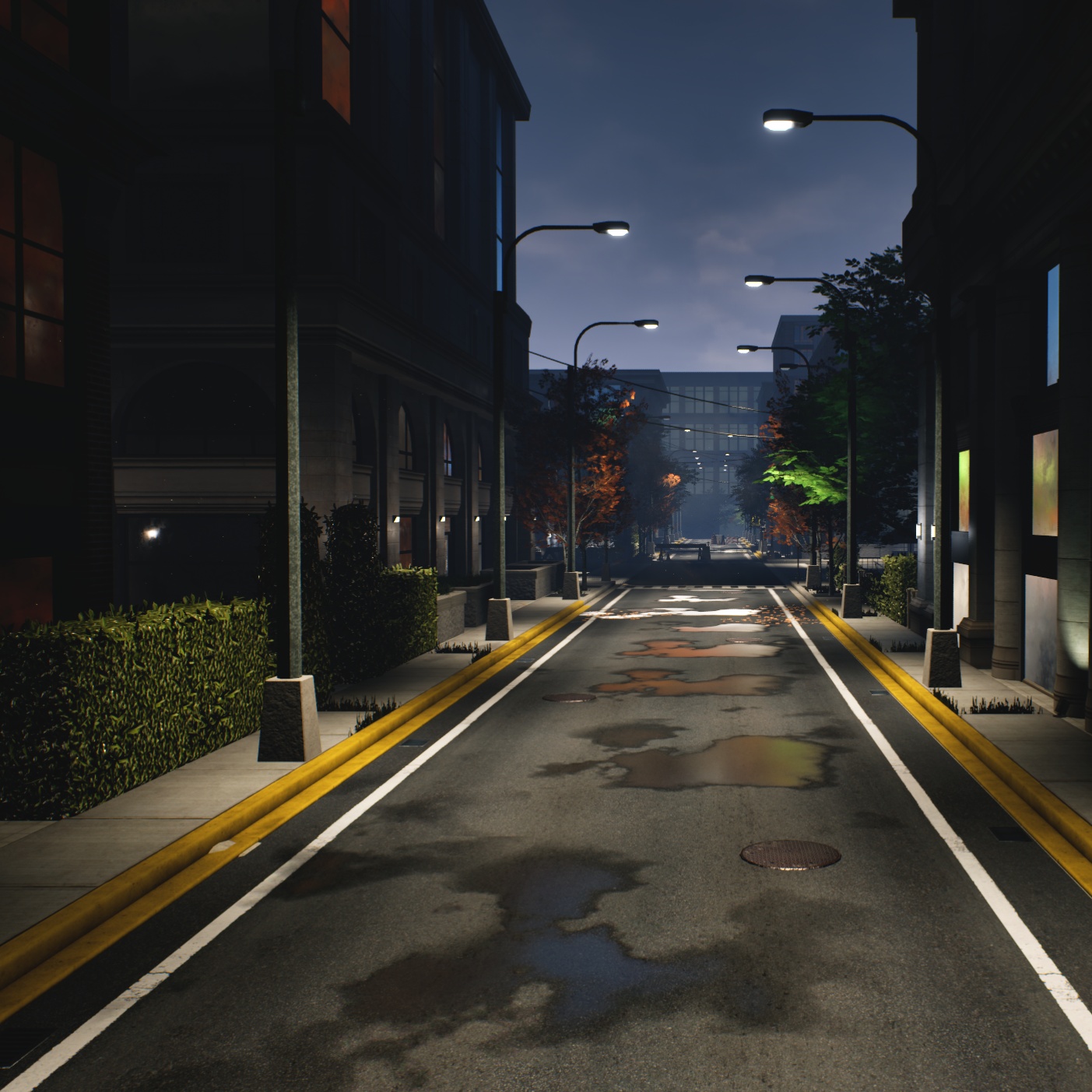}
    \caption{City}\label{subfig:c}
  \end{subfigure}\hfill
  \begin{subfigure}[t]{0.48\linewidth}
    \centering
    \includegraphics[width=1\linewidth]{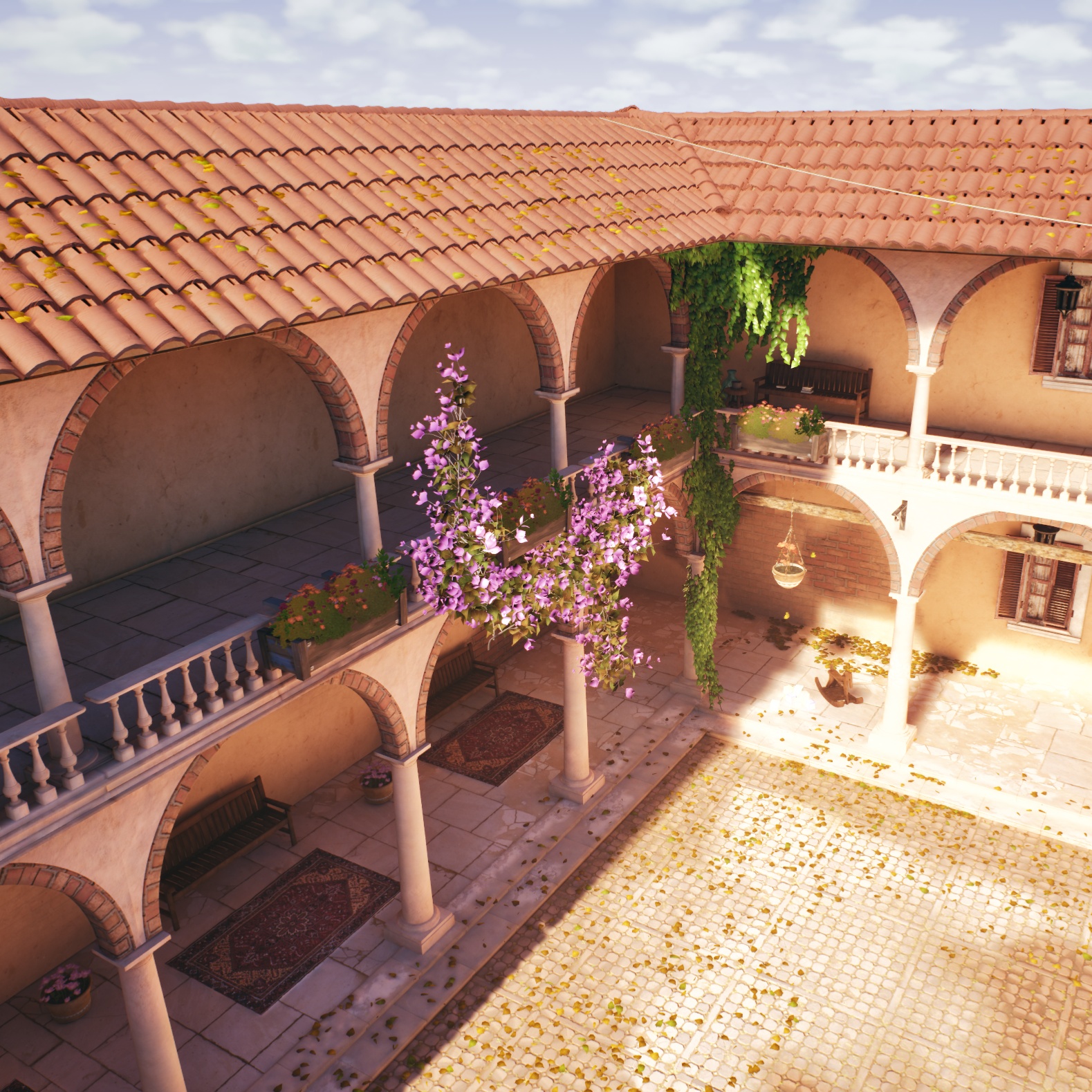}
    \caption{Yard}\label{subfig:d}
  \end{subfigure}
  \caption{Representative examples from our datasets. (a) FarmLand, sourced from ``Arena Breakout: Infinite". (b) An indoor environment with time-varying skylight and sunlight. (c) A city scene where streetlights toggle on and off at specific times. (d) An yard with multiple local light sources.}
  \label{fig:dataset}
\end{figure}

\begin{table*}[t]
\centering
\caption{Average PSNR, 1-SSIM, and LPIPS values over the evaluation data set for the methods used in our comparison. ``L.", ``M." and ``H." is short for ``Low", ``Medium" and ``High". Our method achieves better compression quality while maintaining low bitrates.}
\label{tab:result}
\setlength{\tabcolsep}{6pt}
\renewcommand{\arraystretch}{1.15}
\resizebox{\textwidth}{!}{%
\begin{tabular}{l|ccccc|cccccc}
\toprule
\multicolumn{1}{c|}{%
  \begin{tabular}{@{}c@{}}
  \end{tabular}
} &
\multicolumn{5}{c|}{%
  \begin{tabular}{@{}c@{}}
    Low Bitrates~( $\textless$ 1.0 BPP)\\
  \end{tabular}
} &
\multicolumn{6}{c}{%
  \begin{tabular}{@{}c@{}}
    High Bitrates~( $\textgreater$ 1.0 BPP)\\
  \end{tabular}
}\\
\midrule
 & NDGI L. & NDGI M. & NTC L. & NDGI M.64 & ASTC $12\times12$ & NTC M. & ASTC $10\times10$ & NDGI H. & NTC H. & BC6H & BC7\\
\midrule
BPP (Mean) & 0.50 & 0.68 & 0.78 & 0.86 & 0.89 & 1.10 & 1.28 & 1.39 & 1.78 & 8 & 8 \\
\midrule
\multicolumn{1}{r|}{PSNR ($\uparrow$)}        & 45.96 & 46.69 & 43.61 & \textbf{47.42} & 32.50 & 44.26 & 34.52 & \textbf{48.68} & 45.77 & 44.31 & 42.27\\
\multicolumn{1}{r|}{1 - SSIM ($\downarrow$)}  & 0.014 & 0.012 & 0.026 & \textbf{0.012} & 0.069 & 0.022 & 0.051 & \textbf{0.009} & 0.016 & 0.026 & 0.039\\
\multicolumn{1}{r|}{LPIPS ($\downarrow$)}     & 0.007 & \textbf{0.006} & 0.010 & 0.007 & 0.057 & 0.008 & 0.043 & \textbf{0.004} & 0.007 & 0.010 & 0.016\\
\bottomrule
\end{tabular}%
}
\end{table*}

\begin{figure*}[t]
\centering
\setlength{\tabcolsep}{2pt}
\renewcommand{\arraystretch}{1.0}
\begin{tabular}{c|cccc|c}
\ & NDGI M.~(Ours)  & NTC L. & ASTC $12 \times 12$  & BC7 & Reference \\
\ & \textbf{1.10 MB} & 1.26 MB & 1.44 MB & 13 MB & 156 MB \\
\midrule
\includegraphics[width=0.155\textwidth]{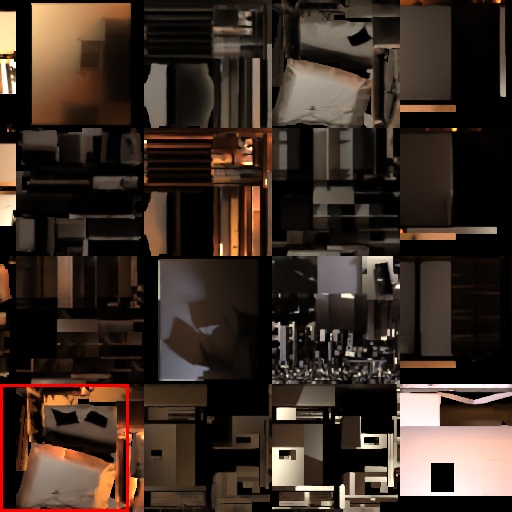} &
\includegraphics[width=0.155\textwidth]{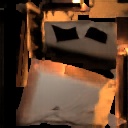} &
\includegraphics[width=0.155\textwidth]{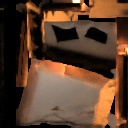} &
\includegraphics[width=0.155\textwidth]{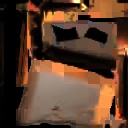} &
\includegraphics[width=0.155\textwidth]{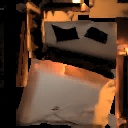} &
\includegraphics[width=0.155\textwidth]{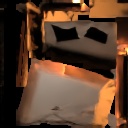} \\
{\scriptsize PSNR ($\uparrow$) }  & {\scriptsize \textbf{49.31 dB} } & {\scriptsize 45.14 dB } & 
{\scriptsize 33.78 dB} & {\scriptsize 44.29 dB} &  \\
\includegraphics[width=0.155\textwidth]{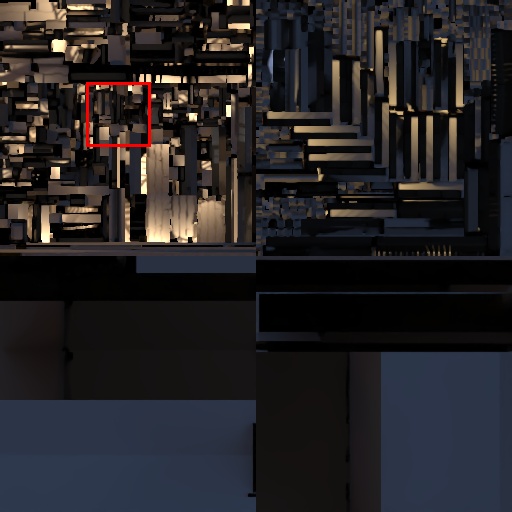} &
\includegraphics[width=0.155\textwidth]{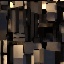} &
\includegraphics[width=0.155\textwidth]{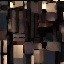} &
\includegraphics[width=0.155\textwidth]{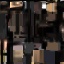}  &
\includegraphics[width=0.155\textwidth]{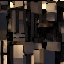} &
\includegraphics[width=0.155\textwidth]{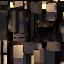} \\
{\scriptsize PSNR ($\uparrow$) } & {\scriptsize \textbf{45.73 dB} }  & {\scriptsize 41.69 dB } &
{\scriptsize 31.13 dB} & {\scriptsize 42.49 dB} & \\
\end{tabular}
\caption{Quality and storage comparison showing PSNR scores for lightmap tiles in the ``FarmLand" scene. Compared with traditional texture compression methods, our approach achieves higher compression ratios and better reconstruction quality. Compared to NTC, NDGI yields obviously less noise with a much smaller decoder, leading to superior decompression performance.}
\label{fig:two_rows_six_each_w_titles}
\end{figure*}

\begin{figure*}[t]
\centering
\setlength{\tabcolsep}{2pt}
\renewcommand{\arraystretch}{1}

\resizebox{\textwidth}{!}
{
\begin{tabular}{c|cccc|c}
Lit Scene &
PRT &
ASTC $12 \times 12$ &
NTC L. &
NDGI M.~(Ours) &
Reference \\
\midrule
\includegraphics[width=0.16\linewidth]{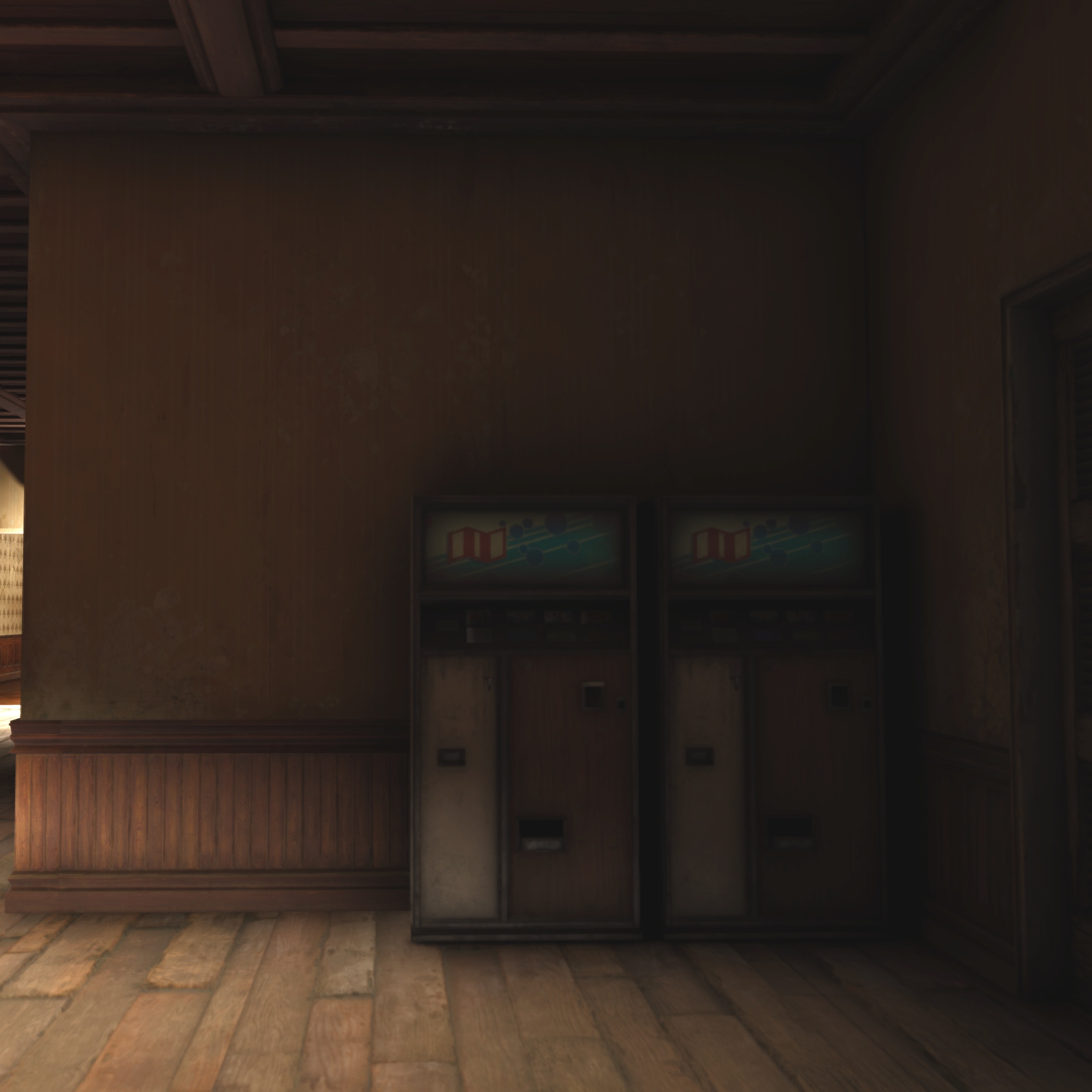} &
\includegraphics[width=0.16\linewidth]{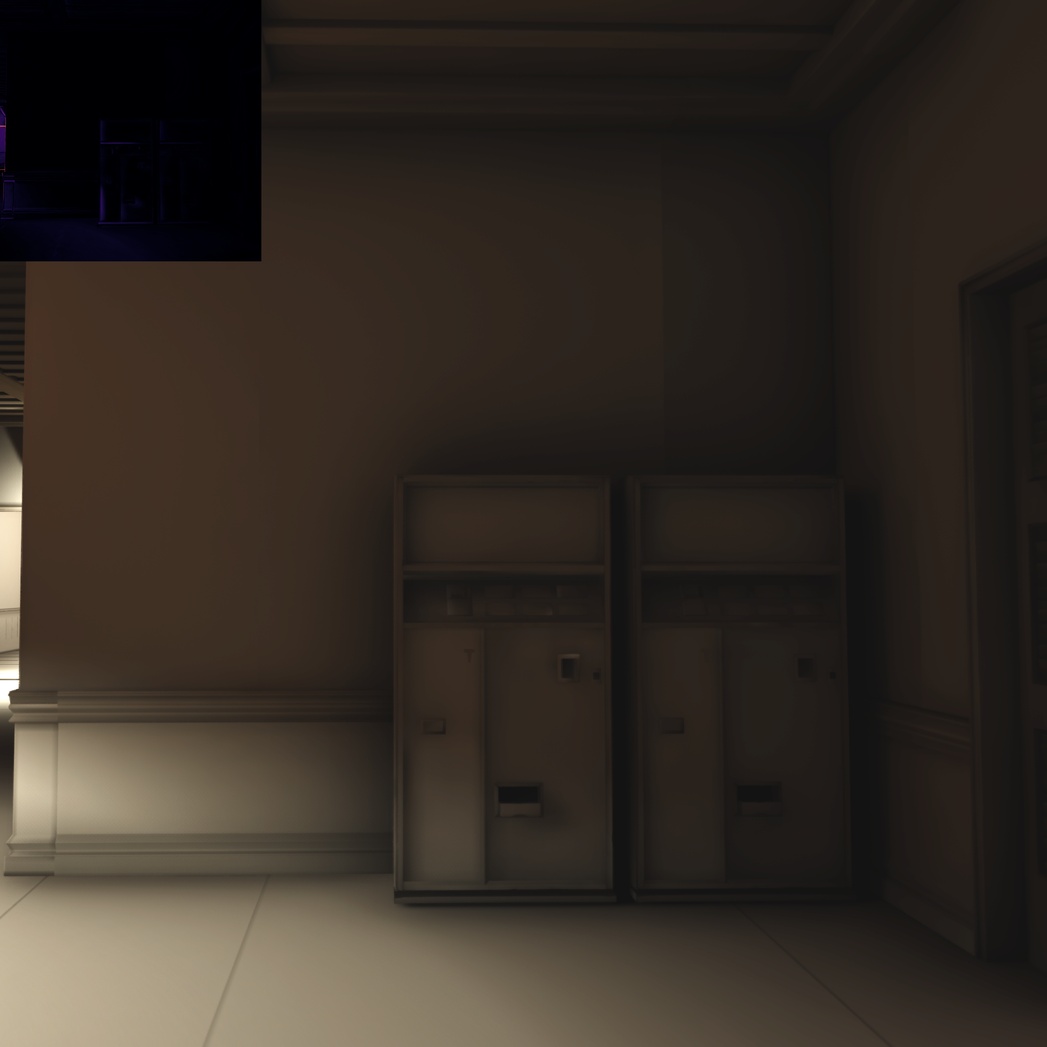} &
\includegraphics[width=0.16\linewidth]{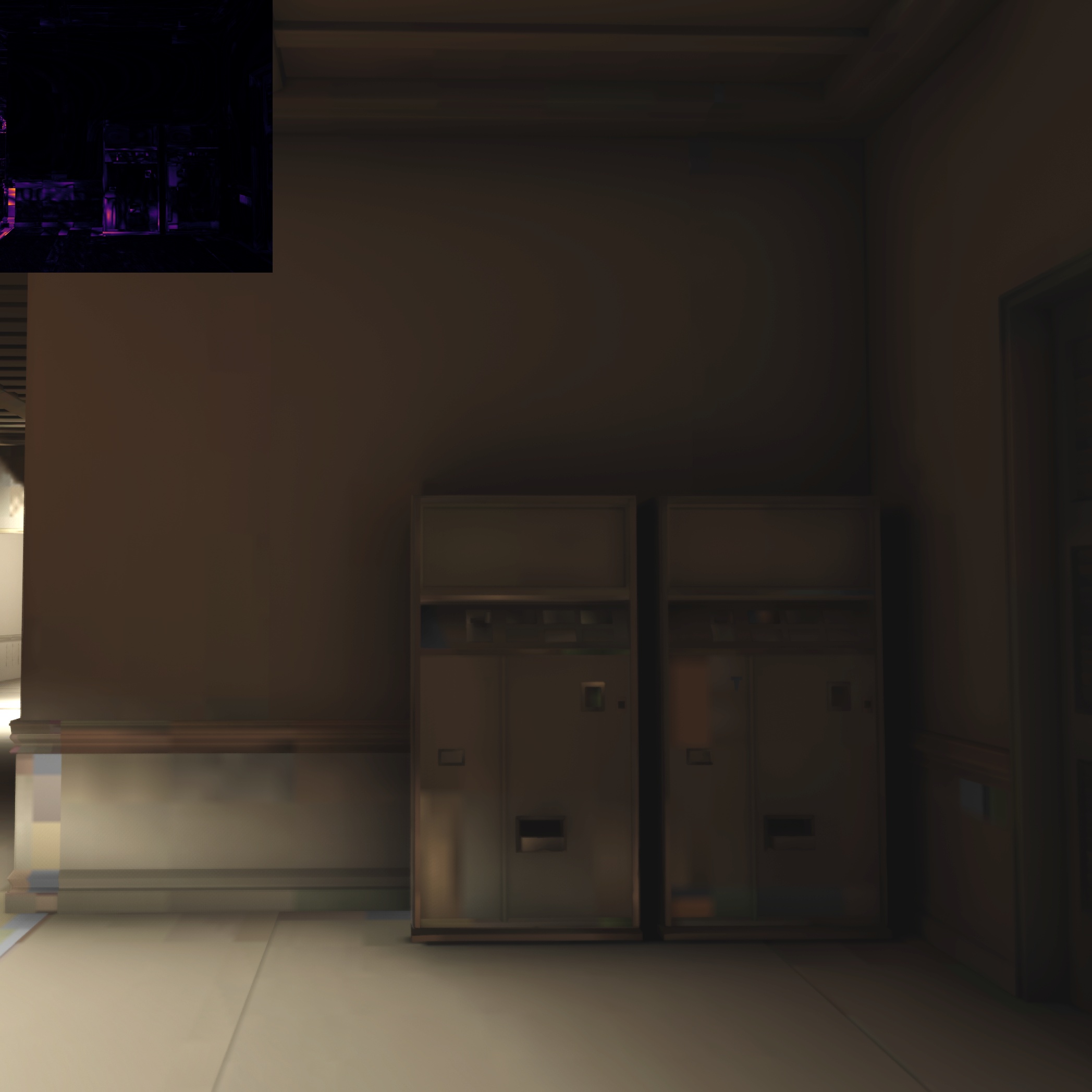} &
\includegraphics[width=0.16\linewidth]{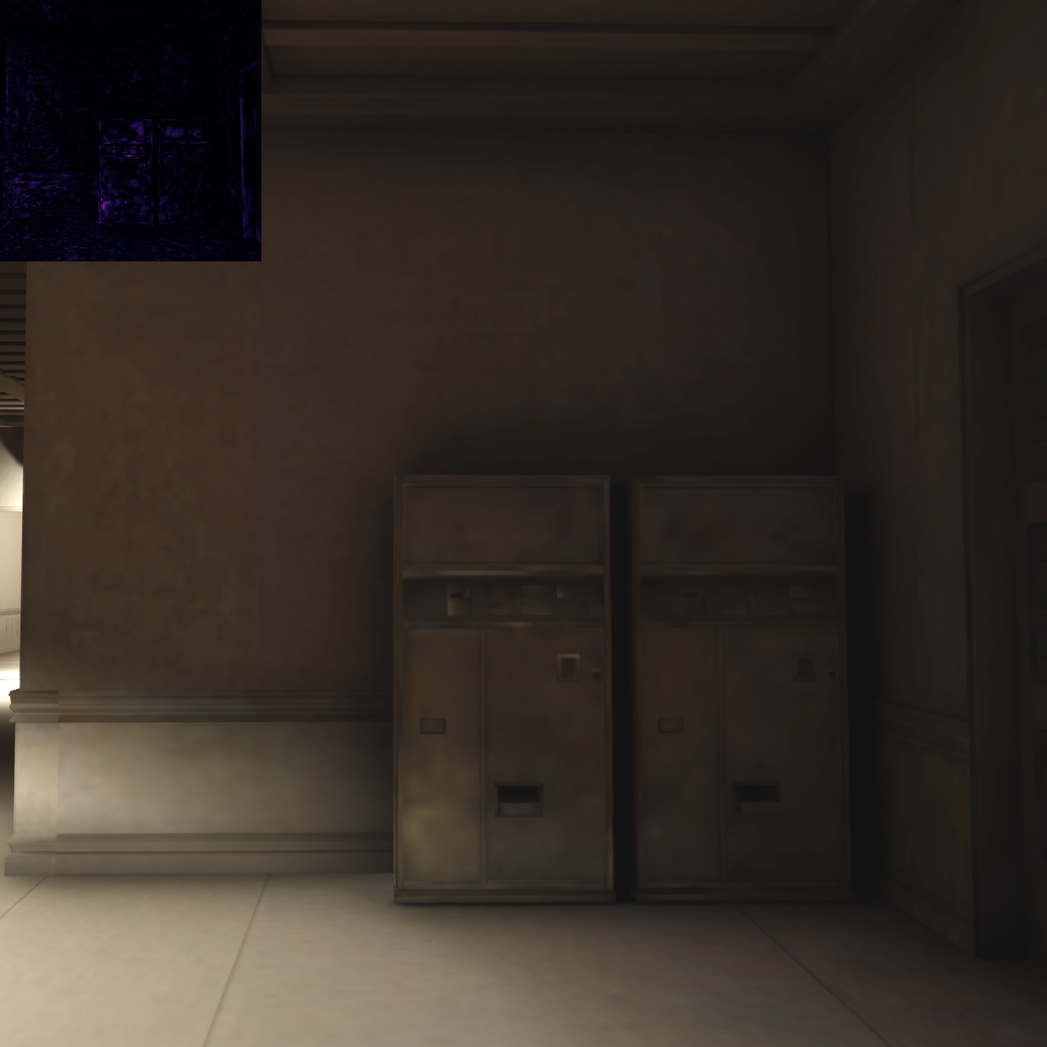} &
\includegraphics[width=0.16\linewidth]{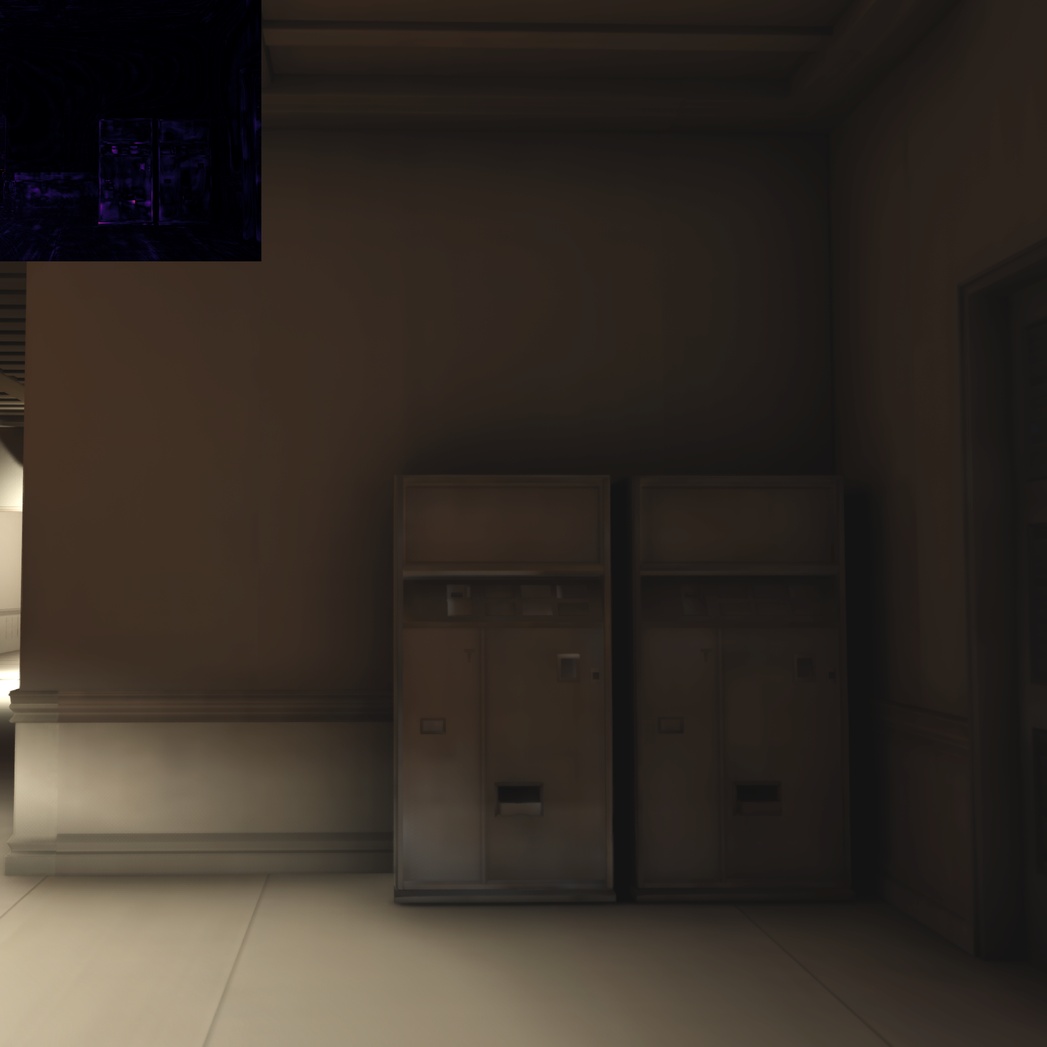} &
\includegraphics[width=0.16\linewidth]{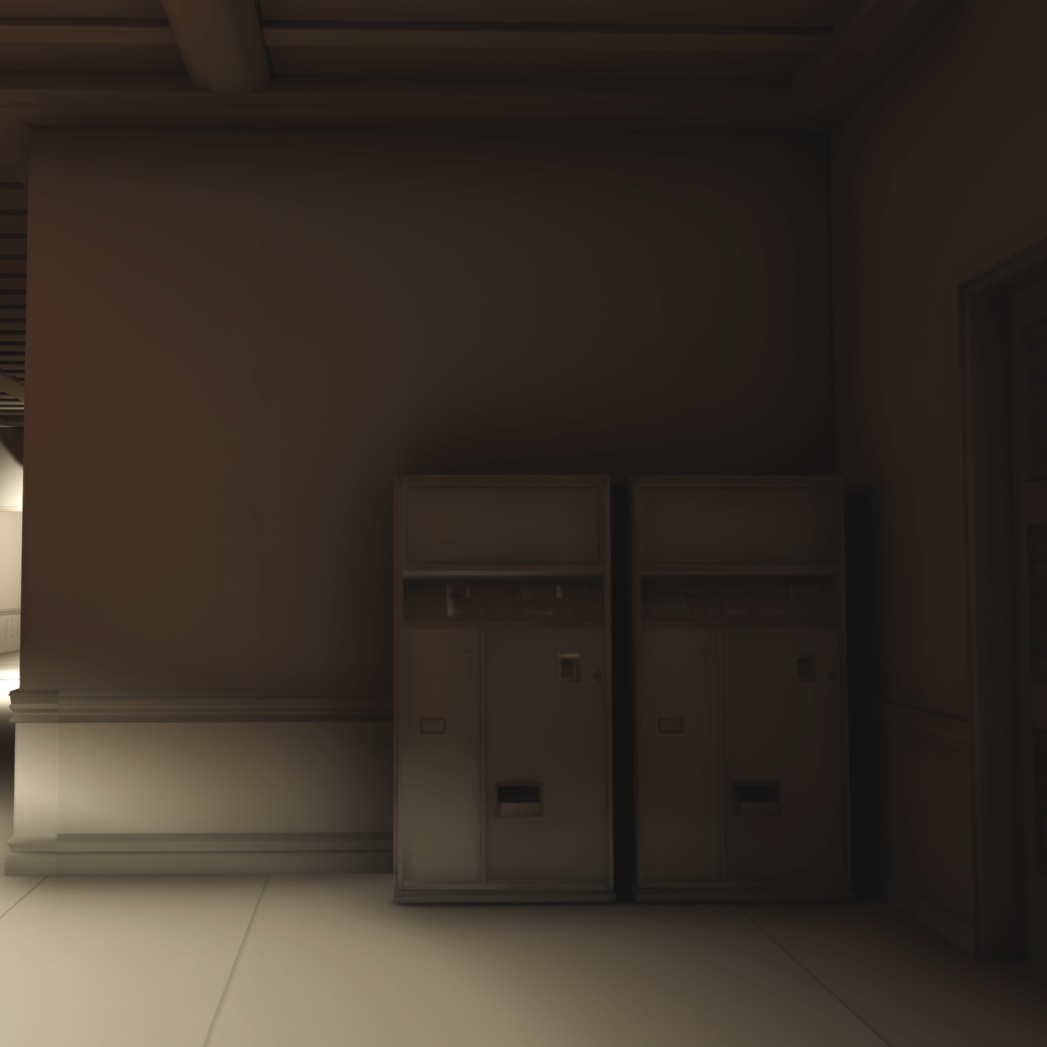} \\
{\scriptsize BPP($\downarrow$), PSNR($\uparrow$), SSIM($\uparrow$)} &
{\scriptsize 0.92, 35.59 dB, 0.993} &
{\scriptsize 0.89, 42.09 dB, 0.998} &
{\scriptsize 0.78, 43.48 dB, 0.998} &
{\scriptsize \textbf{0.67, 45.75 dB, 0.999}} &
{\scriptsize FarmLand} \\
\includegraphics[width=0.16\linewidth]{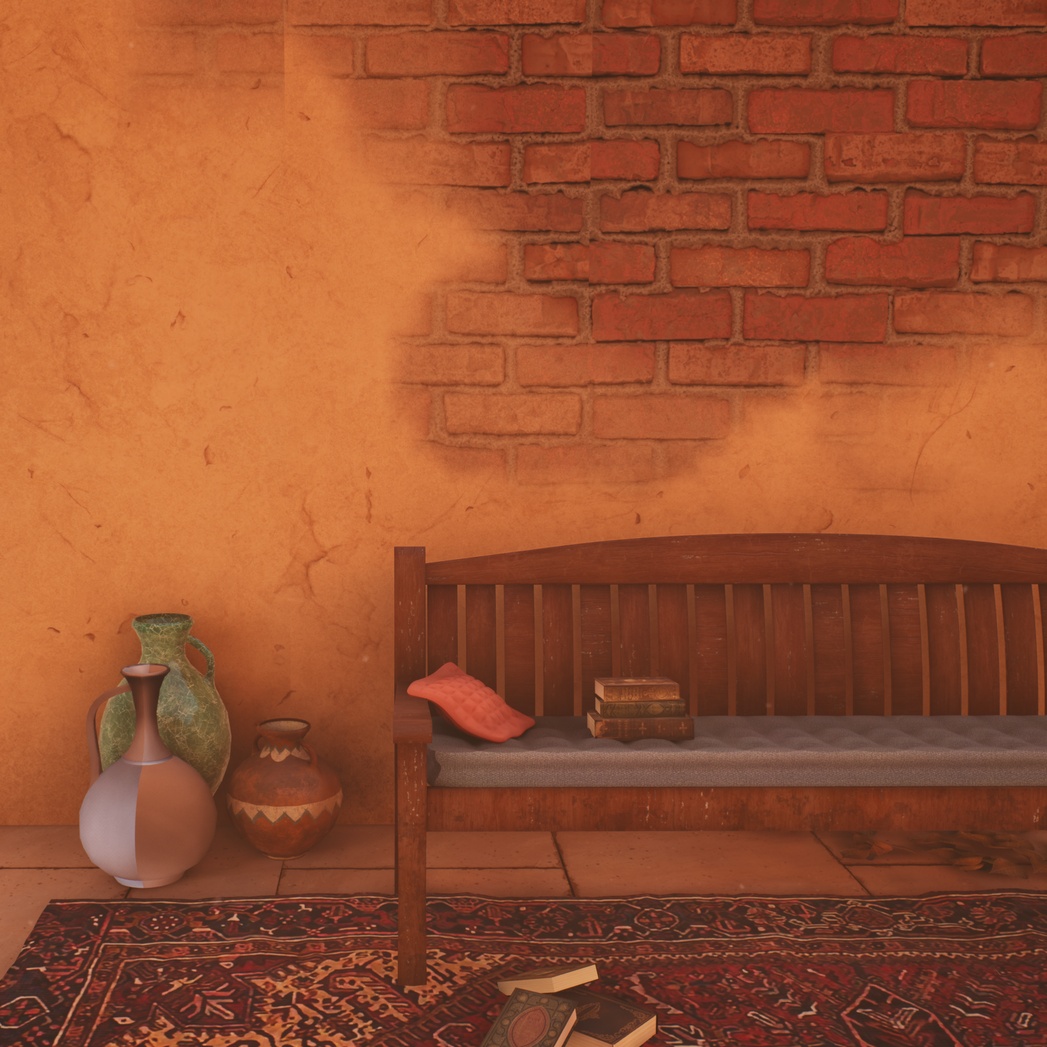} &
\includegraphics[width=0.16\linewidth]{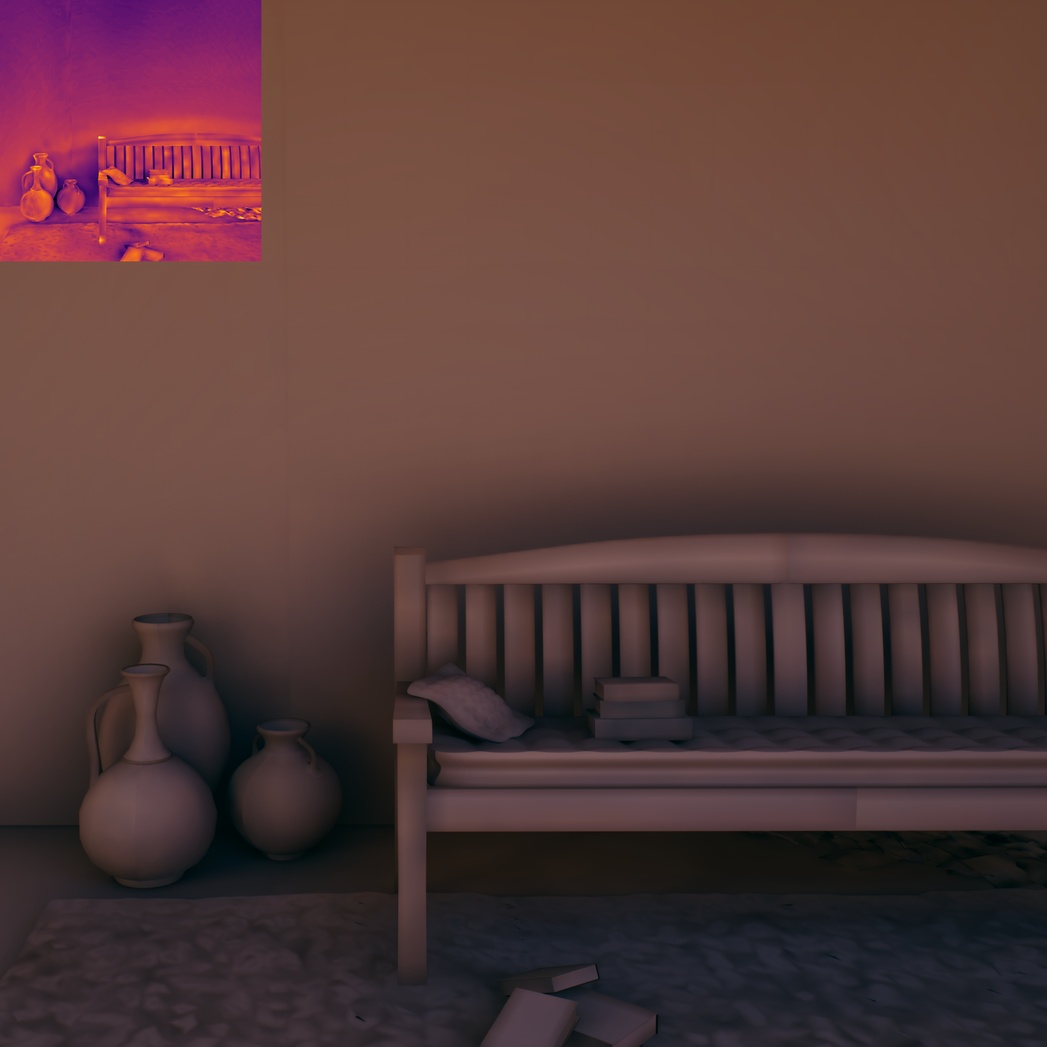} &
\includegraphics[width=0.16\linewidth]{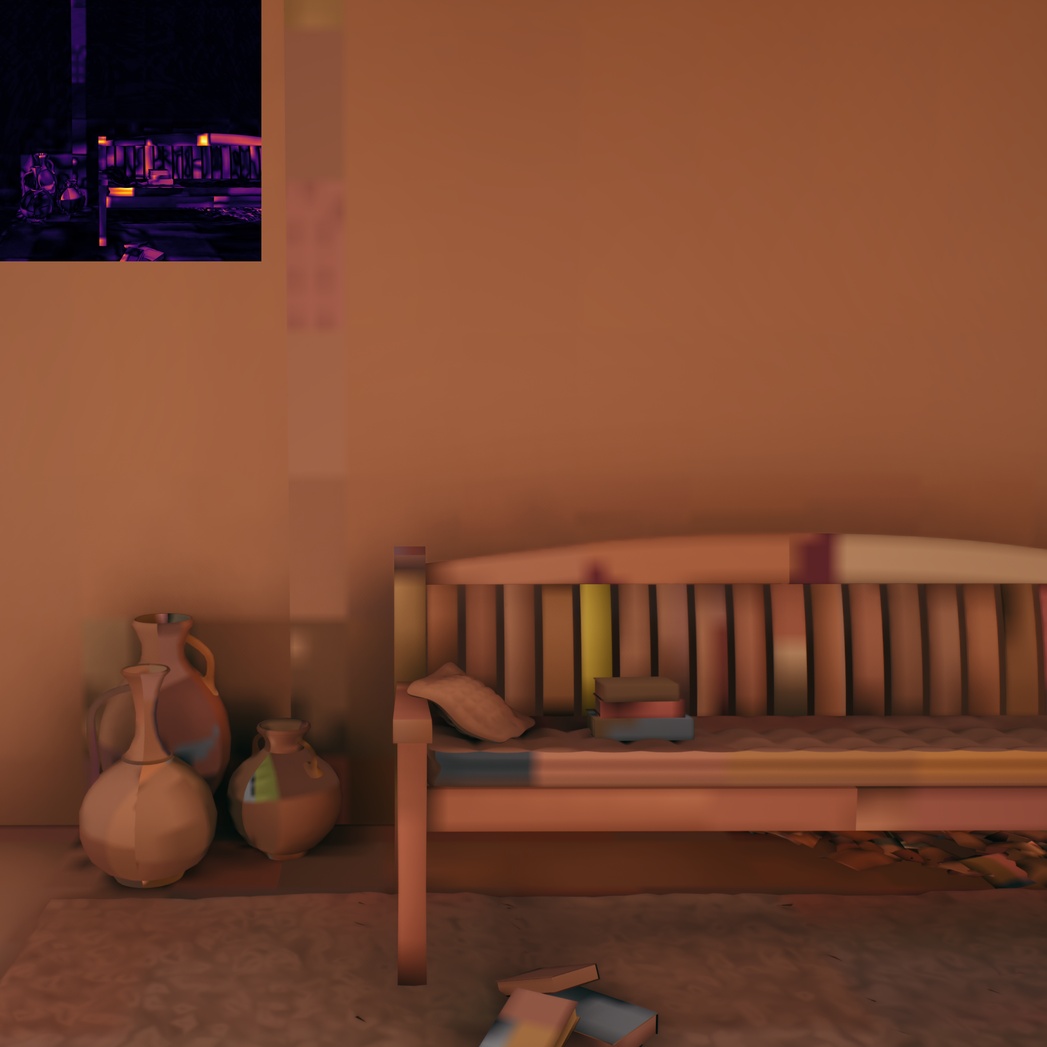} &
\includegraphics[width=0.16\linewidth]{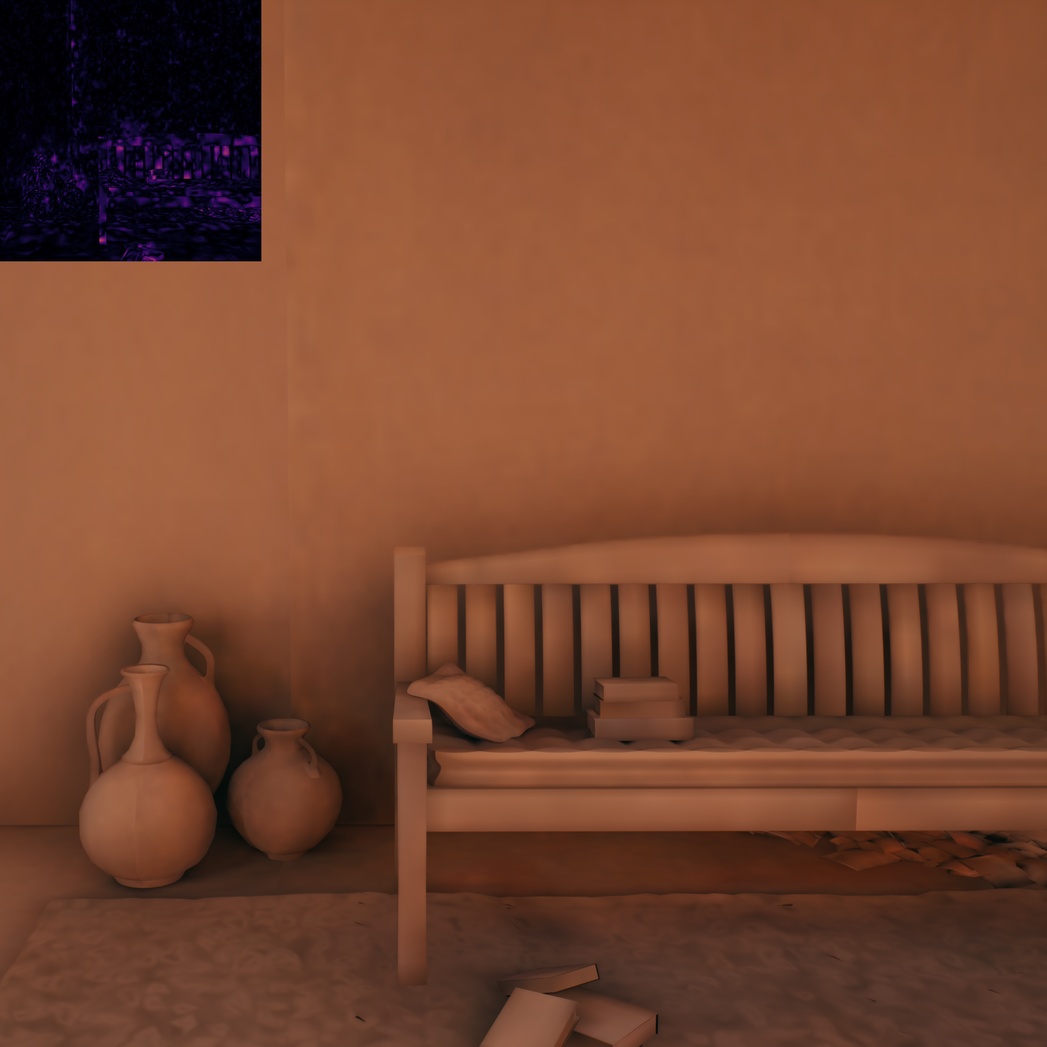} &
\includegraphics[width=0.16\linewidth]{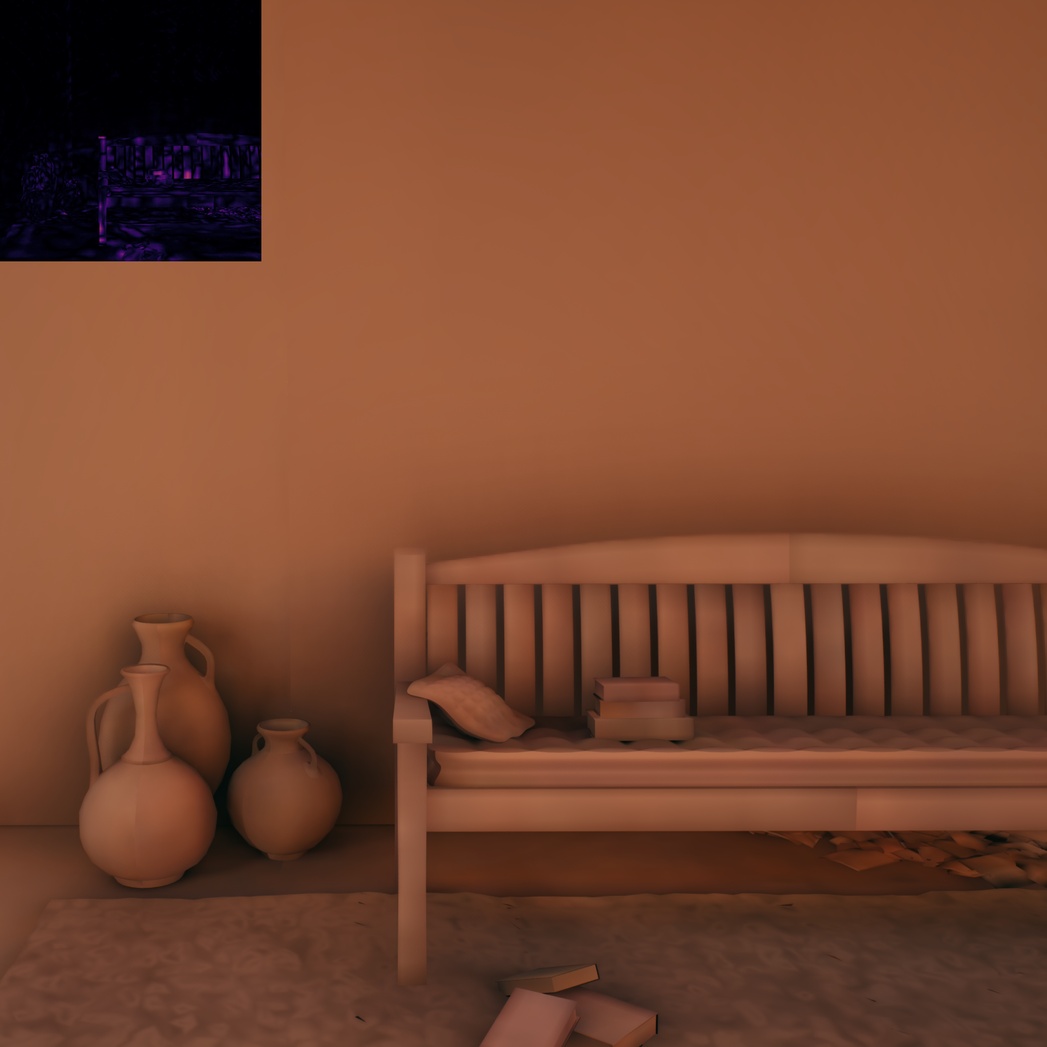} &
\includegraphics[width=0.16\linewidth]{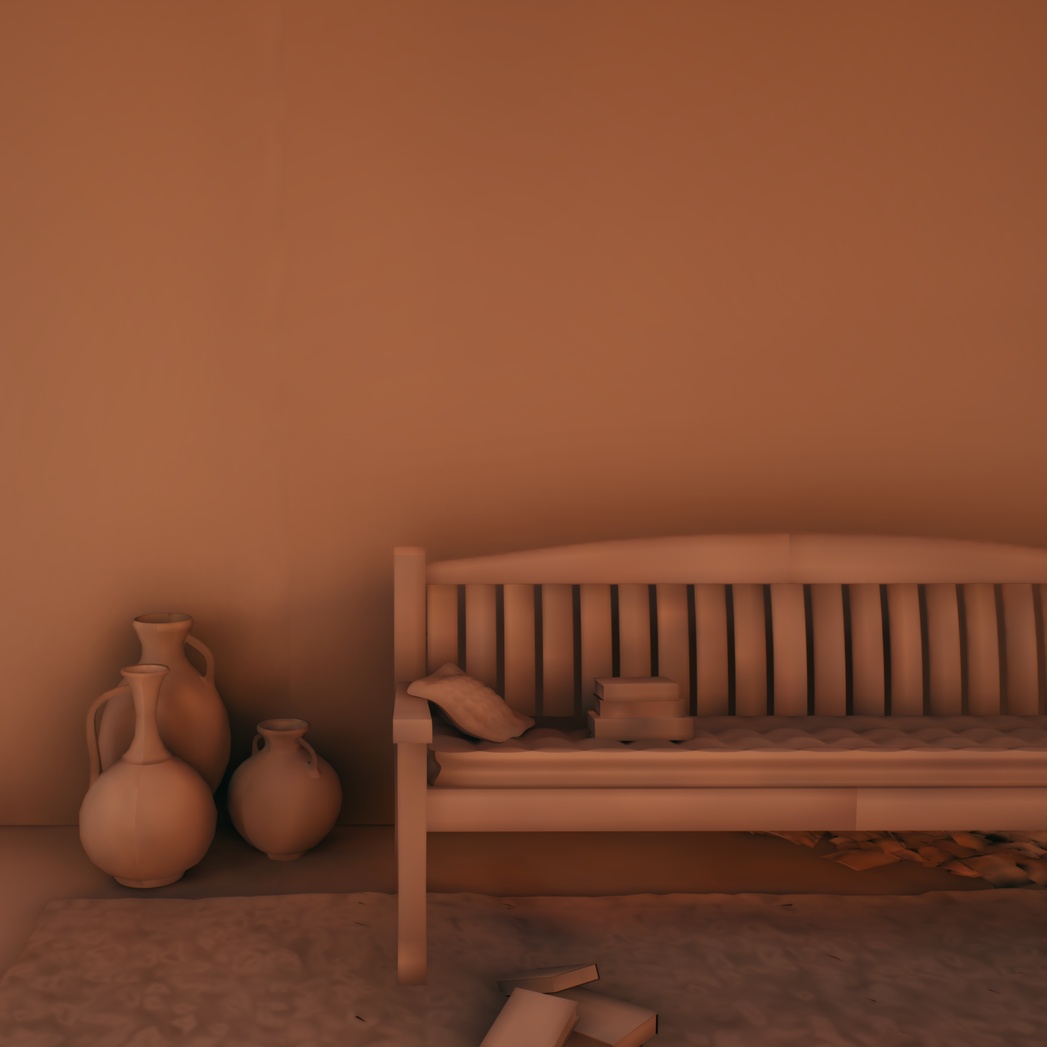} \\
{\scriptsize BPP($\downarrow$), PSNR($\uparrow$), SSIM($\uparrow$)} &
{\scriptsize 1.00, 20.85 dB, 0.935} &
{\scriptsize 0.89, 31.43 dB, 0.967} &
{\scriptsize 0.78, 42.28 dB, 0.997} &
{\scriptsize \textbf{0.71, 43.31 dB, 0.997}} &
{\scriptsize Yard} \\
\end{tabular}
 }
\caption{Comparison of rendered results. The “Lit Scene” displays the final rendered image. Notably, PRT exhibits color tone deviation, ASTC displays obvious blocky artifacts, and NTC produces noticeable noise. Compared with previous methods, NDGI delivers higher-quality GI effects.}
\label{fig:rendered}
\end{figure*}

Our lightmap compression framework supports flexible configurations, allowing users to balance compression quality, storage overhead, and inference performance. We define four compression configurations, as detailed in Table~\ref{tab:config}, each corresponding to varying network costs and bits per pixel (BPP). These configurations are used in our experiments to provide comprehensive comparisons across different trade-offs between quality and performance. More detailed experiment results can be found in supplementary materials.

\subsection{Evaluation Dataset}
We evaluate our method on lightmap datasets precomputed from multiple real game scenes, covering diverse indoor and outdoor environments. The lighting conditions vary over scene time, including changes in skylight, directional light, and local light sources. In particular, switching of local light sources introduces high-frequency variations on the lightmap. Representative scene examples are shown in Figure~\ref{fig:dataset}.

\subsection{Compared Methods}

We compare our method against both traditional GPU texture compression and neural compression approaches to evaluate compression performance and rendering quality.

For traditional methods, GPU texture compression formats such as BC6H, BC7, and ASTC~\cite{10.5555/2383795.2383812, DXTC} are commonly adopted in production, typically chosen according to texture type rather than a unified standard. We evaluate two BC-based configurations: directly applying BC6H to HDR lightmaps, and converting HDR lightmaps to 4-channel 8-bit before applying BC7 compression. In addition, we test ASTC under multiple bit-per-pixel settings for comparison.

For neural methods, we include NTC~\cite{Vaidyanathan_2023}, a representative neural texture compression approach. However, NTC supports at most 16 channels per texture set, which limits its ability to handle our temporal lightmap data. To ensure a fair comparison, we partition our dataset into consecutive temporal segments, each compressed independently within this constraint. We also compare with Precomputed Radiance Transfer~(PRT)~\cite{sloan2023precomputed}, which supports dynamic lighting through precomputed probe data.

We exclude general image compression algorithms such as JPEG~\cite{wallace1991jpeg, alakuijala2019jpeg}, as they do not support random access and are unsuitable for real-time rendering.

\subsection{Lightmap Quality Evaluation}
\begin{figure}[t]
\centering
\includegraphics[width=\linewidth]{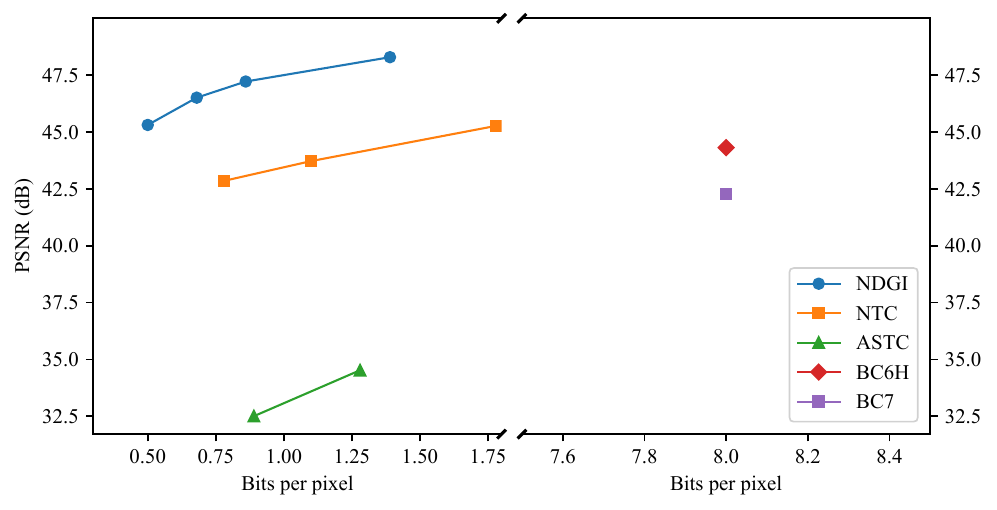}
\caption{Lightmap quality across different BPP. NDGI consistently outperforms other methods across all compression ratios.}
\label{fig:compare}
\end{figure}

Table~\ref{tab:result} summarizes the compression and reconstruction results in our test scene. Overall, NDGI achieves significantly higher lightmap reconstruction quality compared with other methods under similar bitrates. Although our training objective does not directly optimize for perceptual fidelity, our method still consistently outperforms conventional GPU and neural compression techniques under these perceptual metrics. Additionally, as illustrated in Figure~\ref{fig:compare}, NDGI maintains high-fidelity reconstructions across a wide range of compression rates. Notably, even with an extremely lightweight decoder containing only a few hundred parameters, our method preserves fine illumination details and avoids the artifacts commonly observed in other approaches. We attribute this robustness to the efficient representation of spatial-temporal lightmap features learned by our network.

\subsection{Rendered Quality Evaluation}
\begin{table}[b]
  \centering
  \caption{ Decompression time for a $1024 \times 1024$ lightmap. Performance is similar across all lightmaps for a given profile.}
  \label{tab:decompression}
  \setlength{\tabcolsep}{12pt}
  \renewcommand{\arraystretch}{1.15}

  \resizebox{\columnwidth}{!}{%
    \begin{tabular}{c|c|c|c|c}
      \hline
      NDGI L. & NDGI M. & NDGI M.64  & NTC L. & NTC M.\\
      \midrule
      \textbf{0.201ms} & 0.203ms & 0.314ms & 0.886ms & 0.918ms\\
      \hline
    \end{tabular}%
  }
\end{table}

For rendered quality evaluation, we take the original lightmap rendering results as the ground truth and compare our method against representative alternatives, including NTC~\cite{Vaidyanathan_2023} and PRT~\cite{sloan2023precomputed}. To ensure a fair and accurate comparison, we directly measure the differences between the reconstructed lighting and the ground-truth lighting, rather than comparing the final rendered images. This design eliminates potential variations introduced by the rendering pipeline and focuses solely on lighting reconstruction quality. As illustrated in Figure~\ref{fig:head} and Figure~\ref{fig:rendered}, NDGI achieves higher-fidelity lighting reconstruction across multiple scenes. Furthermore, our method exhibits stronger temporal consistency and more accurate reproduction of lighting transitions over time, substantially outperforming PRT in dynamic illumination scenarios.

\subsection{Decompression Performance}
We evaluate real-time performance by decoding a $1024\times1024$ lightmap using a CUDA implementation accelerated by Tensor Cores. Table~\ref{tab:decompression} reports the decoding latency measured on an NVIDIA RTX 4060 GPU. Our method achieves significantly lower latency than NTC~\cite{Vaidyanathan_2023}, primarily due to its compact decoder network and reduced feature dimensionality, which effectively minimize sampling overhead. Although NDGI involves additional computations compared to standard lightmaps, the results demonstrate that our approach remains highly efficient and well suited for real-time rendering scenarios.

\section{Discussion and Conclusion}

In temporally varying scenes, our method supports random access and enables smooth temporal transitions while maintaining a single parameter set in memory. However, abrupt illumination changes cannot be reflected within a single frame and typically exhibit a short latency at common update rates. In addition, the current training process in PyTorch~\cite{paszke2019pytorch} is also relatively slow. We plan to develop a dedicated training framework to improve throughput and to integrate the method more tightly into real-time pipelines through asynchronous execution and better scheduling. 

In conclusion, we present a compact and efficient compression framework for temporal lightmap sets that preserves random access across both space and time with minimal runtime cost. Our method achieves high-quality reconstructions at extremely low bitrates and supports fast online decompression suitable for real-time dynamic lighting. We believe this approach opens new possibilities for neural rendering and we will release our temporal lightmap dataset to encourage further research in this direction.

{
    \small
    \bibliographystyle{ieeenat_fullname}
    \bibliography{main}
}

\clearpage
\setcounter{page}{1}
\maketitlesupplementary

\section{Lightmap}

At a high level, real-time rendering first identifies visible surface points (\eg via rasterization) and then shades each fragment by combining emitted radiance with reflected radiance integrated over the incident hemisphere. Formally, the rendering equation~\cite{10.1145/15886.15902} relates outgoing radiance to emission and reflection:
\begin{equation}
L_o(\mathbf{p},\mathbf{v}) = L_e(\mathbf{p},\mathbf{v}) + \int_{\Omega} f(\mathbf{l},\mathbf{v})\, L_i(\mathbf{p},\mathbf{l})\, (\mathbf{n}\!\cdot\!\mathbf{l})^{+}\, d\mathbf{l}.
\label{eq:sup1}
\end{equation}
Here, $ L_o(\mathbf{p},\mathbf{v}) $ is the outgoing radiance at point $ \mathbf{p} $ toward view $ \mathbf{v} $. $ L_e(\mathbf{p},\mathbf{v}) $ denotes self emission from $ \mathbf{p} $ toward $ \mathbf{v} $. $ \Omega $ is the visible hemisphere. $ f(\mathbf{l},\mathbf{v}) $ is the bidirectional reflectance distribution function~(BRDF), and $ L_i(\mathbf{p},\mathbf{l}) $ is the incident radiance from direction $ \mathbf{l} $. The reflected term is often decomposed as $ f = f_{\mathrm{diff}} + f_{\mathrm{spec}} $, where the diffuse term is direction independent. This yields:
\begin{equation}
\begin{aligned}
L_o^{\mathrm{diff}}(\mathbf{p}) &= f_{\mathrm{diff}} \int_{\Omega} L_i(\mathbf{p},\mathbf{l})\, (\mathbf{n}\!\cdot\!\mathbf{l})^{+}\, d\mathbf{l}, \\
L_o^{\mathrm{spec}}(\mathbf{p},\mathbf{v}) &= \int_{\Omega} f_{\mathrm{spec}}(\mathbf{l},\mathbf{v})\, L_i(\mathbf{p},\mathbf{l})\, (\mathbf{n}\!\cdot\!\mathbf{l})^{+}\, d\mathbf{l},
\end{aligned}
\label{eq:sup2}
\end{equation}
and thus:
\begin{equation}
\begin{aligned}
L_o(\mathbf{p},\mathbf{v}) \approx L_e(\mathbf{p},\mathbf{v}) + L_o^{\mathrm{diff}}(\mathbf{p}) + L_o^{\mathrm{spec}}(\mathbf{p},\mathbf{v}).
\end{aligned}
\label{eq:sup3}
\end{equation}

Lightmap baking targets the view independent diffuse component. Concretely, we store the lighting information in lightmaps:
\begin{equation}
L_{\mathrm{LM}}(\mathbf{p}) = \int_{\Omega} L_i(\mathbf{p},\mathbf{l})\, (\mathbf{n}\!\cdot\!\mathbf{l})^{+}\, d\mathbf{l}.
\label{eq:sup4}
\end{equation}

A high-quality lightmap typically includes direct illumination as well as multi-bounce indirect illumination. At runtime, we fetch the baked diffuse term using fragment UVs and apply it during shading.

Generating lightmaps requires UVs. Only texels that map to visible surfaces are valid, so we use a binary mask to flag valid regions. Resolution sets the trade-off between detail and memory.

\section{Block Compression}

Block compression~(BC) is a family of fixed-rate, lossy texture compression formats designed for real-time GPU decoding. The core idea originates from Block Truncation Coding~(BTC)~\cite{1094560}: the image is partitioned into small blocks (\eg $4\times4$ texels), and the color values within each block are approximated by a compact set of representative colors together with per texel selection indices.

In the simplest and most representative case~\cite{DXTC}, each $4\times4$ block stores two color \emph{endpoints} $e_1, e_2$ and a set of per texel interpolation weights $\{w_p\}$. At decode time, the color of each texel $p$ is reconstructed by linearly interpolating between the endpoints:
\begin{equation}
c_p = (1 - w_p)\,e_1 + w_p\,e_2,\qquad p=1,\dots,16.
\label{eq:bc_interp}
\end{equation}
More advanced variants (\eg BC7) extend this scheme by supporting multiple partitions within a single block, each with its own pair of endpoints, thereby improving quality at the cost of a more complex encoding.
This design enables constant-time random access to any texel, which is critical for GPU texture sampling. Different BC variants target different use cases.

Adaptive Scalable Texture Compression~(ASTC)~\cite{10.5555/2383795.2383812} generalizes this framework with flexible block sizes, which provides a continuous trade-off between quality and bitrate. ASTC supports LDR, HDR, and 3D textures and is widely adopted on modern mobile platforms.

A key advantage of block compression is that decoding is performed entirely in hardware during texture sampling, imposing small computational overhead on the shader pipeline. In NDGI, we leverage this property by applying BC7 to our learned feature maps $\mathbf{F}^{2D}_{uv}$ and $\mathbf{F}^{3D}_{uvt}$.

\section{Virtual Texturing}

Virtual texturing~(VT)~\cite{vt}, also known as megatexture or sparse virtual texturing, is a streaming technique that decouples the logical texture space from the physical GPU memory. It enables applications to reference texture data far exceeding the available video memory, loading only the portions that are actually visible.

The key data structures of a VT system are:
\begin{itemize}
\item \textbf{Virtual texture.} The entire logical texture, which may be many gigabytes and is divided into fixed-size \emph{tiles} (\eg $128\times128$ or $256\times256$ texels). Tiles are the atomic unit of streaming.
\item \textbf{Physical texture.} A GPU-resident texture atlas that holds only the currently needed tiles. Its capacity is bounded by available video memory.
\item \textbf{Page table.} An indirection texture that maps each virtual tile to its location in the physical texture. During shading, the shader first samples the page table with the fragment's UV coordinates to obtain a physical tile address, and then samples the physical texture to retrieve the actual texel data.
\end{itemize}

At runtime, the VT system operates in a feedback-driven loop. First, the GPU renders a low-resolution \emph{feedback buffer} that records which virtual tiles are required for the current view. The CPU then reads back the feedback, identifies missing tiles, and transfers them from disk into the physical texture. Finally, the page table is updated to reflect the new mapping. Tiles that have not been referenced within a interval may be evicted to reclaim capacity for newly requested tiles.

This architecture benefits NDGI in two ways. First, only the tiles visible in the current frame need to be decoded, reducing runtime decompression cost. Second, once decoded, a tile remains cached in the physical texture across frames until the lighting state changes, avoiding repeated inference. In our pipeline, each lightmap is partitioned into fixed-size tiles with a dedicated NDGI model per tile. The VT system identifies required tiles each frame and invokes neural decompression via a compute shader. Decoded results are written into the physical texture and invalidated only when re-decoding is needed. To support hardware texture filtering, each tile is stored with a small border (\eg 4 pixels per edge). We apply the same mirrored padding during training, ensuring seamless tile boundaries.

\section{Dataset}
\label{sec:dataset}
\subsection{Dataset Description}
Our dataset comprises baked lightmap data across multiple scenes. For each scene, we provide two types of files: lightmap data files and mask files. The lightmap file stores 3-channel (RGB) lighting textures. We bake 24 sets per day at hourly intervals aligned to the top of the hour. For scenes that exhibit light-switching behavior, we additionally bake two extra sets around the on/off transition times to better capture fast lighting changes.

The mask is a single-channel image with the same spatial resolution as the lightmaps. Each texel indicates whether the corresponding lightmap texel is valid. In real-time rendering, only valid regions are sampled, and compression methods can leverage this information to further improve compression ratio.

We also provide a per scene configuration file that records the spatial resolution of each lightmap set and the correspondence among lightmaps, masks, and time indices. Unless otherwise noted, all lightmap values are treated as linear HDR intensities for evaluation and visualization.

\subsection{Dataset Evaluation}
We evaluate lightmap reconstruction quality using PSNR, SSIM~\cite{wang2004image}, and LPIPS~\cite{zhang2018unreasonable}. Metrics are computed in a tiled manner by partitioning each lightmap into non-overlapping \(128\times128\) tiles. For each tile, we compute the metric within the valid region defined by the mask and report the result by averaging over all tiles across all lightmaps of that scene.

Because the data are HDR, we determine the dynamic range per tile: the minimum and maximum values within the tile’s valid region set the range (\eg the peak value for PSNR), rather than adopting fixed global constants. Unless specified otherwise, metrics are computed in linear RGB. This tiled, mask-aware protocol yields stable and comparable scores across scenes with different UV packings and sparsity patterns.

\section{Ablation Study}
\begin{figure}[!t]
\centering
{\setlength{\tabcolsep}{2pt}%
\begin{tabular}{ccc}
\includegraphics[width=0.33\linewidth]{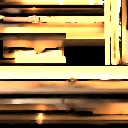} &
\includegraphics[width=0.33\linewidth]{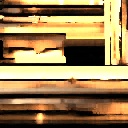} &
\includegraphics[width=0.33\linewidth]{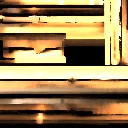} \\
{\scriptsize Reference} & {\scriptsize 2D Feature Map} & {\scriptsize Hybrid Feature Map} \\
{\scriptsize PSNR($\uparrow$)} & {\scriptsize 41.12 dB} & {\scriptsize \textbf{44.77 dB}} \\
\end{tabular}}
\caption{At comparable bitrates and with the same decoder, our hybrid feature maps outperform 2D-only feature maps, producing less noise.}
\label{fig:feature_ablation}
\end{figure}

\begin{figure}[!t]
\centering
{\setlength{\tabcolsep}{2pt}%
\begin{tabular}{ccc}
\includegraphics[width=0.33\linewidth]{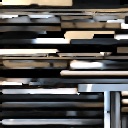} &
\includegraphics[width=0.33\linewidth]{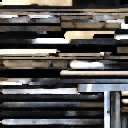} &
\includegraphics[width=0.33\linewidth]{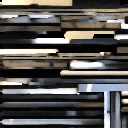} \\
{\scriptsize Reference} & {\scriptsize w/o BC simulation} & {\scriptsize w/ BC simulation} \\
{\scriptsize PSNR($\uparrow$)} & {\scriptsize 39.82 dB} & {\scriptsize \textbf{44.92 dB}} \\
\end{tabular}}
\caption{We compare variants with and without the BC simulation strategy and report PSNR. Parameterizing the feature map with BC endpoints and weights yields less noisy final reconstructions than directly optimizing feature parameters.}
\label{fig:bc_ablation}
\end{figure}

\begin{table}[b]
  \centering
  \caption{PSNR comparison for feature maps with different structures at comparable bitrates.}
  \label{tab:featuremap_abla}
  \setlength{\tabcolsep}{10pt}
  \renewcommand{\arraystretch}{1.15}
  \resizebox{\columnwidth}{!}{%
  \scriptsize
  \begin{tabular}{c|c|c}
    \hline
    Feature Map Structure & BPP($\downarrow$) & PSNR($\uparrow$) \\
    \midrule
    2D Feature Maps & 0.73 & 42.1 dB \\
    3D Feature Maps & 0.69 & 36.6 dB \\
    Hybrid Feature Maps & \textbf{0.67} & \textbf{44.9 dB}  \\
    \hline
  \end{tabular}%
  }
\end{table}

\begin{table}[b]
  \centering
  \caption{PSNR comparison for modeling feature maps with endpoints and weights or not.}
  \label{tab:bc_ablatable}
  {\renewcommand{\arraystretch}{1.15}%
  { \small
    \begin{tabular}{c|c|c|c}
    \hline
    Feature Map Strategy & PSNR($\uparrow$) & SSIM($\uparrow$) & LPIPS($\downarrow$) \\
    \midrule
     w/o BC Simulation & 37.96 dB & 0.975 & 0.013\\
     w/ BC Simulation & \textbf{44.20 dB} & \textbf{0.992} & \textbf{0.004}\\
    \hline
  \end{tabular}
  }
  }
\end{table}

We validate the effectiveness of our hybrid feature representation. Figure~\ref{fig:feature_ablation} compares a baseline that uses only 2D feature maps with our hybrid features at comparable bitrates under the same decoder configuration. The hybrid representation captures lightmap content more faithfully, yielding smoother shading transitions and less aliasing, and it consistently improves reconstruction quality across viewpoints. Table~\ref{tab:featuremap_abla} reports PSNR on a test scene across several feature map architectures at matched bitrates, showing that the hybrid representation attains higher accuracy with similar storage.

We also evaluate a BC simulation strategy and observe clear gains across test scenes and metrics. The results are shown in Figure~\ref{fig:bc_ablation} and Table~\ref{tab:bc_ablatable}. To ensure a fair comparison, both variants apply BC compression after training, while they differ in how the feature map is parameterized during optimization. In the first variant, we model the feature map with BC endpoints and weights during training, which better matches the target representation, encourages piecewise linear structure, and improves rate distortion behavior. In the second variant, we directly optimize the full feature map and compress it with BC at the end, which creates a mismatch with the BC quantization grid and introduces high frequency noise and block inconsistency. Empirically, the endpoints and weights modeling yields consistently higher fidelity and fewer artifacts, whereas direct optimization introduces substantial noise that degrades the final quality.

\begin{table}[t]
  \centering
  \caption{Tri-plane vs.\ 8-channel $\mathbf{F}^{2D}_{uv}$ on NDGI.M. Tri-plane yields smaller storage at higher quality.}
  \label{tab:triplane_vs_fuv}
  \setlength{\tabcolsep}{9pt}
  \renewcommand{\arraystretch}{1.15}
  \resizebox{\columnwidth}{!}{%
  \begin{tabular}{l|c|c|c}
    \hline
    Method & PSNR $\uparrow$ & BPP $\downarrow$ & Size (MB) $\downarrow$\\
    \midrule
    8-channel $\mathbf{F}^{2D}_{uv}$ & 41.32 & 0.86 & 358 \\
    Tri-Plane & \textbf{41.50} & \textbf{0.67} & \textbf{279} \\
    \hline
  \end{tabular}%
  }
\end{table}

We further compare our tri-plane hybrid feature representation against a single 8-channel $\mathbf{F}^{2D}_{uv}$ baseline on NDGI.M. As shown in Table~\ref{tab:triplane_vs_fuv}, the tri-plane design achieves comparable or higher PSNR at a noticeably lower bitrate and smaller storage footprint. We attribute this to the additional $\mathbf{F}^{2D}_{ut}$ and $\mathbf{F}^{2D}_{vt}$ planes, which more effectively capture higher-frequency temporal variations that a single spatial feature map cannot represent as compactly.

\section{Rendered Results}
We provide additional qualitative comparisons of rendered results in the supplementary. Under matched bitrates, Figure~\ref{fig:rendered} compares PRT~\cite{sloan2023precomputed}, NTC L.~\cite{Vaidyanathan_2023} and our NDGI~M.~(Ours) against the reference. Our method delivers cleaner global illumination, fewer color shifts and less noise, and better preservation of fine details, which leads to higher visual fidelity relative to the reference. Row annotations report BPP, PSNR, and SSIM, and our method consistently achieves a better trade-off between quality and bitrate across scenes.

\begin{figure*}[t]
\centering
\setlength{\tabcolsep}{2pt}
\renewcommand{\arraystretch}{1}

\resizebox{\textwidth}{!}
{
\begin{tabular}{c|ccc|c}
Lit Scene &
PRT &
NTC L. &
NDGI M.~(Ours) &
Reference \\
\midrule
\includegraphics[width=0.16\linewidth]{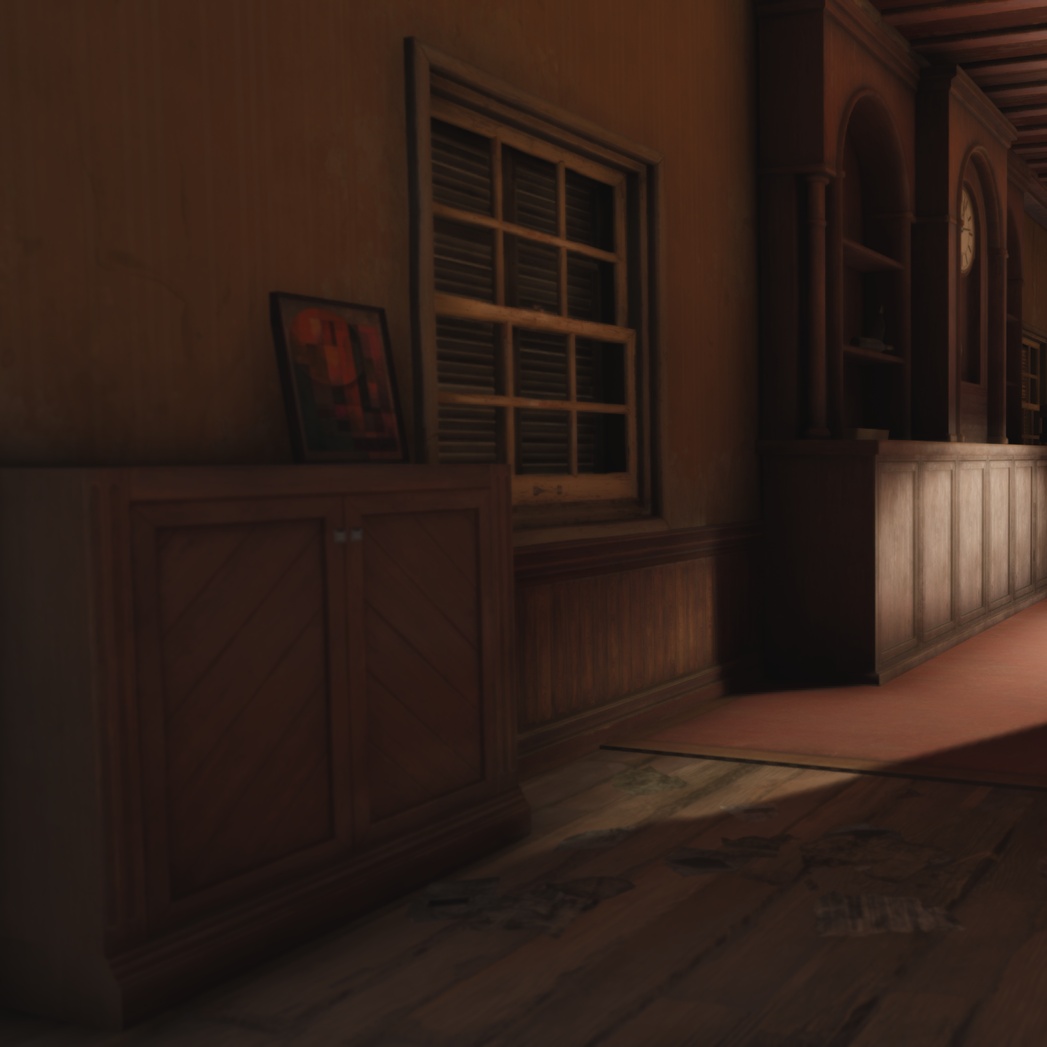} &
\includegraphics[width=0.16\linewidth]{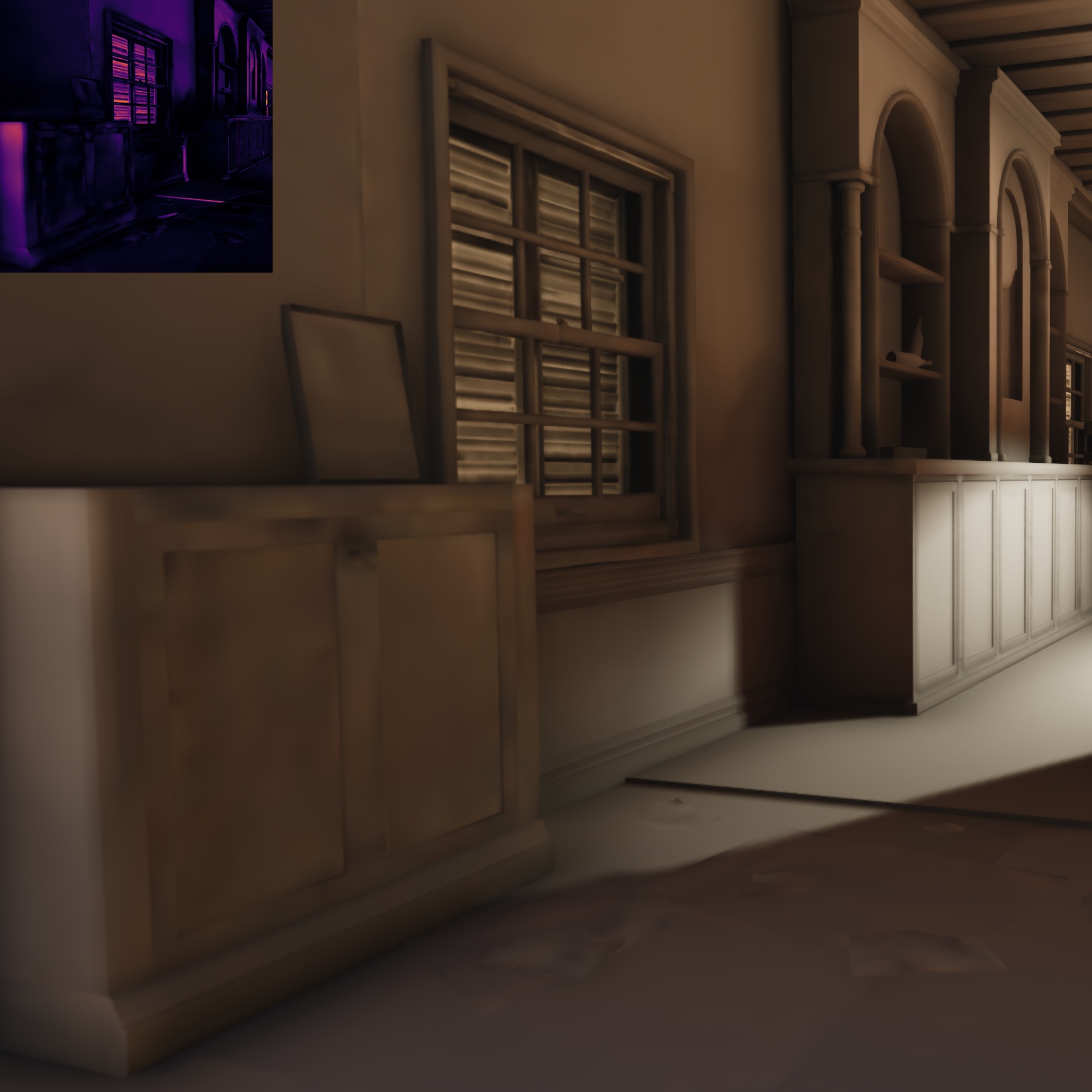} &
\includegraphics[width=0.16\linewidth]{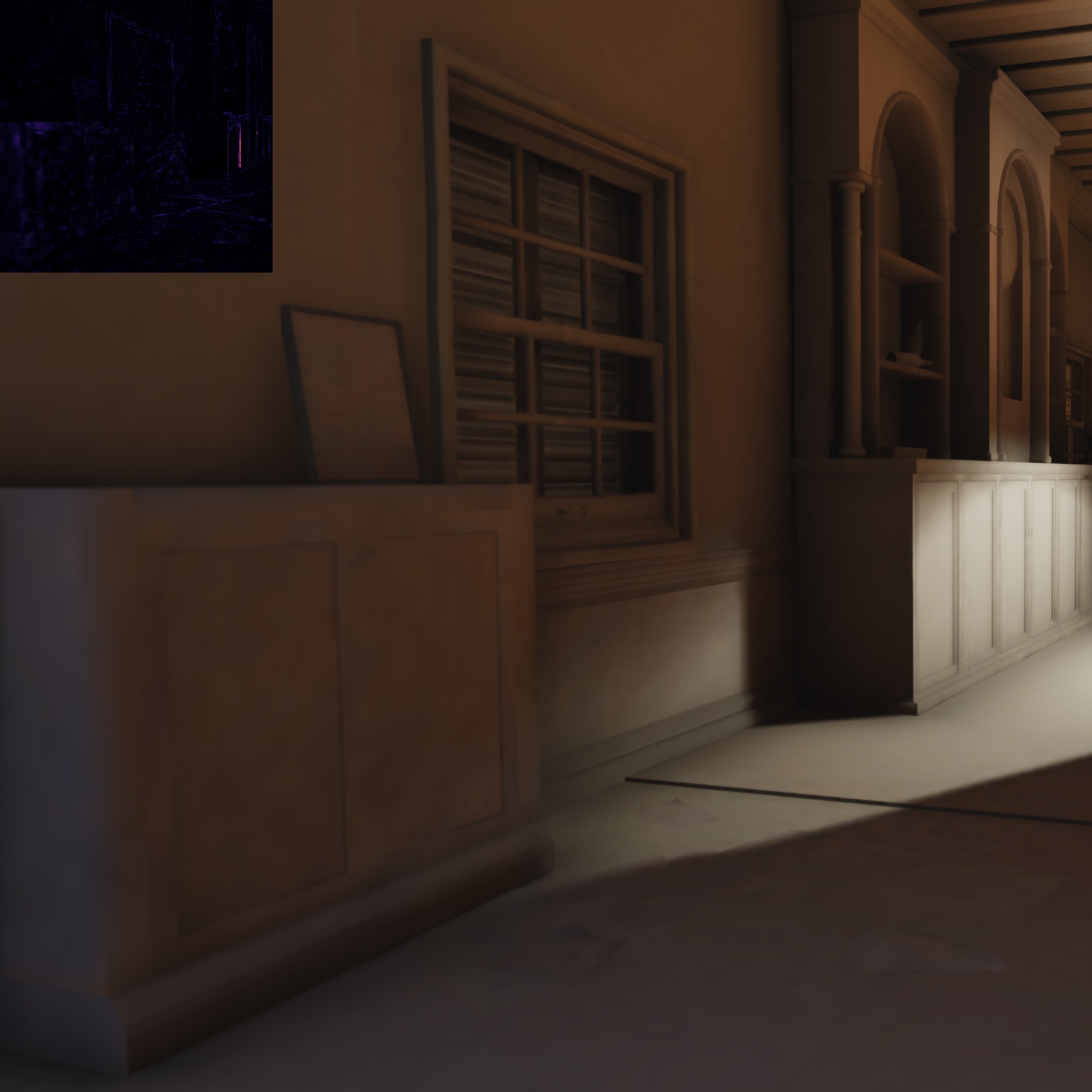} &
\includegraphics[width=0.16\linewidth]{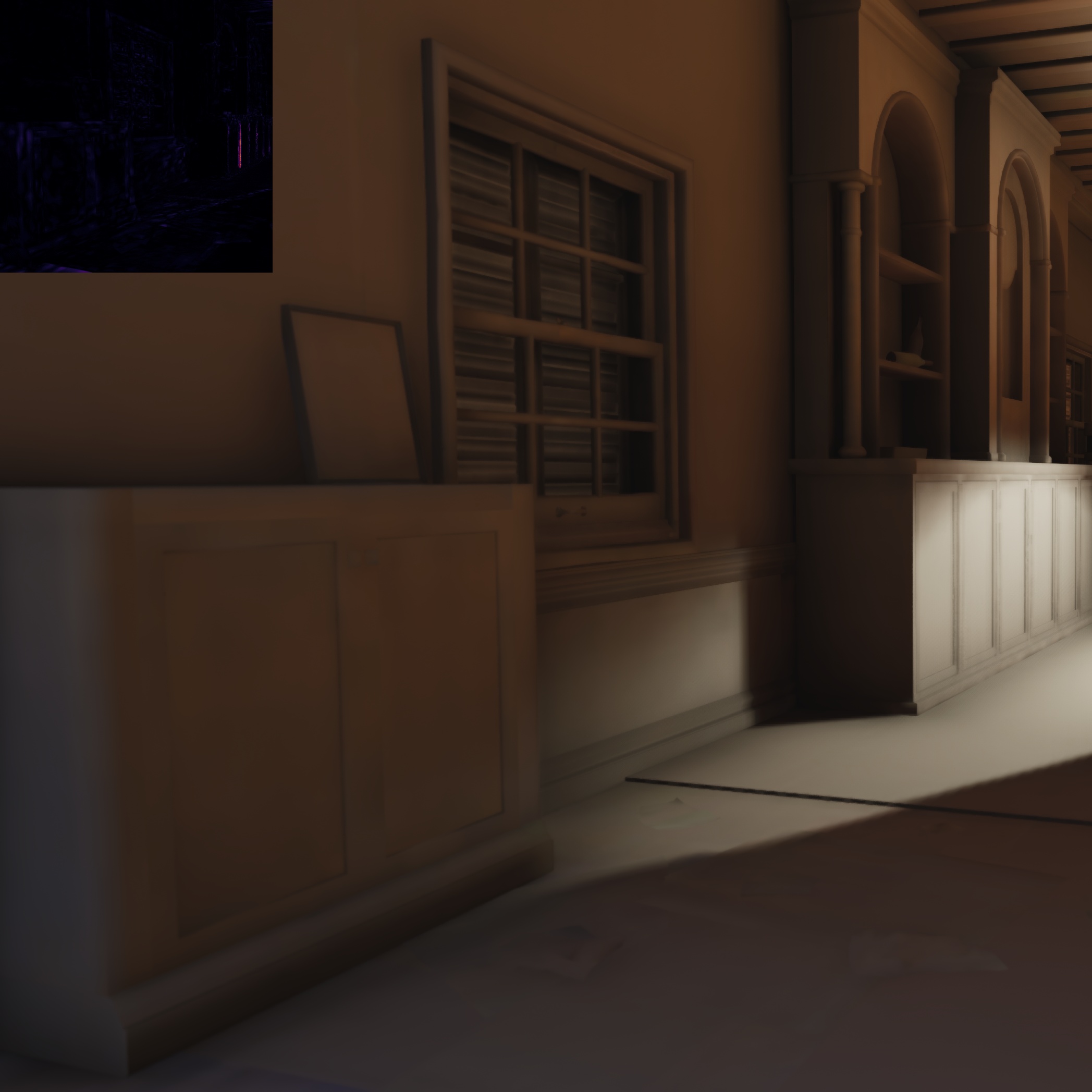} &
\includegraphics[width=0.16\linewidth]{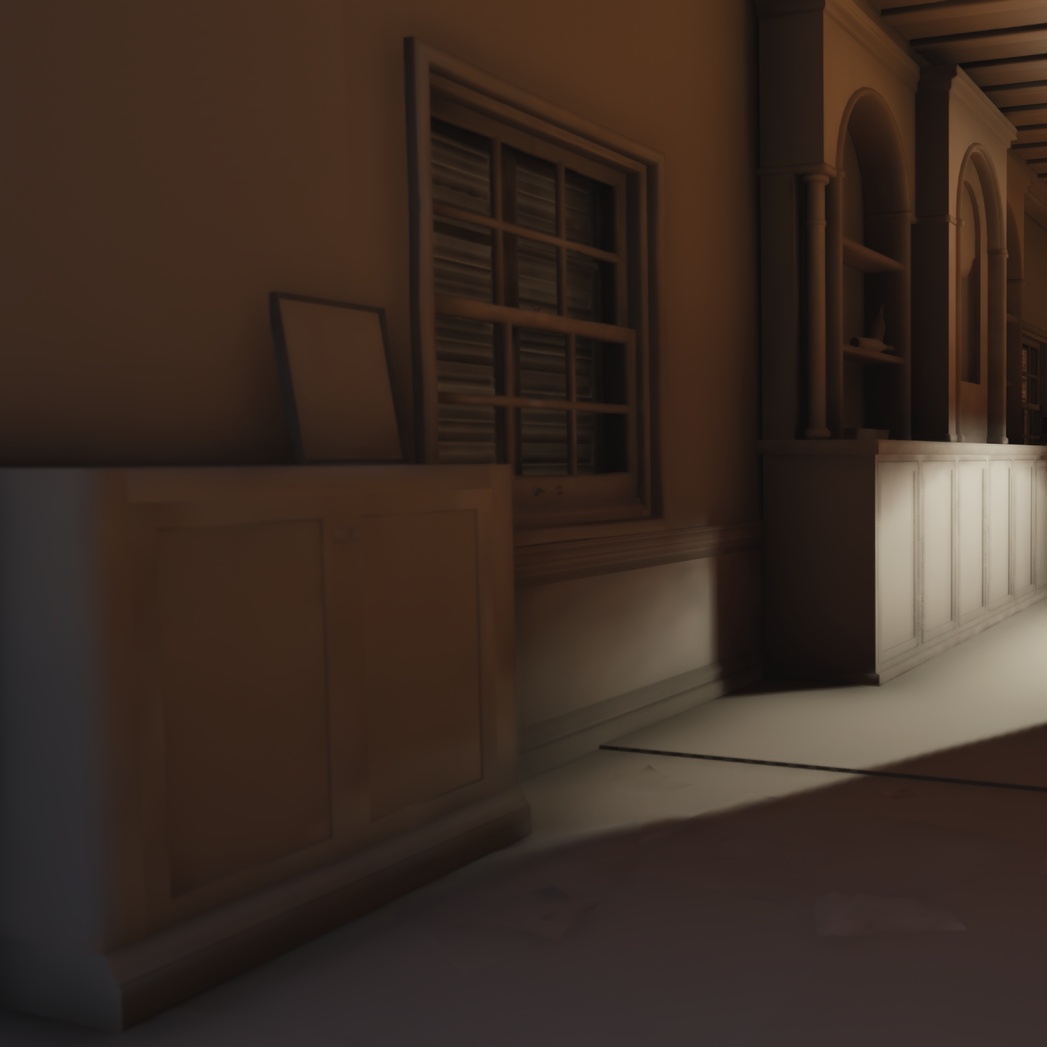} \\
{\scriptsize BPP($\downarrow$), PSNR($\uparrow$), SSIM($\uparrow$)} &
{\scriptsize 0.92, 26.09 dB, 0.921} &
{\scriptsize 0.78, 45.56 dB, 0.998} &
{\scriptsize \textbf{0.67, 47.53 dB, 0.999}} &
{\scriptsize FarmLand} \\
\includegraphics[width=0.16\linewidth]{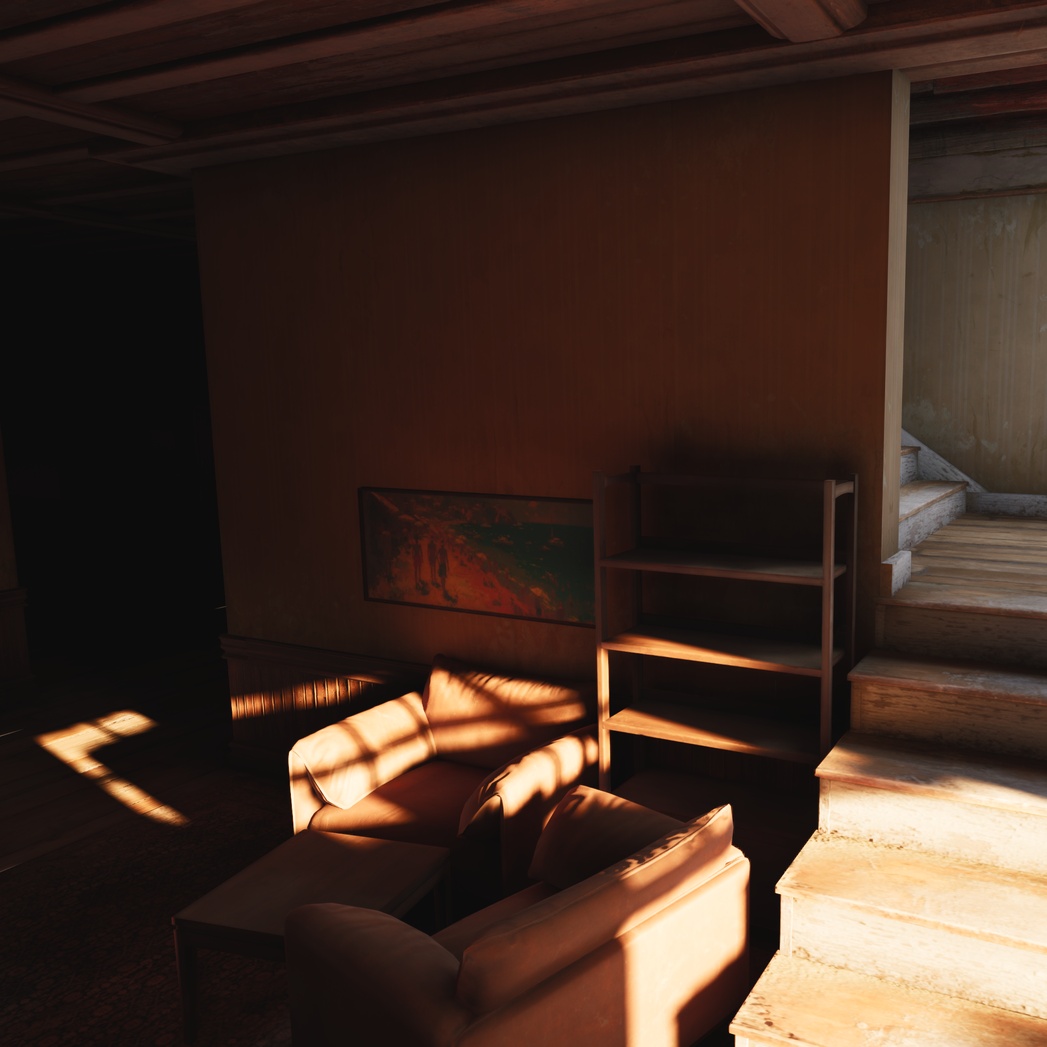} &
\includegraphics[width=0.16\linewidth]{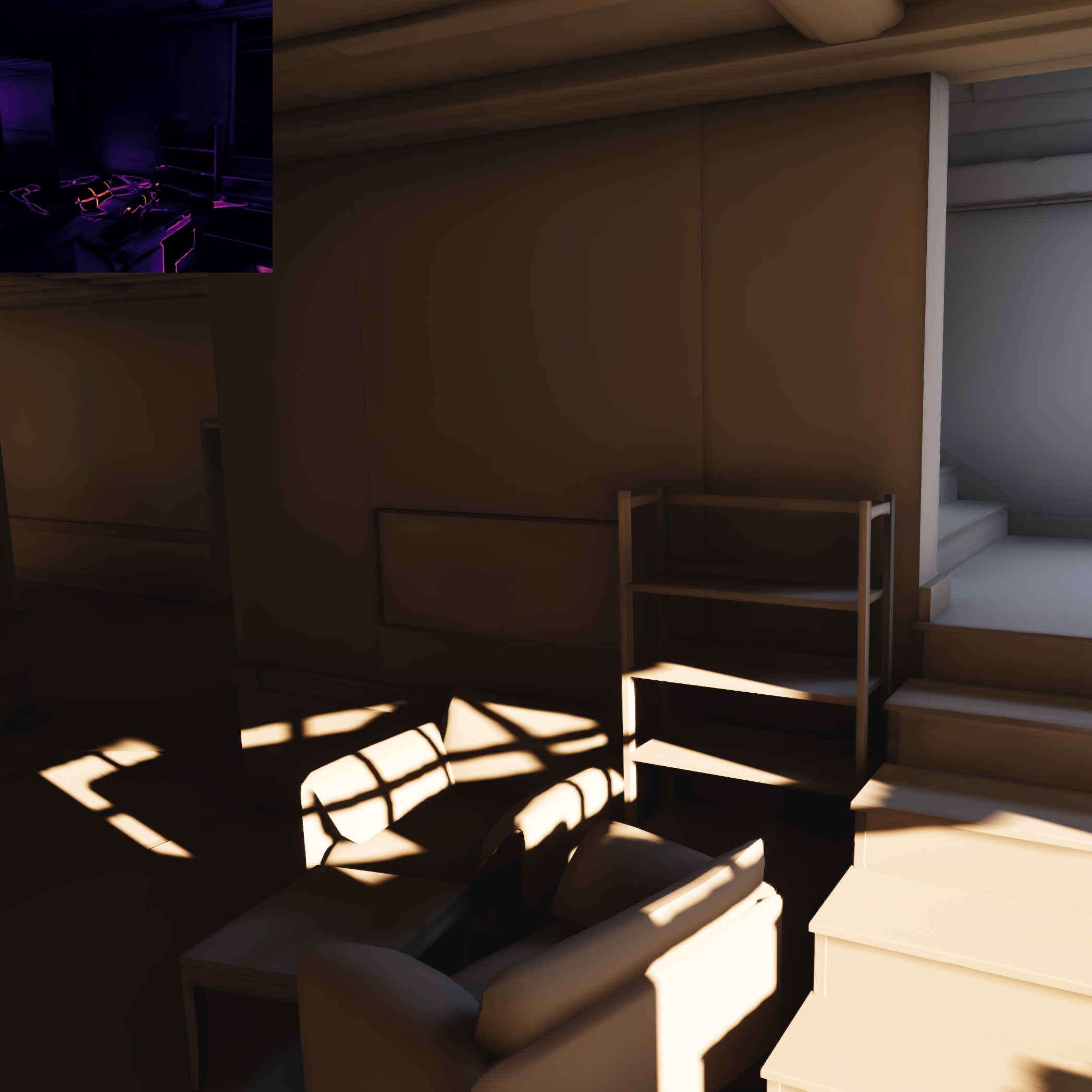} &
\includegraphics[width=0.16\linewidth]{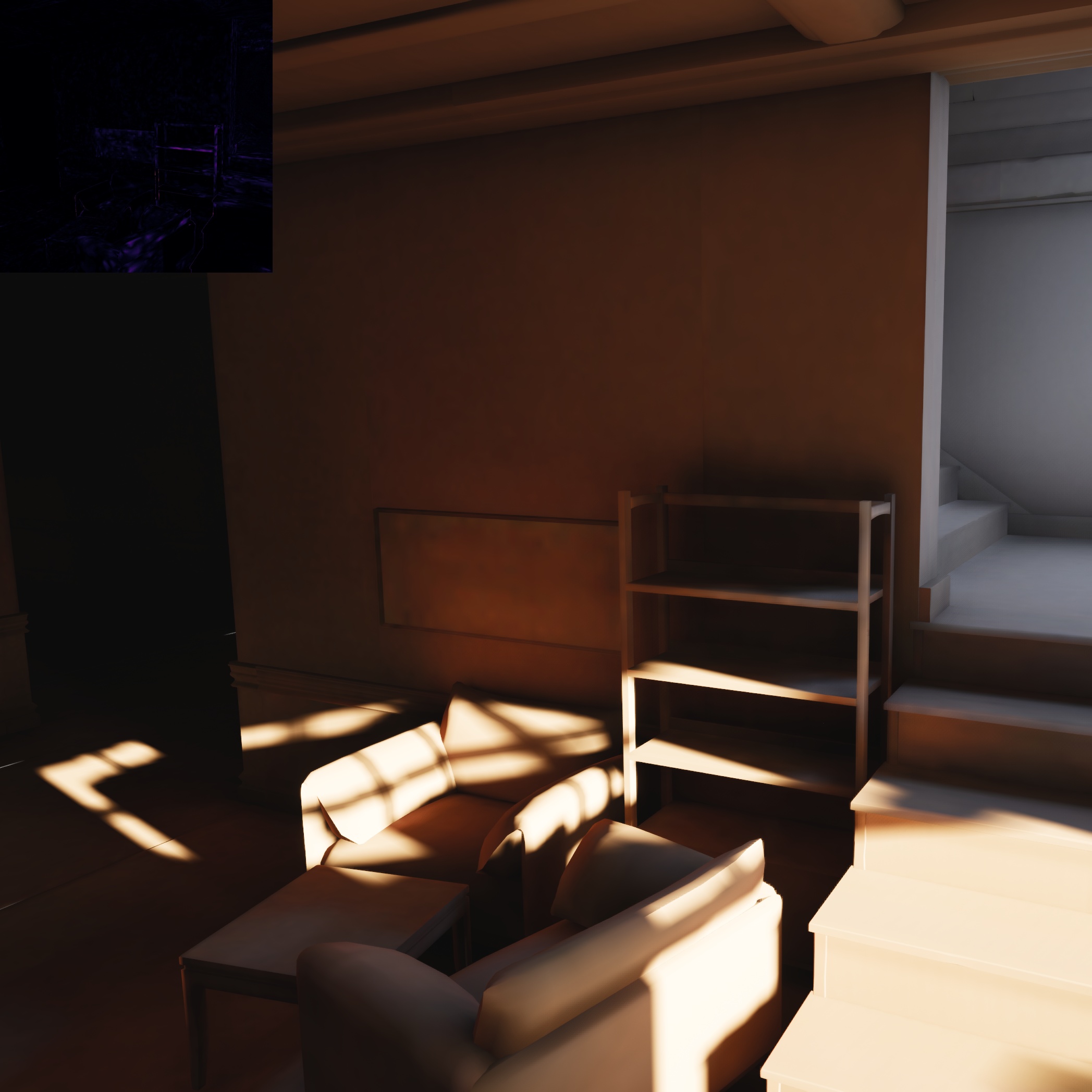} &
\includegraphics[width=0.16\linewidth]{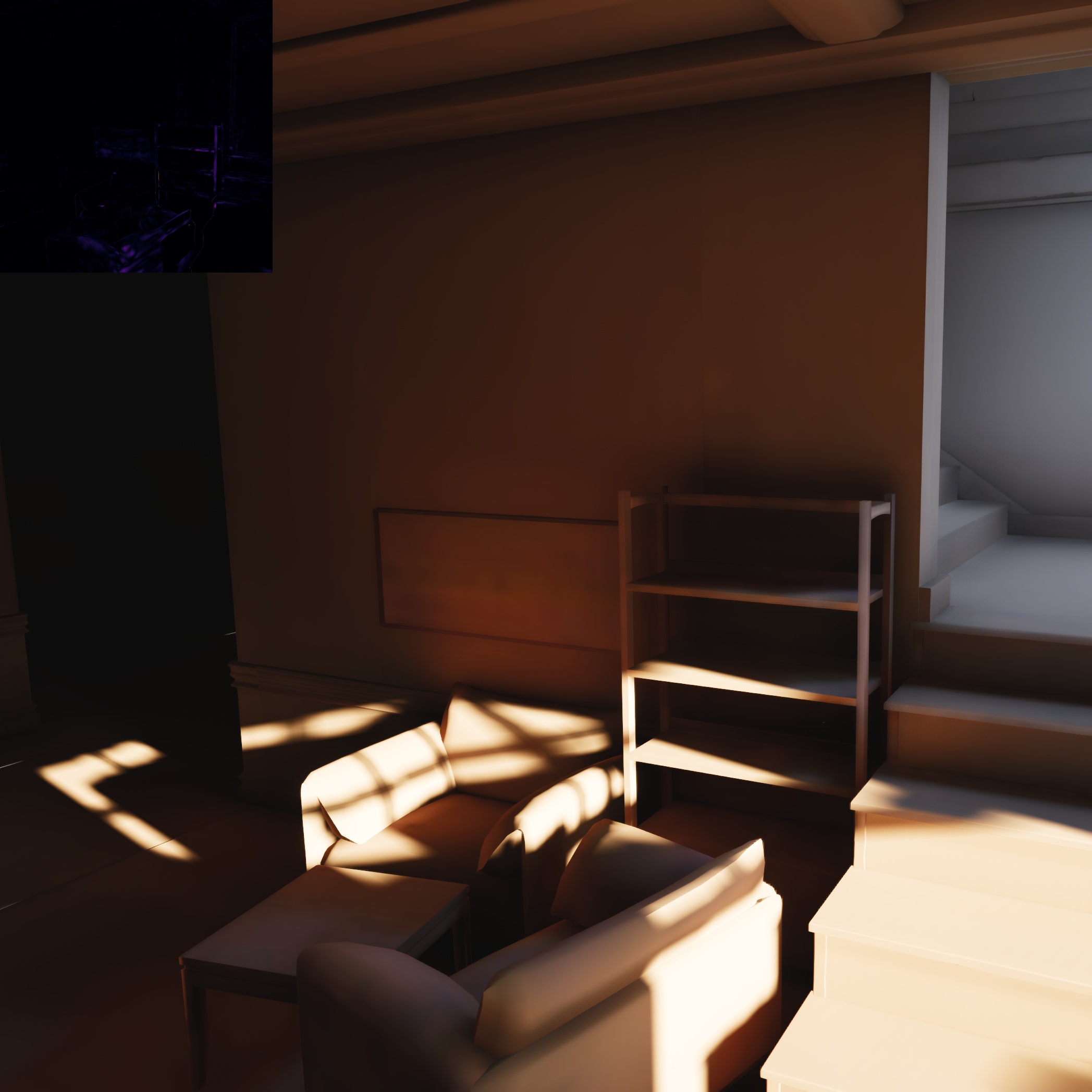} &
\includegraphics[width=0.16\linewidth]{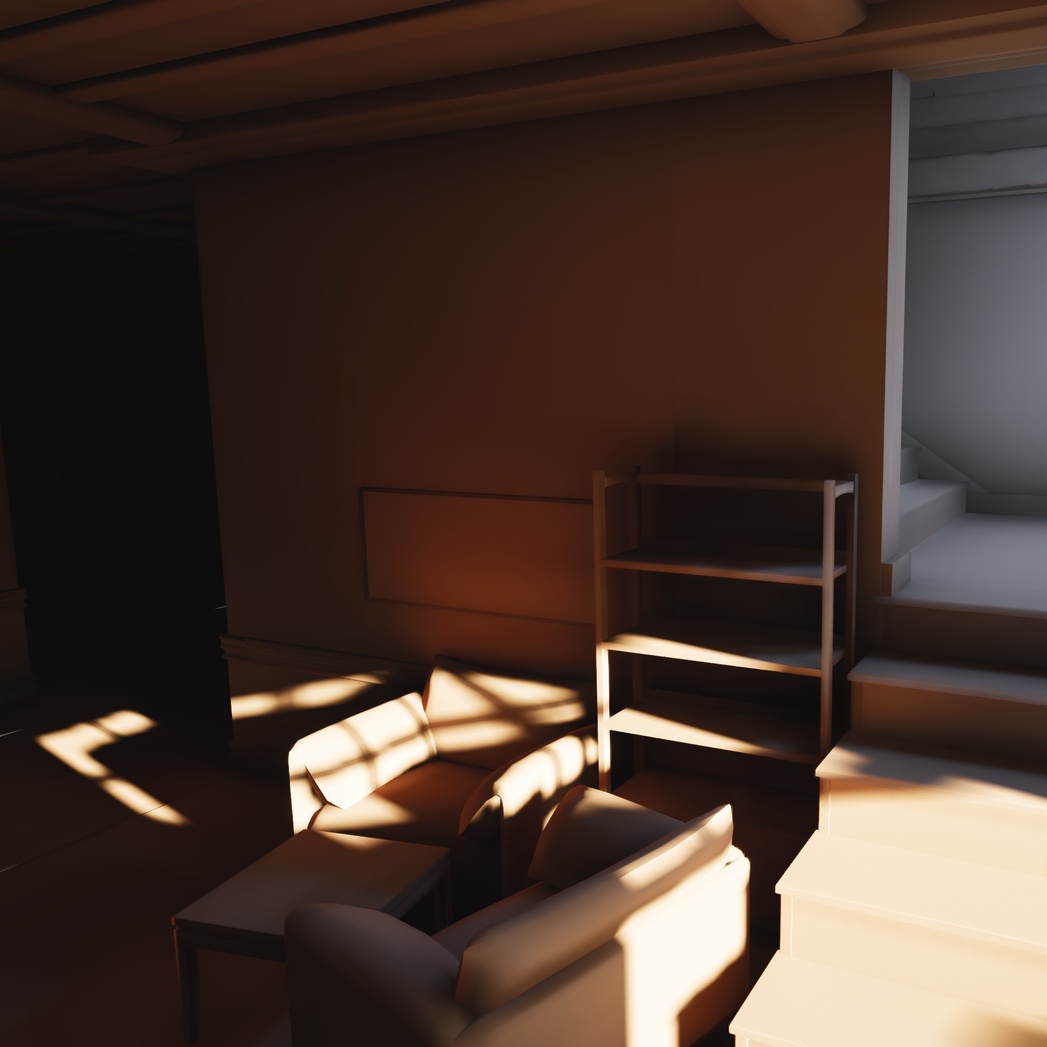} \\
{\scriptsize BPP($\downarrow$), PSNR($\uparrow$), SSIM($\uparrow$)} &
{\scriptsize 0.92, 24.39 dB, 0.971} &
{\scriptsize 0.78, 46.28 dB, 0.998} &
{\scriptsize \textbf{0.67, 47.55 dB, 0.999}} &
{\scriptsize FarmLand} \\
\includegraphics[width=0.16\linewidth]{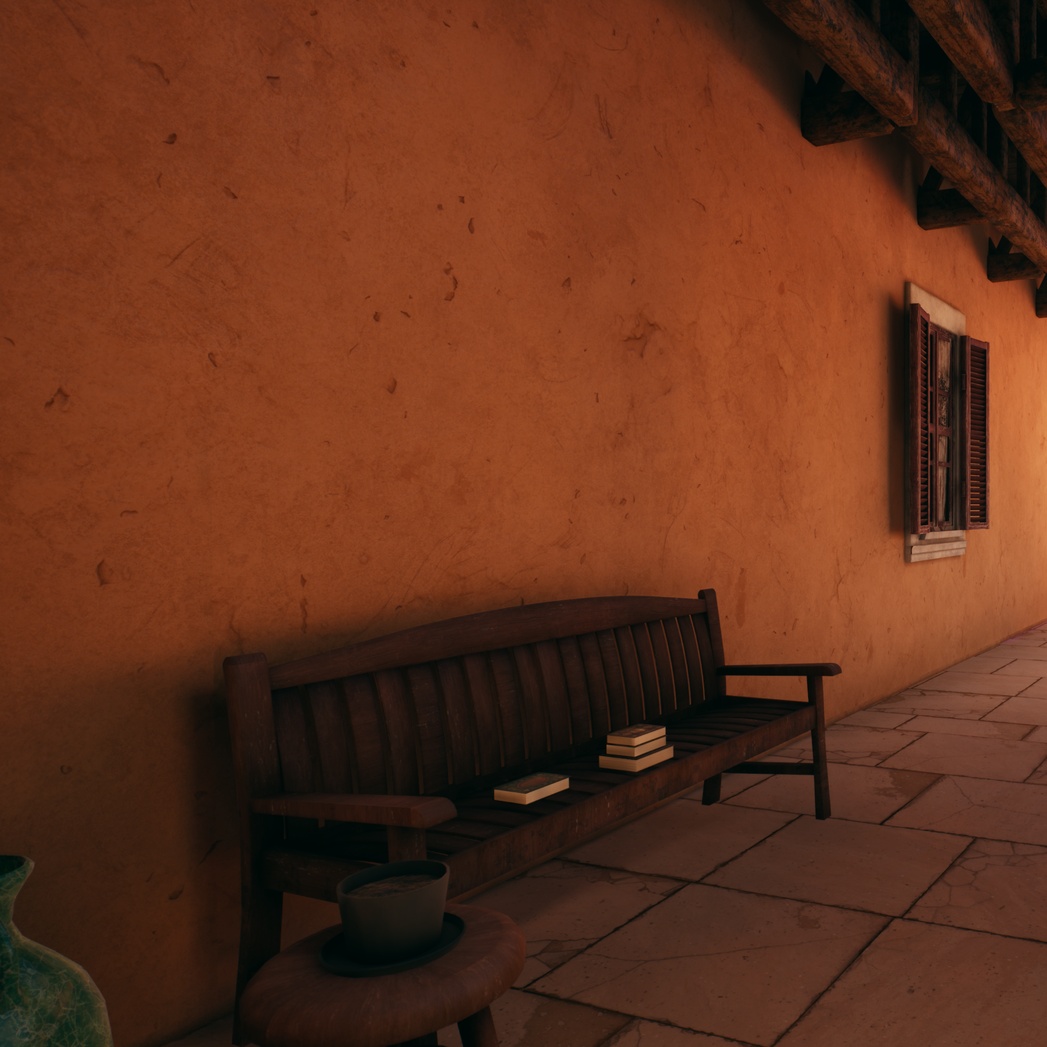} &
\includegraphics[width=0.16\linewidth]{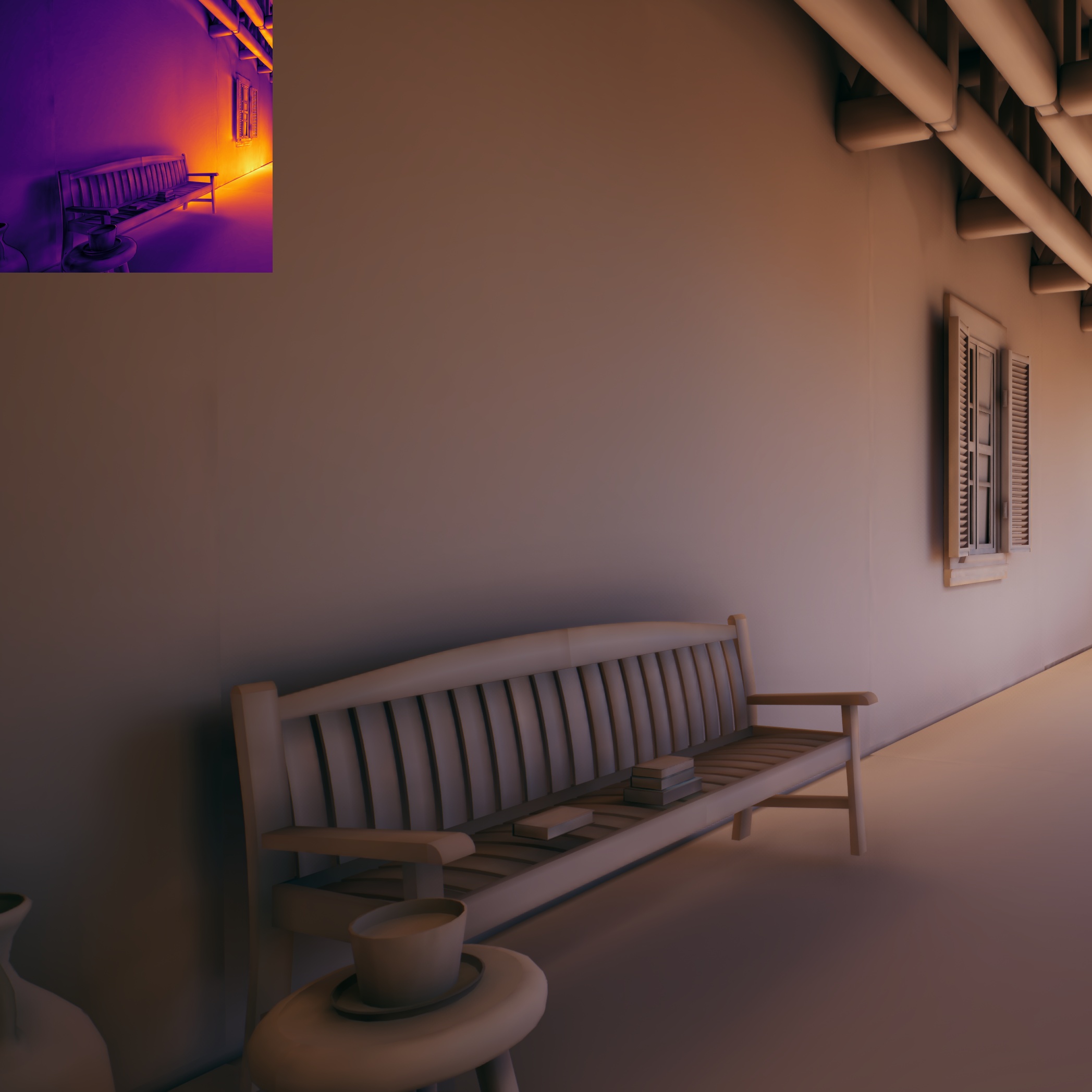} &
\includegraphics[width=0.16\linewidth]{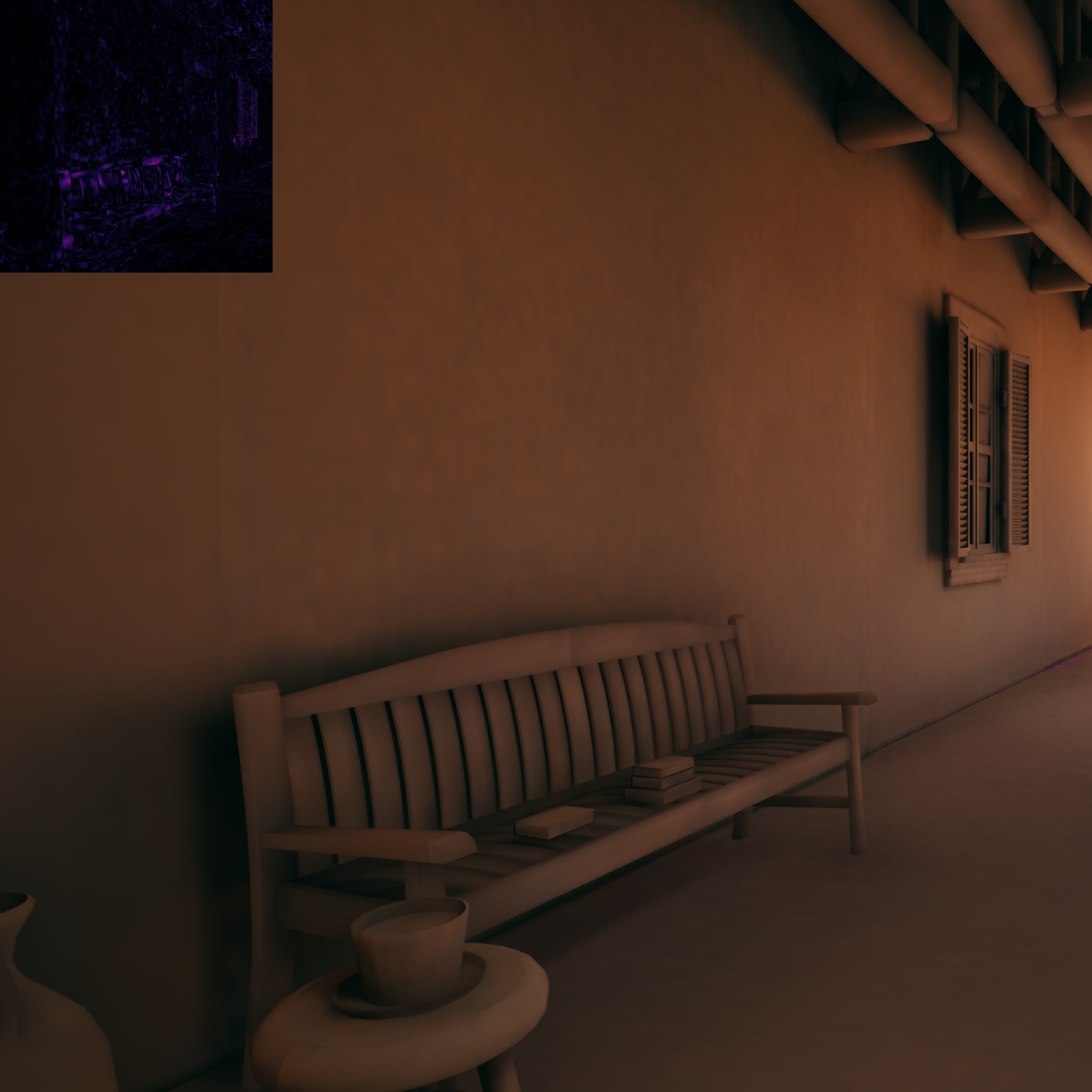} &
\includegraphics[width=0.16\linewidth]{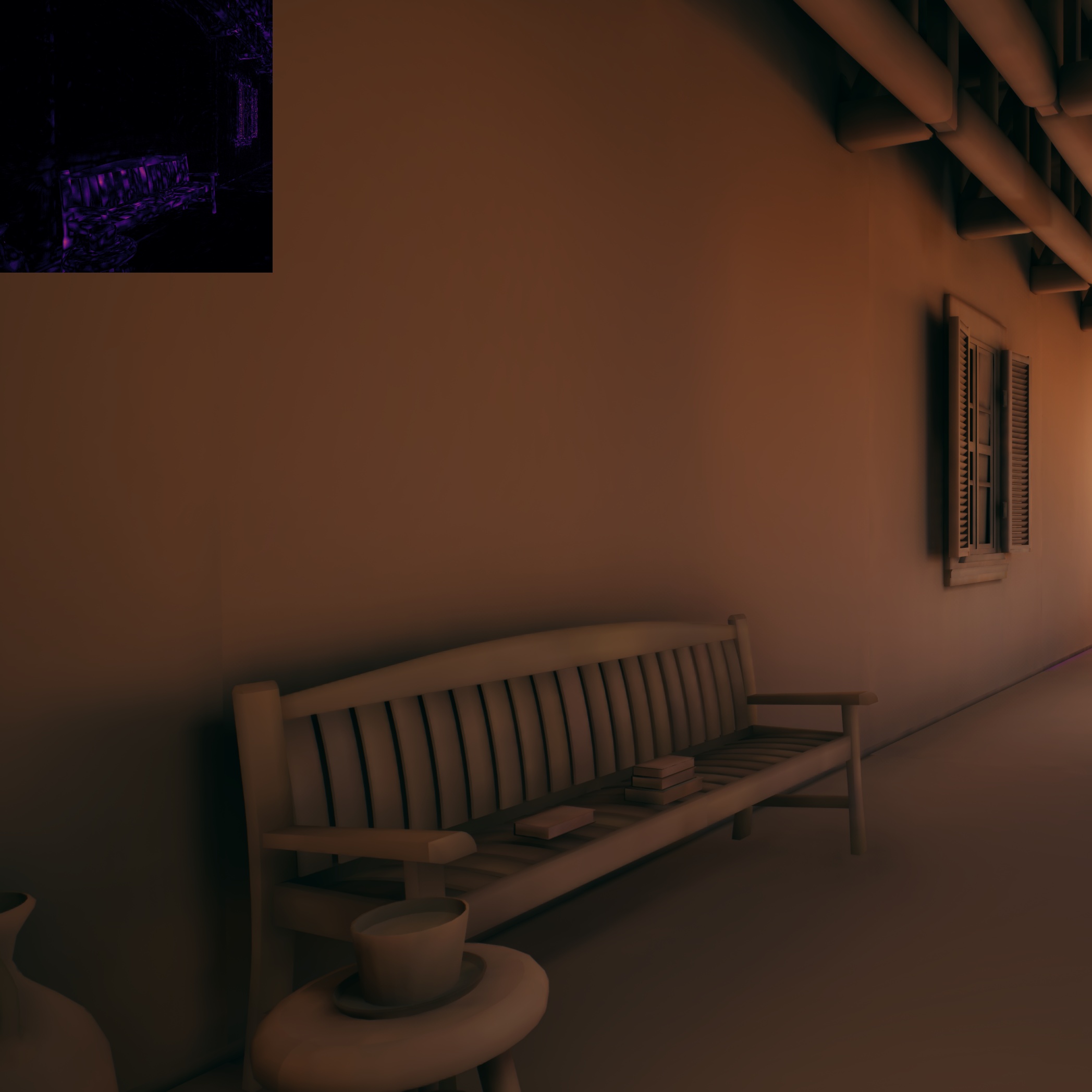} &
\includegraphics[width=0.16\linewidth]{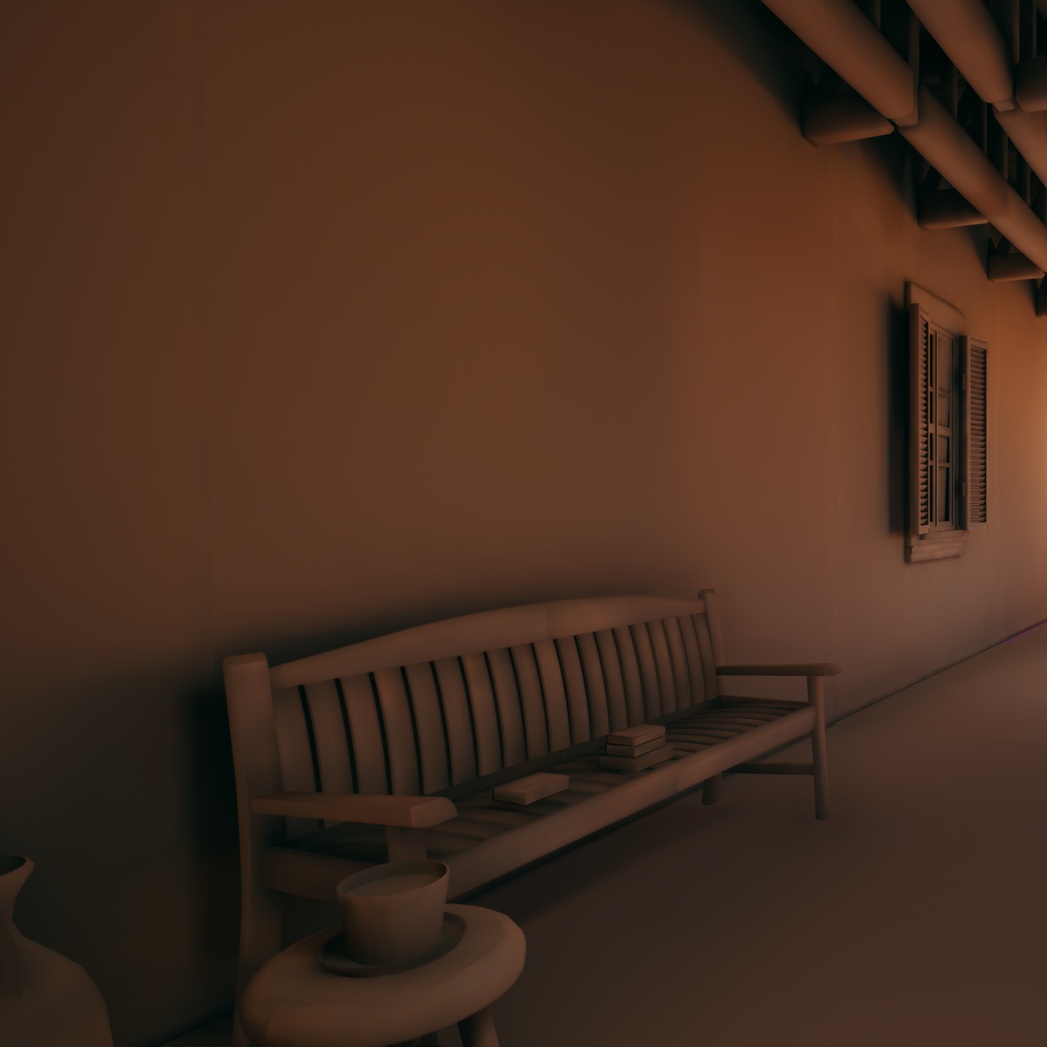}  \\
{\scriptsize BPP($\downarrow$), PSNR($\uparrow$), SSIM($\uparrow$)} &
{\scriptsize 1.00, 17.45 dB, 0.719} &
{\scriptsize 0.78, 45.41 dB, 0.997} &
{\scriptsize \textbf{0.71, 47.42 dB, 0.998}} &
{\scriptsize Yard} \\
\includegraphics[width=0.16\linewidth]{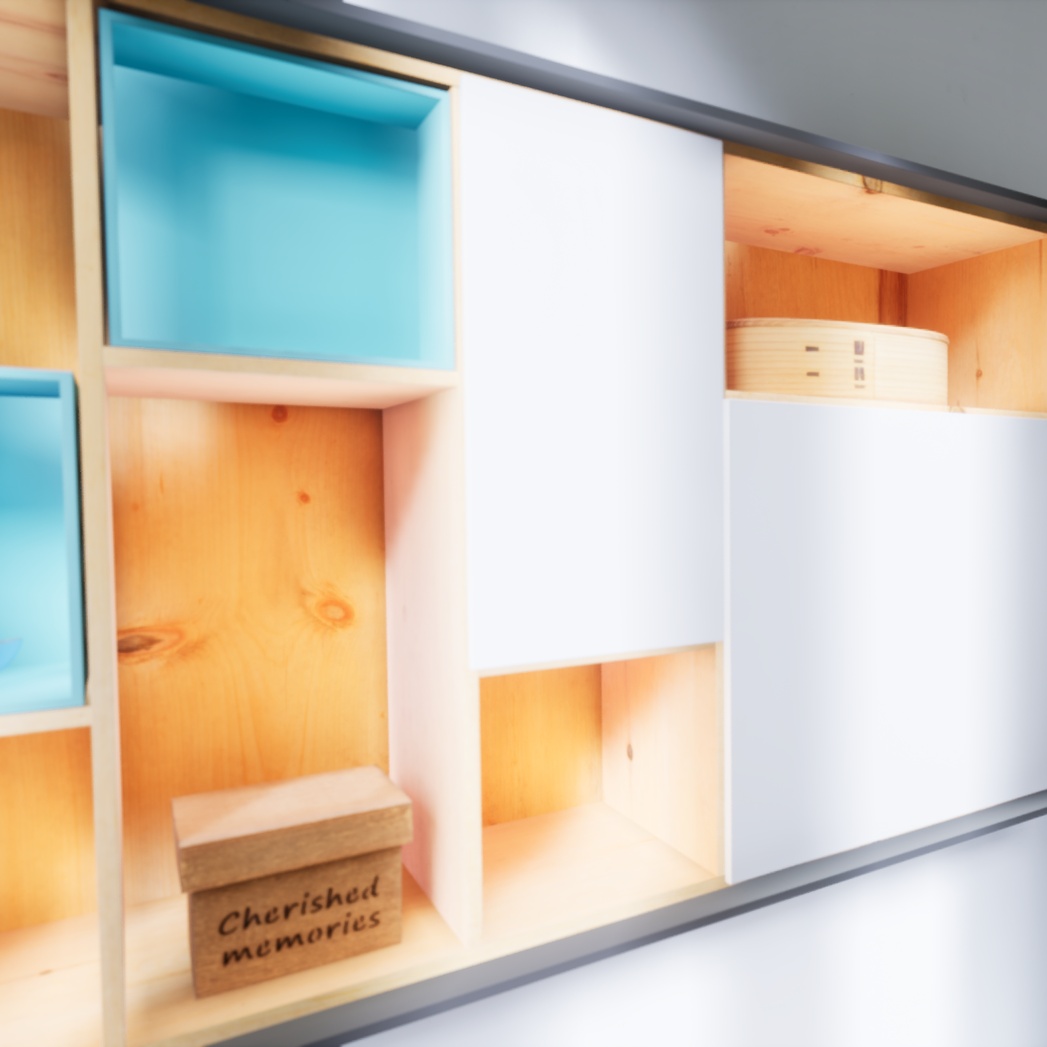} &
\includegraphics[width=0.16\linewidth]{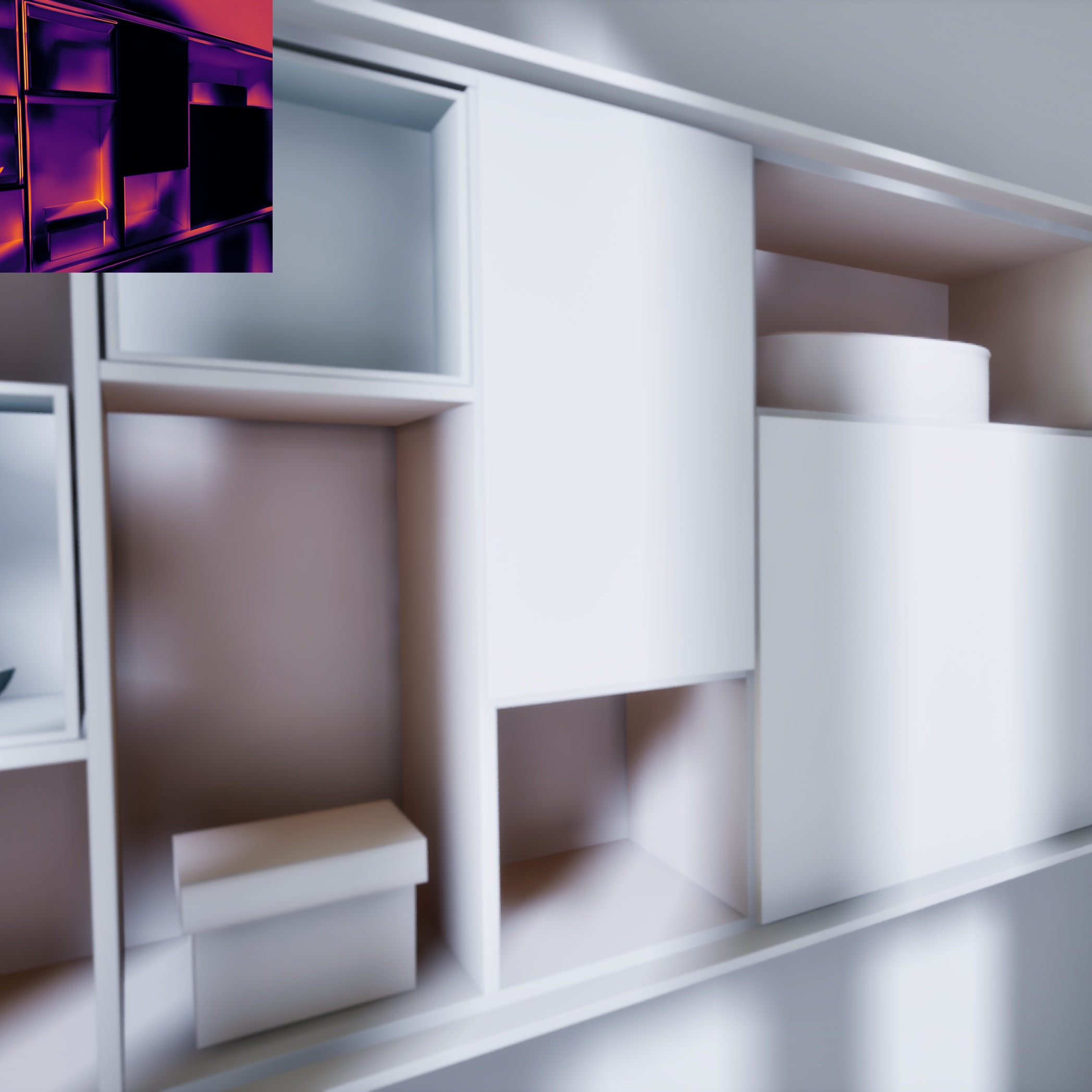} &
\includegraphics[width=0.16\linewidth]{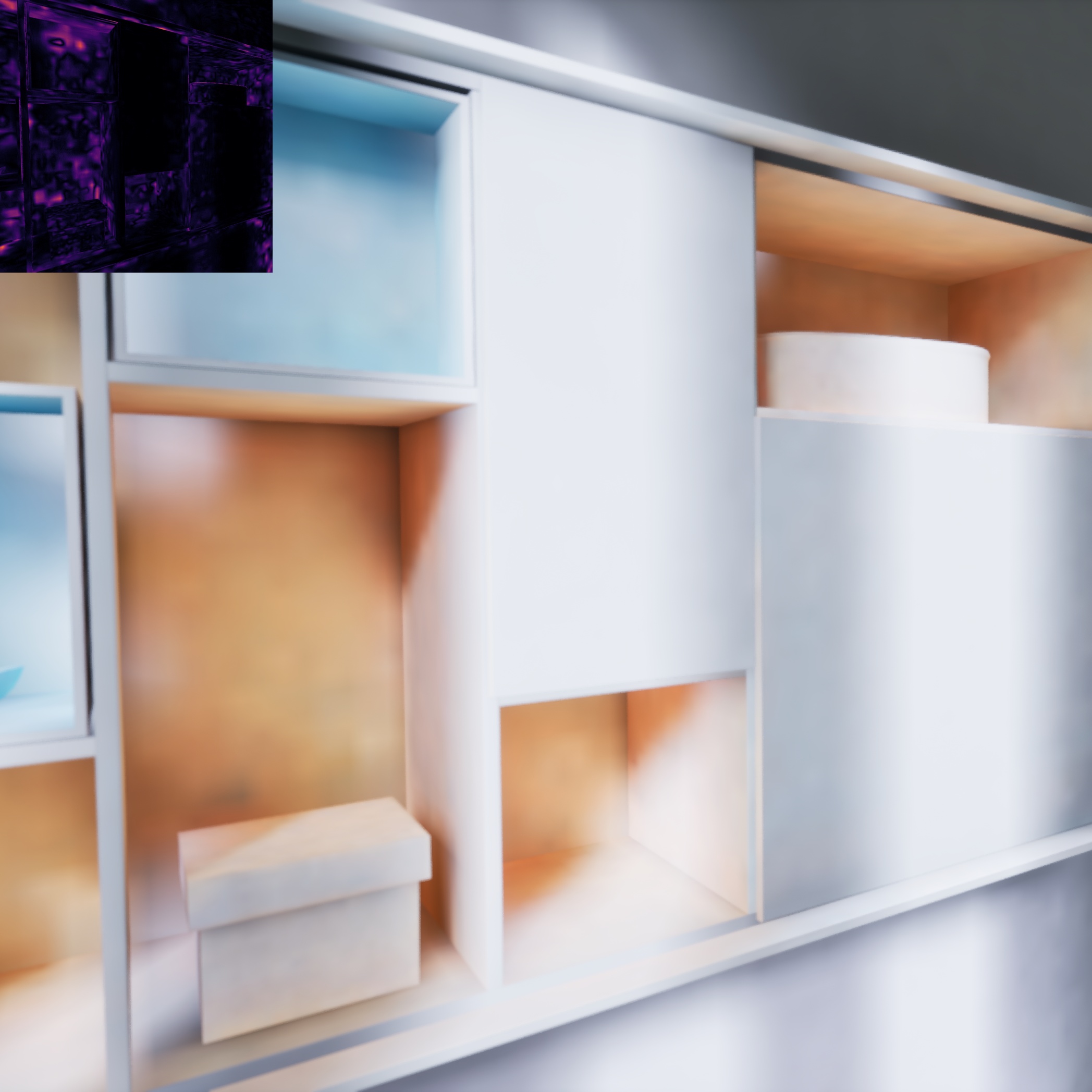} &
\includegraphics[width=0.16\linewidth]{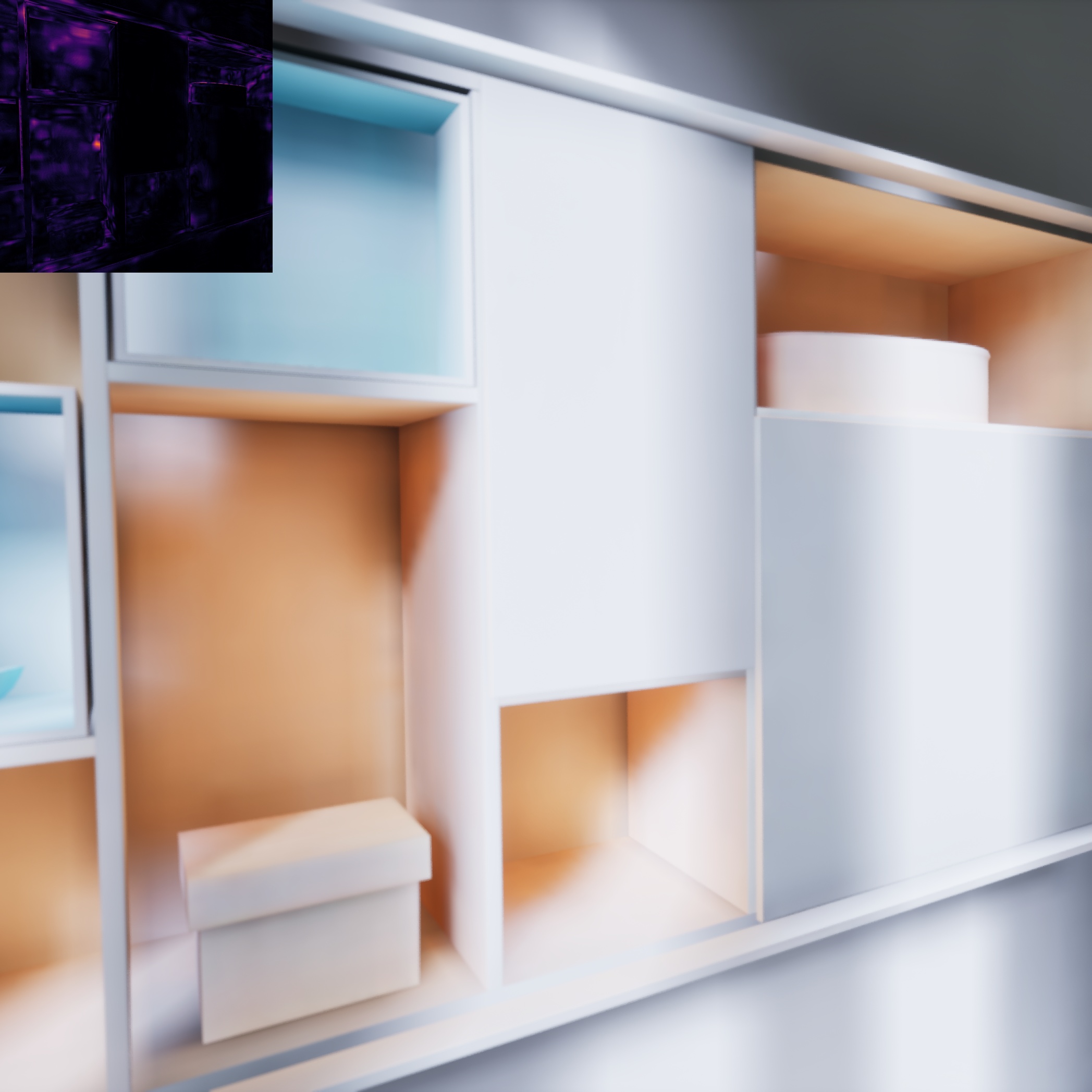} &
\includegraphics[width=0.16\linewidth]{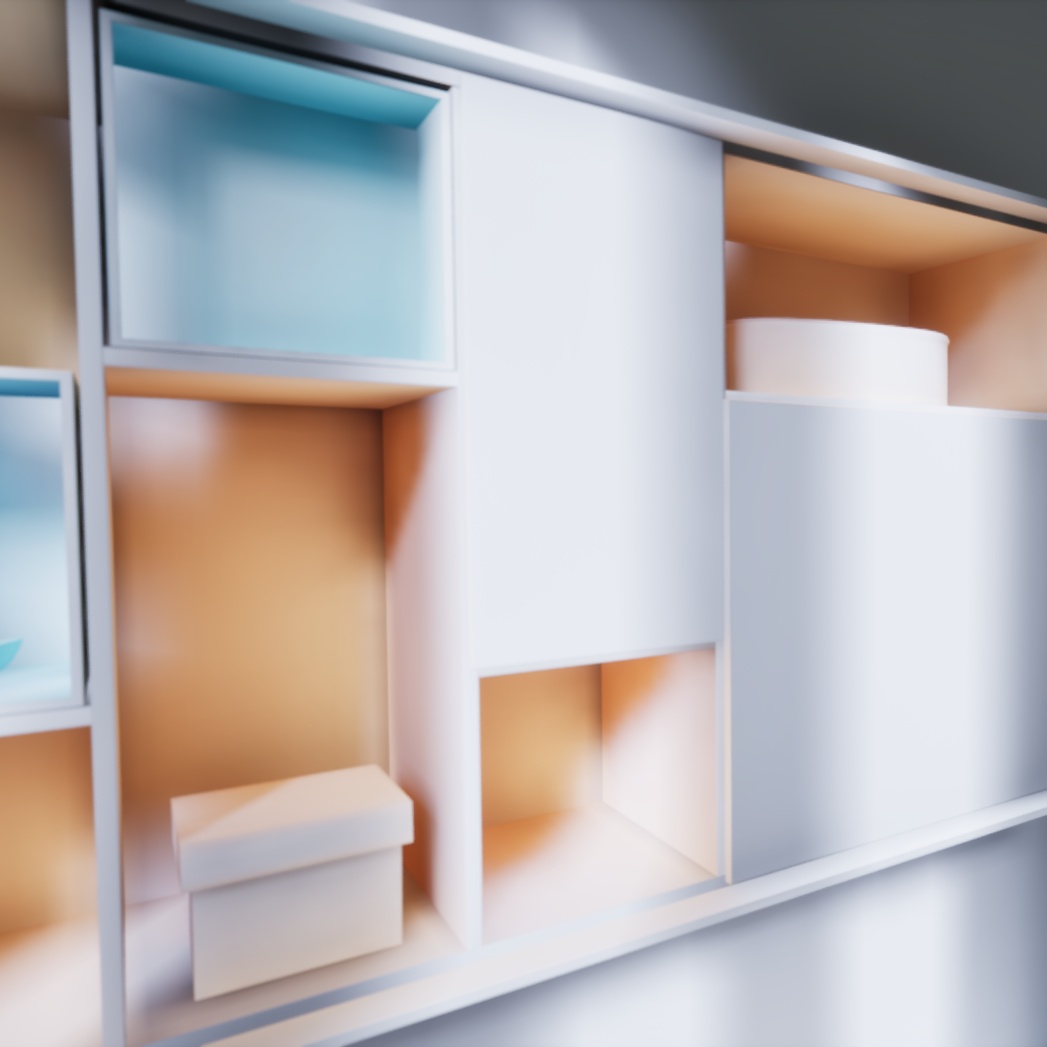}  \\
{\scriptsize BPP($\downarrow$), PSNR($\uparrow$), SSIM($\uparrow$)} &
{\scriptsize 1.00, 15.18 dB, 0.667} &
{\scriptsize 0.78, 38.50 dB, 0.998} &
{\scriptsize \textbf{0.71, 41.07 dB, 0.999}} &
{\scriptsize Room} \\
\includegraphics[width=0.16\linewidth]{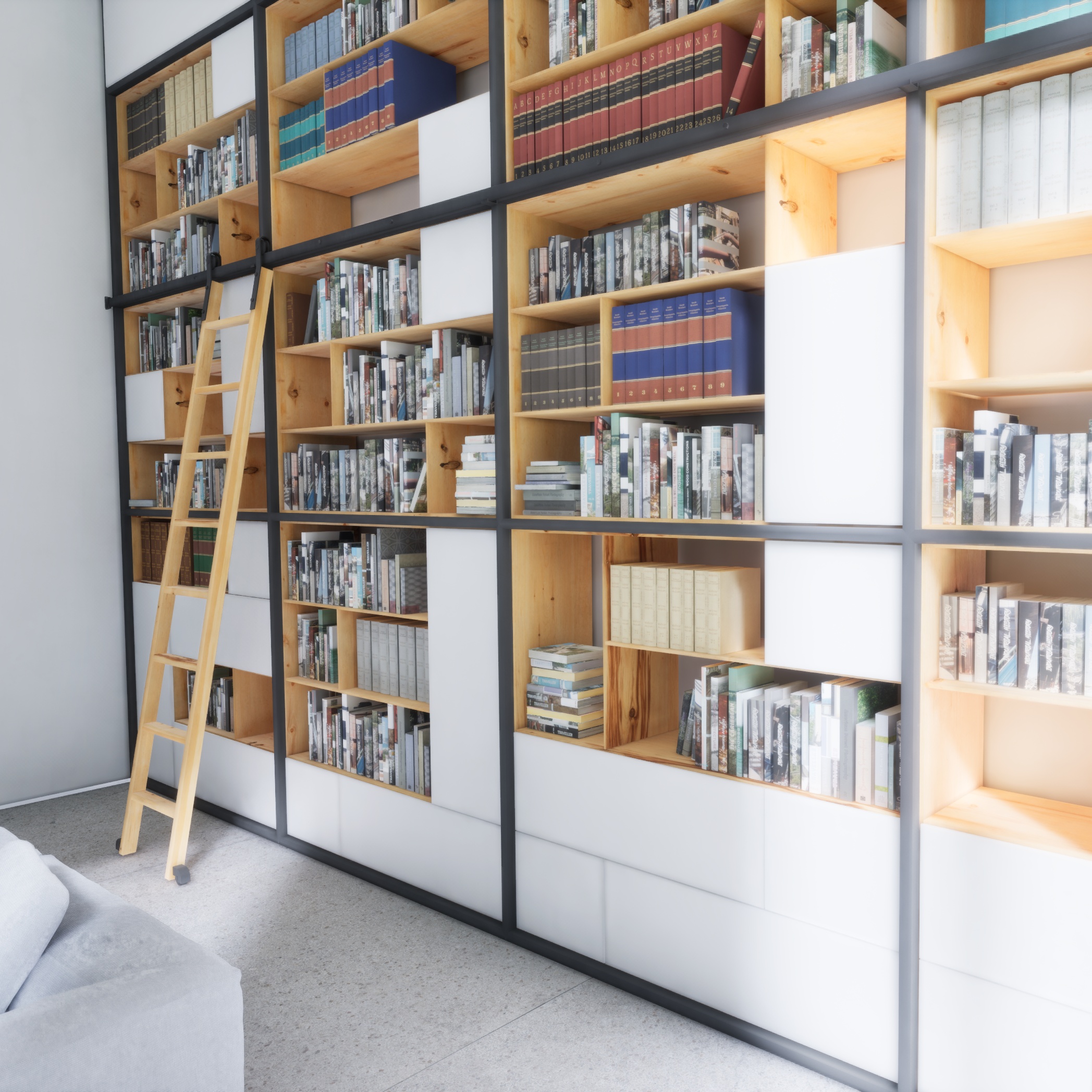} &
\includegraphics[width=0.16\linewidth]{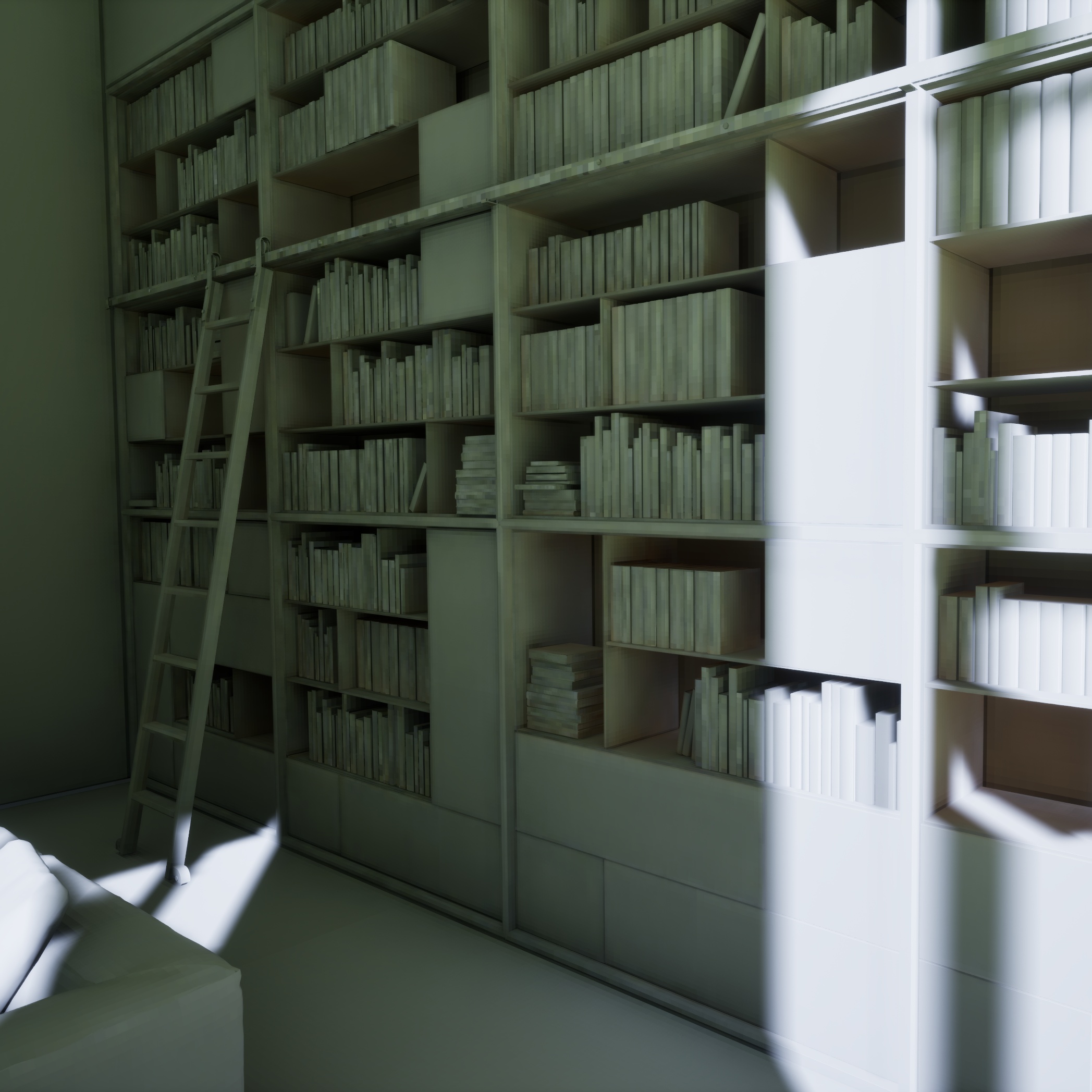} &
\includegraphics[width=0.16\linewidth]{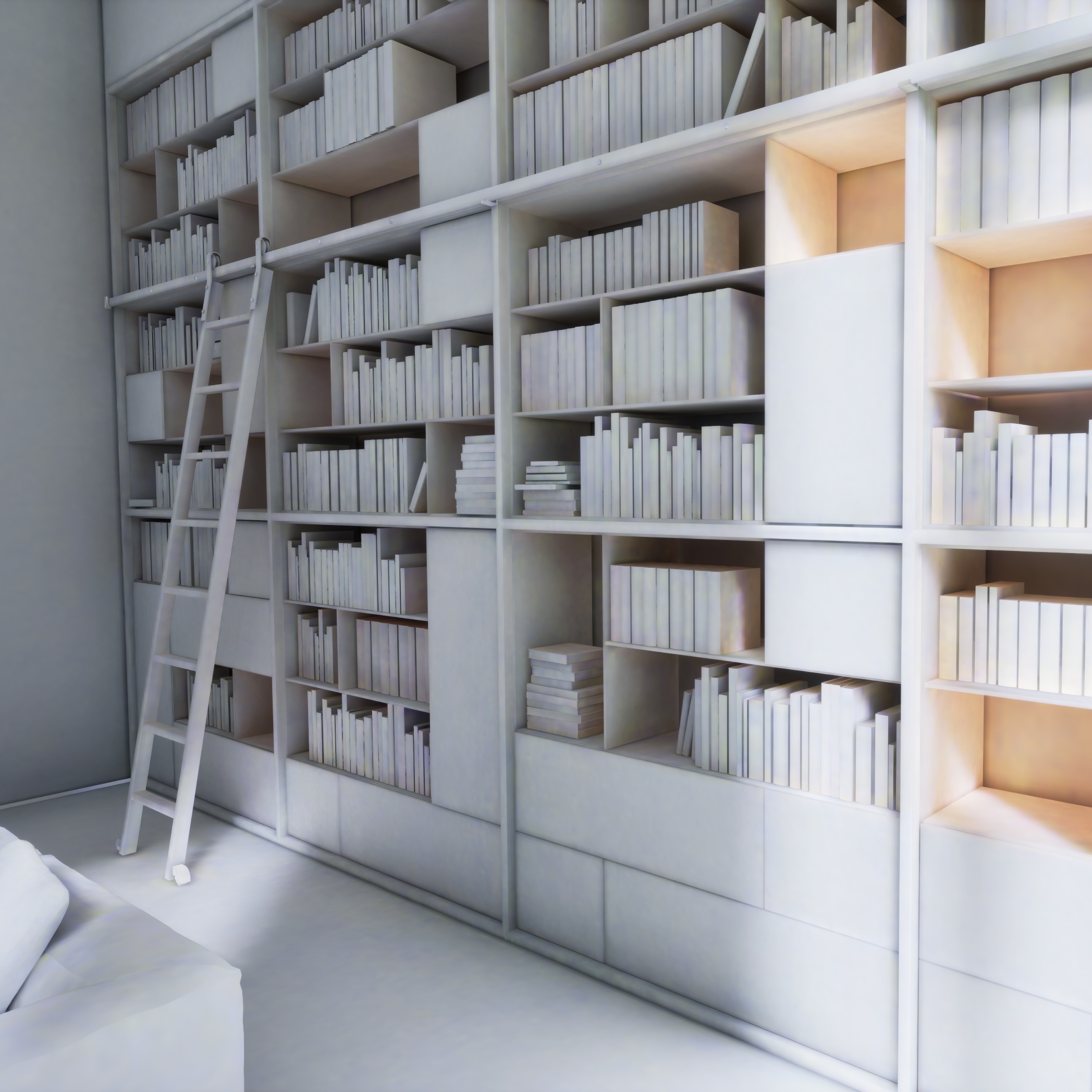} &
\includegraphics[width=0.16\linewidth]{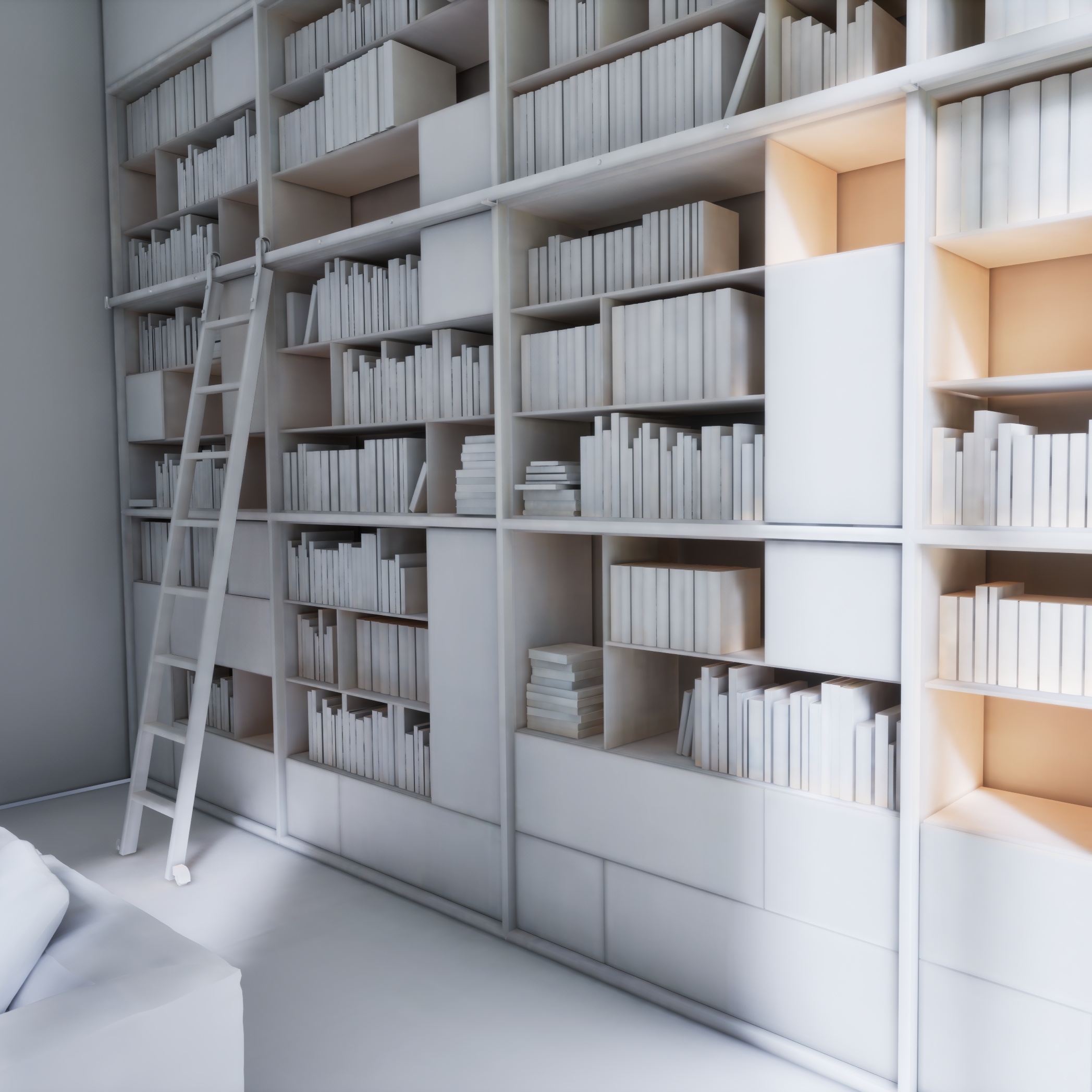} &
\includegraphics[width=0.16\linewidth]{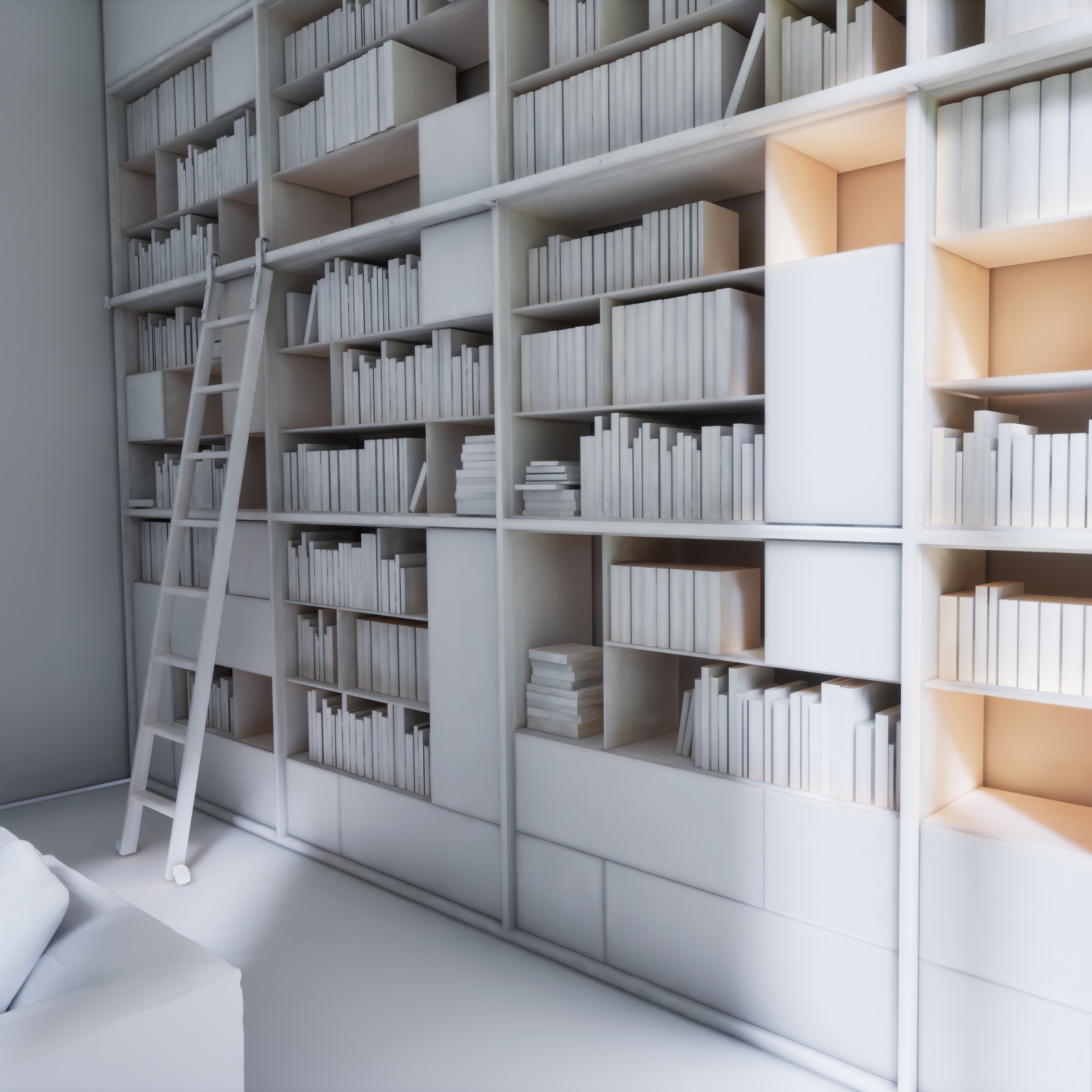} \\
{\scriptsize BPP($\downarrow$), PSNR($\uparrow$), SSIM($\uparrow$)} &
{\scriptsize 1.00, 10.37 dB, 0.688} &
{\scriptsize 0.78, 39.97 dB, 0.999} &
{\scriptsize \textbf{0.71, 42.25 dB, 0.999}} &
{\scriptsize Room} \\
\end{tabular}
 }
\caption{Additional comparisons of rendering quality. Compared to PRT~\cite{sloan2023precomputed}, our method better captures global illumination tone variations, such as multi-bounce interreflections. Compared to NTC~\cite{Vaidyanathan_2023}, it reconstructs lightmaps with higher fidelity and noticeably less noise.}
\label{fig:rendered}
\end{figure*}


\end{document}